%% file: hartlep.tex
\documentclass[twocolumn,times,tighten]{aastex63}
\newcommand{\reducedlinewidth}{\linewidth}
\usepackage{color,ulem}
\usepackage{multirow}
\usepackage{amsmath}
\usepackage{tabularx}
\newcolumntype{L}[1]{>{\raggedright\arraybackslash}p{#1}}
\newcolumntype{C}[1]{>{\centering\arraybackslash}p{#1}}
\newcolumntype{R}[1]{>{\raggedleft\arraybackslash}p{#1}}

\received{2019 November 12}
\revised{2020 February 6}
\accepted{2020 February 13}

\submitjournal{The Astrophysical Journal}

\shorttitle{Cascade Model for Planetesimal Formation}
\shortauthors{Hartlep \& Cuzzi}

\begin{document}

\sloppy

\title{\Large Cascade Model for Planetesimal Formation by Turbulent Clustering}

\correspondingauthor{Thomas Hartlep}
\email{hartlep@baeri.org}

\author[0000-0002-5062-9507]{Thomas Hartlep}
\affiliation{\rm{Bay Area Environmental Research Institute, NASA Research Park, Moffett Field, CA 94035, USA}}
\affiliation{\rm{NASA Ames Research Center, Moffett Field, CA 94035, USA}}

\author{Jeffrey N. Cuzzi}
\affiliation{\rm{NASA Ames Research Center, Moffett Field, CA 94035, USA}}



\begin{abstract}
We use a newly developed cascade model of turbulent concentration of particles in   protoplanetary nebulae to calculate several properties of interest to the formation of primitive planetesimals and to the meteorite record. The model follows, and corrects, calculations of the primary planetesimal Initial Mass Function (IMF) by \citet{Cuzzietal2010}, in which an incorrect cascade model was used. Here we use the model of \citet{Hartlepetal2017}, which has been validated against several published numerical simulations of particle concentration in turbulence. We find that, for a range of nebula and particle properties, planetesimals may be ``born big", formed as sandpiles with diameters in the $10-100$\,km range, directly from freely floating particles. The IMFs have a modal nature, with a well-defined peak rather than a powerlaw size dependence. Predictions for the inner and outer nebula behave similarly in these regards, and observations of inner and outer nebula primitive bodies support such modal IMFs. Also, we present predictions of local particle concentrations on several lengthscales in which particles ``commonly" find themselves, which have significance for meteoritical observations of the redox state and isotopic fractionation in regions of chondrule formation. An important difference between these results, and those of \citet{Cuzzietal2010}, is that particle growth-by-sticking must proceed to at least the $1-$few\,cm radius range for the IMF and meteoritical properties to be most plausibly satisfied. That is, as far as the inner nebula goes, the predominant ``particles" must be aggregates of chondrules (or chondrule-size precursors) rather than individual chondrules themselves.

\end{abstract}




\section{Introduction} \label{sec:intro}

The formation of the first 100\,km size planetesimals -- primitive asteroids in the inner solar system, and Kuiper Belt Objects (KBOs, or Trans-Neptunian Objects -- TNOs) in the outer solar system -- remains one of the major unsolved problems in the standard planet formation scenario \citep{Johansenetal2015}.

{\it Incremental growth of planetesimals:} For decades, planetesimal formation was modeled as ``incremental growth" -- slow growth by sticking all the way from mm-size particles like those commonly seen in meteorites, up to 100\,km size where gravitational effects become important \citep[most recently][and references therein]{Weidenschilling2011,Weidenschilling2019}.
However, this process is successful only in the context of a protoplanetary nebula that is essentially nonturbulent, because it has been shown that if the nebula gas is turbulent, several sequential barriers to growth arise, due to bouncing, fragmentation, and rapid radial drift of mm-to-m size particles \citep{Braueretal2008,Zsometal2010,Birnstieletal2011,Estradaetal2016}. The most recent incremental growth models \citep{Weidenschilling2019} merely assume $15$\,m radius objects as an initial condition and focus on growth to larger objects. However, collisions between even $1-10$\,km size bodies due to gravitational scattering by density fluctuations in the turbulent gas, much like giant molecular clouds scatter stars in the galaxy, provide yet another erosion/fragmentation barrier for incremental accretion \citep{Idaetal2008,Gresseletal2012,OrmelOkuzumi2013}. This final barrier to  growth has not been included in the incremental (nonturbulent) models of \citet{Weidenschilling2011,Weidenschilling2019,SchlichtingSari2011} and \citet{KenyonBromley2012}.

{\it Turbulence:} After a long-lasting consensus in the 1970s and 80s that nebula turbulence was common, a perspective arose in the 1990s and 2000s that it could only be triggered by ``magnetorotational" instabilities (MRI), and then only in the low-density upper reaches of protoplanetary disk atmospheres \citep[e.g.,][and references therein]{Gammie1996}, leaving most of the nebula dynamically ``dead". More recently, in rapid succession, complications seemed to render the MRI inoperative even in upper regions of disks \citep[et seq.]{Bai2013}, just as other studies showed that turbulence can be triggered and maintained in most or all regions of the disk by one or more purely hydrodynamical processes \citep{Nelsonetal2013, Lyra2014, Turneretal2014, StollKley2014, Marcusetal2015}.  These hydrodynamical processes depend primarily on the thermal opacity that controls local temperature variations (mostly provided by particles); some operate in rapidly cooled (nearly isothermal) regions and some operate in slowly cooled (nearly adiabatic) regions, so in principle all nebula regions could be susceptible to one or more turbulent instabilities of this kind \citep{LyraUmurhan2019}. While our understanding of protoplanetary nebula turbulence will continue to evolve, a reasonable expectation seems to be that most of the nebula was mildly turbulent for extended periods.

The properties of turbulence can be captured by the turbulent Reynolds number $Re = \nu_T/\nu_m = U_L L/\nu_m$, where the turbulent viscosity $\nu_T=U_L L$, $U_L$ and $L$ are the scales containing most of the energy and with the highest velocities, and $\nu_m$ is the gas molecular viscosity. For most numerical simulations, $Re \sim 10^4$ at most, while in the nebula, $Re \sim 10^7 - 10^9$. More familiar to astrophysicists is the $\alpha$ notation, in which the turbulent viscosity $\nu_T = \alpha c H$, with $c$ the sound speed and $H$ the gas vertical scale height; thus $Re = \alpha c H / \nu_m$. Typical values of $\alpha$ associated with the new hydrodynamical instabilities are $10^{-4}-10^{-3}$. The uncomfortable current paradox is that these moderate values, which characterize most of the disk gas within a scale height or two of the midplane and are the most relevant for the evolution of particles and planetesimals, seem to be too small to actually produce observed disk accretion rates and evolve disks away on the observed disk lifetimes of a few million years, there may be other processes that contribute, such as magnetically driven disk winds \citep[see, e.g.,][for more discussion]{Turneretal2014}.

{\it Shortcuts to big planetesimals:} To avoid the barriers to incremental growth posed by turbulence, the idea that planetesimals can be ``born big" due to collective processes that transform dense clumps of small particles directly into 100\,km size objects became popular about a decade ago \citep{Johansenetal2007, Cuzzietal2008, Morbidellietal2009BornBig, Johansenetal2015}. To some extent, these scenarios can be grouped into two different pathways (see below), but they may, in the end, be two aspects of the same process. The relative importance of the two pathways is strongly dependent on how turbulent the nebula was.

{\it Streaming Instability:} One pathway is represented by the popular ``streaming instability" or SI \citep{GoodmanPindor2000, YoudinGoodman2005, Jacquetetal2011, SquireHopkins2018}, in which (to simplify the effect drastically) regions that are overdense in solids can drive the local gas towards Keplerian velocity. This diminishes the headwind drag felt by the particles, which drift inwards more slowly. These dense regions then continue to accrete more rapidly drifting particles from less dense, surrounding regions in a sort of peloton effect, until the denser region becomes gravitationally bound or even unstable to collapse. A number of numerical simulations illustrate this effect in action \citep{Johansenetal2007, Balsaraetal2009, Carreraetal2015, Yangetal2017, Simonetal2017}.

The main challenge facing SI is the precondition of a sizeable local region where the ratio of particle volume mass density $\rho_p$ to gas volume mass density $\rho_g$ exceeds unity (enhancement of about 100 times over cosmic abundance). Sometimes the requirement is obscured by emphasizing the vertically integrated {\it surface mass density}, but the key parameter is really the local {\it volume mass density} \citep{YoudinGoodman2005, Johansenetal2015}. Sometimes one hears that SI can occur in so-called ``stratified turbulence", which means the thin layer of extremely weak turbulence ($\alpha \sim 10^{-6}$) generated around a settled midplane solids layer in a globally {\it laminar} nebula  \citep{Weidenschilling1980, Cuzzietal1993, BaiStone2010}; however, a {\it globally} turbulent nebula with $\alpha = 10^{-4}-10^{-3}$ is a clearly distinguishable and much more challenging situation.

For example, \citet{Estradaetal2016} showed that, for the above values of $\alpha$ and using the most realistic lab-based prescriptions  for incremental growth by sticking in plausible weak global turbulence, particles -- even the rare ``lucky particles" \citep{Windmarketal2012, Garaudetal2013} or {\it sticky} icy particles -- can't grow large enough to settle into a midplane layer where $\rho_p/\rho_g$ is large enough for SI to occur as formulated, at least until well after $2\times10^5$ years after the first solids formed. By this time however, there is good evidence that sizeable planetesimals and perhaps even a proto-jovian core had formed \citep{Kruijeretal2017}. More recent work \citep{Umurhanetal2019} derives the necessary conditions and growth times for SI in detail, and shows {\it why} all numerical simulations to date in which SI occurs had to assume either much lower levels of global turbulence than suggested by recent theoretical work (above), or much larger particles than self-consistently allowed by recent incremental growth models. There is nothing wrong with the SI physics -- the issue is one of using appropriate, self-consistent initial conditions.

{\it Preferential concentration or turbulent clustering of particles:} The work discussed here follows a second pathway. It has been known since the 1990s, from laboratory experiments and numerical simulations, that particles of certain sizes are locally concentrated by homogeneous, isotropic turbulence to different degrees \citep{SquiresEaton1991, WangMaxey1993, Hoganetal1999, HoganCuzzi2001, HoganCuzzi2007, Becetal2007, Becetal2010, Calzavarinietal2008, Toschietal2009, Panetal2011, BraggCollins2014, Irelandetal2015, GustavssonMehlig2016}; for a brief review see \citet{Johansenetal2015}. The exact mechanisms leading to this effect are still under debate \citep{BraggCollins2014, Irelandetal2015}. Also, until recently, even how the concentration varies with spatial scale and particle stopping time has been unresolved \citep[see below]{HoganCuzzi2007, Becetal2007, Panetal2011}.

The so-called inertial range of turbulence is key to the process under discussion. The inertial range extends from the smallest lengthscales comparable to the dissipation or Kolmogorov scale $\eta$ where molecular viscosity becomes effective, to the largest, energy-containing scale $L$ where the eddy velocity is $U_L$. This range of scales is related to the Reynolds number as $L/\eta \sim Re^{3/4}$. Within the inertial range, scale-invariant effects are seen as energy cascades losslessly from the largest scales to the smallest scales, where it is dissipated by molecular viscosity. The energy spectrum of high-$Re$ turbulence is usually taken as the Kolmogorov scaling, in which the turbulent kinetic energy at scale $\ell$ per unit scale length $E_{\ell}= (U_L^2/2L)(L/\ell)^{1/3}$, and from this the characteristic eddy velocity $U_{\ell}=(2\ell E_{\ell})^{1/2}$ and frequency $\omega_{\ell}= U_{\ell}/\ell = \Omega(L/\ell)^{2/3}$ easily follow \citep{TennekesLumley, Cuzzietal2001, Cuzzietal2010}. A good example of an inertial range is shown in \citet[their Figure~2]{Becetal2010}, from which dataset the cascades used here were derived; see also \citet[chapter 8]{TennekesLumley} for more discussion.

The original work on ``turbulent concentration" or ``turbulent clustering" (in the astrophysics context) made the assumption that a certain kind of scale invariance observed in atmospheric observations and numerical simulations of dissipation of turbulent kinetic energy could be used on particle concentrations throughout the inertial range \citep{Hoganetal1999, Cuzzietal2001, HoganCuzzi2007, Cuzzietal2008, Cuzzietal2010, Chambers2010}. This assumption was used to derive a {\it cascade model} to describe the {\it statistical properties} of particle concentration in turbulence, along the lines of established cascade models for the statistical properties of kinetic energy {\it dissipation} in turbulence \citep[][see Section~\ref{sec:model} below]{MeneveauSreenivasan1987, SreenivasanStolovitsky1995}.

The primary accretion scenario of \citet{Cuzzietal2008, Cuzzietal2010}, also semi-independently derived by \citet{Chambers2010}, applied a combination of the \citet{HoganCuzzi2007} cascade model, and certain thresholds (Section~\ref{sec:model:thresholds}) to predict the {\it Initial Mass Function (IMF)} of planetesimals in the inner and outer solar nebula, finding IMFs with distinct modes at planetesimal diameters between tens and hundreds of km for a range of plausible parameter values.  An additional feature of these primary accretion scenarios was the prediction of the total planetesimal {\it mass} produced in both the inner and outer solar system over the likely planetesimal formation period, taken to be about 2 Myr. For a range of plausible parameters, the agreement was not bad but the results were parameter dependent and consequently not highly predictive. In this paper we repeat that general approach (Section~\ref{sec:model}), but correct what turns out to have been an incorrect cascade model.

One of the conclusions of this early work was that the most effectively concentrated particles have a gas drag stopping time $t_s$ comparable to the Kolmogorov (smallest) eddy overturn time $\tau_\eta$. The aerodynamic properties of particles are captured by their {\it Stokes number $St_{\ell}$}, which can be referenced to any general lengthscale $\ell$ using the eddy time $\tau_\ell$ at the scale $\ell$: $St_\ell = t_s /\tau_\ell$. Thus, it was believed that the optimum Stokes number was $St_\eta = 1$, and it was especially intriguing that such particles seemed to have the typical size (roughly $0.1-1$\,mm diameter) of meteoritic ``chondrules", a ubiquitous but poorly understood constituent of primitive chondrites \citep{Cuzzietal2001}. Several studies since then have found that particles of different sizes become optimally concentrated at different lengthscales \citep{Becetal2007, ZaichikAlipchenkov2003, ZaichikAlipchenkov2009, Hartlepetal2017}. In this work we find that nebula particles significantly larger than individual chondrules are needed to lead to the formation of planetesimals with roughly $100$\,km diameter.

{\it The planetesimal IMF itself} is not perfectly known, of course (see below for more discussion). Observations of the number $N$ of objects in the current population at diameter $D$ are usually presented as cumulative distributions $N(>D)$, which are often approximated by powerlaws of different slopes. Flat slopes at small sizes are separated from steep slopes at large sizes by a so-called ``knee" in the cumulative distribution. While some studies focus on the specific values of these slopes, they can include confusing effects such as fragmentation debris on the small-size end and small-number statistics on the large-size end both for the asteroids and the KBOs.  Fortunately however, because the flat cumulative powerlaw slope at small sizes has most of the mass at its large end, and the steep powerlaw at large sizes has most of its mass at the small end, most of the mass lies in objects with diameters at the knee.  So while the details of all of these powerlaws are uncertain, the location and meaning of the knees remain robust: they point to modal values of diameter containing most of the mass (with a slightly larger modal value if {\it weighted} by mass). This is discussed for the asteroids by \citet{Cuzzietal2010}, who showed an incremental (not cumulative) distribution that has a well-defined {\it  mass} mode (the diameter where most of the mass lies, somewhat larger than the straight modal diameter) at slightly more than 100\,km diameter.

For the asteroids, Bottke et al (2005) claim the knee (or the mass mode) represents the primary or ``fossil" asteroids, with nearly all smaller objects being fragments from subsequent collisions. The more recent WISE data for asteroids \citep{Masieroetal2011}, and recent removal of background objects to refine the distribution of primordial objects \citep{Delboetal2017}, also give results consistent with a mass mode around 80-100\,km diameter (unweighted by mass, as we report here). For the KBOs, the data are mostly given in brightness (visual magnitudes), and also usually shown as cumulative powerlaw distributions \citep{Bernsteinetal2004,Morbidellietal2009KBOs}. Like the asteroids, most of the mass is found at a ``knee" at between $20-100$\,km diameter. In the end, it seems the data are most clearly telling us to seek a primary accretion process that creates ``most of the mass" with a fairly well-defined modal diameter - {\it not} a powerlaw. Indeed it was one of the intriguing results of \citet{Cuzzietal2010} that both in the inner and outer solar system, turbulent concentration/clustering led naturally to distributions with such modal shapes, {\it not} powerlaws, and the modal diameter was in the range observed.

{\it However,} a major problem arose with this line of study when subsequent direct numerical simulations of particle concentration at higher $Re$ \citep{Panetal2011} disagreed with the initial cascade model predictions \citep{HoganCuzzi2007, Cuzzietal2008} that underlay the IMF calculations of \citet{Cuzzietal2010}, casting the validity of their results into question.  Therefore, applying and extending the current state of the art to the protoplanetary nebula has been frustrating. This paper and its companion \citet{Hartlepetal2017} attempt to place at least the latter question on firmer ground.

{\it New and improved Turbulent Concentration model:} To explore this issue, \citet{Hartlepetal2017} revisited the statistics of particle concentration in turbulence, using careful statistical analyses of 3D numerical simulations of particles with a range of $t_s$, in homogeneous, isotropic turbulence. These simulations had been run at the highest values of $Re$ available to date, and posted online by \citet{Becetal2010}. \citet{Hartlepetal2017} found that the scale invariance inferred and extended to the nebula by \citet{Cuzzietal2008, Cuzzietal2010} was an artifact of the dissipation range covering most of the spectral range in the (lower-$Re$) direct numerical simulations used by \citet{HoganCuzzi2007} to derive the cascade, and was {\it not} applicable to the inertial range.  Those results may ultimately be useful for studies {\it directed to} the dissipation range of scales. Most importantly though, \citet{Hartlepetal2017} found another kind of scale invariance that {\it is} valid in the inertial range. Specifically, they found a concentration {\it function} that is scale-invariant when expressed in terms of the particle Stokes number at each eddy scale $\ell$: $St_{\ell} = t_s /\tau_\ell$.

In this paper, we apply the scale-invariant ``universal curve" of \citet{Hartlepetal2017} to recalculate planetesimal IMFs, generalizing the approach of \citet{Cuzzietal2010} (see next section). The new results produce similar-looking ({\it ie} modal, not powerlaw) planetesimal IMFs with much the same range of diameters and planetesimal mass production rates, {\it but} critically, only if starting with {\it larger particles than before}. In the inner nebula, this means particles with at least the mass of more than $\sim 10^4$ chondrules -- cm-size aggregates of chondrules with $St_\eta$ of roughly $10$ to few $100$. In Sections~\ref{sec:trends} and \ref{sec:conclusions} we discuss recent evidence for such aggregates.

As we will see, the planetesimal IMFs depend on the low-probability, high-concentration parts of the particle concentration Probability Distribution Functions (PDFs). In this paper we will also show the PDFs in {\it more probable} ranges of concentration (conditions in which particles spend most of their time) as functions of lengthscale, turbulent intensity, and particle size. These ``typical" local particle densities are important for understanding the mineralogical properties of the once-molten chondrules that dominate primitive meteorites -- properties that are influenced by the local solids densities in the regions where the chondrules were melted and cooled.

Finally, we note that something like one of these processes, or perhaps a combination of them \citep{Cuzzietal2017a}, must produce the sizeable ``seeds" that are required to trigger the second-stage sweepup process called pebble accretion  \citep{OrmelKlahr2010, LambrechtsJohansen2012}; that is, pebble accretion is {\it unimportant} in even very weak turbulence, until planetesimals form that are larger (200\,km diameter) than the current ``fossil" asteroids we see today \citep{VisserOrmel2016}. It is intriguing that one, or maybe both, of the two pathways  mentioned above may provide an almost ideal environment in which pebble accretion can {\it extend} growth rapidly to embryo or planetary core mass - a small number of large seeds embedded in a sea of pebbles, with no intermediate size objects to complicate the accretion. For this reason we call the first formation of 100\,km diameter objects directly from small, freely floating nebula particles {\it primary accretion}, and sometimes refer to these collective pathways that jump over the various barriers as {\it leapfrog} processes. A number of properties of primitive chondrite parent bodies argue in favor of this kind of primary accretion \citep{Johansenetal2015}.

\section{Model} \label{sec:model}

Our model of planetesimal formation is based on the observation that turbulent clustering produces spatial and temporal fluctuations in the volume density of solids carried by the gas, as well as in the gas vorticity (enstrophy).
Under the right conditions in the protoplanetary nebula, particle densities can reach values high enough for such clusters to become gravitationally bound, and then to sediment under their own self-gravity into actual planetesimals, while resisting disruption by ram pressure or local vorticity.

The model works in the following way: A statistical model of turbulent clustering is used to predict the joint PDF of particle density and enstrophy in the protoplanetary disk at all relevant spatial scales.
Simple thresholds derived on physical grounds then describe the various disruptive effects, and are used to identify the part of the joint PDF where planetesimal formation is possible.
Integration of the PDF above the thresholds (both are scale {\it dependent}) yields a size distribution of planetesimals formed (Initial Mass Function, or IMF), assuming each sedimenting clump forms a single planetesimal\footnote{
However, a clump might fragment into some number of smaller sub-clumps and ultimately smaller objects. This might happen due to strong density substructure within the overall clump, or due to fission by a rotating clump, in which case the assumption of “one object per clump” would fail. The degree to which this might happen will depend on $St_L$ and local conditions, and is hard to foresee, but is well worth future study.}, and the rates at which they are produced.
The present work improves on~\cite{Cuzzietal2010} and similar work \citep{Chambers2010}, from which we carry over the threshold description but use a new, more sophisticated and more realistic cascade model for the turbulent clustering statistics. \citet{Hopkins2016simple,Hopkins2016IMF} has presented a similar approach, which is more analytical and provides useful insight into the process, but makes simplifying assumptions that limit its application.

\begin{table}[t]
\footnotesize
\centering
\caption{Frequently used symbols and parameters.} \label{tab:SymbolsAndParameters}
\begin{tabular}{ll}
\hline
\hline
Symbol & Description \\
\hline

$A/A_o$ \dotfill & Solids enhancement factor (Section~\ref{sec:model:nebula}) \\
$F_{\tilde\beta}$ \dotfill & Pressure gradient factor (Section~\ref{sec:model:nebula}) \\
$F_\rho$ \dotfill & Gas density enhancement factor (Section~\ref{sec:model:nebula}) \\
$H$ \dotfill & Vertical density scale height (Section~\ref{sec:model:nebula})  \\
$L, \tau_L$ \dotfill & Largest turbulent length and time scales (Section~\ref{sec:intro})    \\
$Re$ \dotfill & Reynolds number (Section~\ref{sec:intro}) \\
$S$ \dotfill & Normalized gas enstrophy (Section~\ref{sec:model:enstrophy}) \\
$St_L$ \dotfill & Stokes number based on $\tau_L$ (Section~\ref{sec:model:simulation})  \\
$St_\eta$ \dotfill & Stokes number based on $\tau_\eta$ (Section~\ref{sec:intro}) \\
$t_s$ \dotfill & Particle stopping time (Section~\ref{sec:intro}) \\
\hline
$\alpha$ \dotfill & Turbulence intensity (Section~\ref{sec:intro}) \\
$\tilde\beta$ \dotfill & Pressure gradient (Section~\ref{sec:model:nebula}) \\
$\eta, \tau_\eta$ \dotfill & Kolmogorov length and time scales (Section~\ref{sec:intro})    \\
$\Phi$ \dotfill & Mass loading factor $=\rho_p / \rho_g$ (Section~\ref{definePhi}) \\
$\rho_g$ \dotfill & Gas density (Section~\ref{sec:model:nebula}) \\
$\rho_p$ \dotfill & Mass density in particles (Section~\ref{sec:intro}) \\
$\rho_s$ \dotfill & Mass density of collapsed planetesimals (Section~\ref{sec:model:integration}) \\
$\sigma$ \dotfill & Surface gas density (Section~\ref{sec:model:nebula}) \\

\hline
\end{tabular}
\end{table}

\subsection{Statistical Model of Turbulent Clustering} \label{sec:model:clustering}

\citet{Hartlepetal2017} have developed a new cascade model to describe the statistics of particle concentrations and enstrophy in particle-laden flows.
In a cascade model, a partition function or {\sl multiplier} $0 \le m \le 1$ describes how a quantity ${\cal P}$ in some volume is partitioned into subvolumes.
Cascade models are widely used in studying the statistical properties of turbulent dissipation, and can take several forms \citep{MeneveauSreenivasan1987,Meneveauetal1990,SreenivasanStolovitsky1994}. In particular, \citet{Hartlepetal2017} considered a binary cascade where ${\cal P}$ is partitioned into two equal sized subvolumes\footnote{Cascades with different subdivisions were studied by \citet{SreenivasanStolovitsky1994} who found two- and three-subvolume cascades to yield nearly identical results. Higher numbers of sub-partitions however led to mathematical problems.}, and the process is continued to increasingly smaller subvolumes.
In turbulence, the multipliers are stochastic quantities with PDFs that can be approximately described by $\beta$-distribution functions
\begin{equation}
\label{Eqn:BetaPDF}
  f_{\cal P}(m;\beta_{\cal P}) = \left( m - m^2 \right)^{\beta_{\cal P}-1} \frac{\Gamma(2\beta_{\cal P})}{2\Gamma(\beta_{\cal P})}
\end{equation}
where $\Gamma$ denotes the Gamma function.
The parameter $\beta_{\cal P}$ describes the width of the distribution, with small $\beta_{\cal P}$ corresponding to wide distributions causing strong spatial intermittency\footnote{A property such as particle density or turbulent dissipation is {\it intermittent} when its spatial distribution becomes increasingly variable, rather than increasingly well-defined, going to smaller scales.} in property ${\cal P}$, and {\sl vice versa}.

\subsubsection{Particle Concentration Model} \label{sec:model:concentration}

Analyzing direct numerical simulations (DNS) of particle-laden, homogeneous, isotropic turbulence, \citet{Hartlepetal2017} found that the distribution functions $f_{\Phi}$ giving the partition fractions or multipliers for the particle number densities, and correspondingly the mass loading factor $\Phi= \rho_p / \rho_g \label{definePhi}$, have scale-invariant properties within the inertial range of turbulence.
In particular, the width parameter $\beta_\Phi$ is solely a function of $St_\ell = t_s / \tau_\ell$, the Stokes number based on the eddy time $\tau_\ell$ at scale $\ell$. $\beta_\Phi (St_\ell)$ forms a ``universal curve'' that can be approximately described by a sum of two power laws:
\begin{equation}
   \beta_\Phi(St_\ell) \approx \left(  \left( \frac{St_\ell}{a_1} \right)^{b_1}  +   \left( \frac{St_\ell}{a_2} \right)^{b_2}   \right) \beta_{\Phi, \textrm{min}},
   \label{Eqn:BetaModel}
\end{equation}
with parameters $a_1$, $a_2$, $b_1$, $b_2$ determining the slopes and positions of the power laws, respectively, and $\beta_{\Phi, \textrm{min}}$ setting the minimum value.
The curve reaches large values of $\beta_{\Phi}$ at low and high $St_\ell$ (narrow PDF, little clustering) and a minimum (widest PDF, strong clustering) at intermediate values of $St_\ell$.

This result opens the possibility of extending the model derived from numerical simulations at moderately low $Re$ to the conditions in the protoplanetary nebula, assuming the scale-invariant properties continue to hold for the much higher nebula $Re$.
This is not an unreasonable assumption. The observed scale-invariance (in the fashion described above) extends through the not-insignificant inertial range of the 3D simulations it is based on \citep{Becetal2010}. Multiplier distributions for turbulent dissipation have been found scale-invariant over far wider ranges of scales in the inertial range \citep{Meneveauetal1990, SreenivasanStolovitsky1994}; specifically, see Figure~1 of \citet{Hartlepetal2017}. In addition, \citet{Hartlepetal2017} show remarkably good agreement between their cascade model results and independent, analytical model predictions for particle concentration in an infinitely wide inertial range \citep{ZaichikAlipchenkov2003, ZaichikAlipchenkov2009}, and give a discussion of the implications. That is, application of the universal function described above is not an {\it extrapolation} to larger $Re$; it is a {\it property} that is thought to be valid throughout the inertial range at arbitrary $Re$.

For our present model, we use Equation~\ref{Eqn:BetaModel} with the asymptotic parameter values at small scales based on \citet{Hartlepetal2017}:
\begin{align}
   a_1 &= 0.15, & a_2 &= 0.45, &
   b_1 &= -1.2, & b_2 &= 1.55.
   \label{Eqn:AsymptoticParameters}
\end{align}
The functional form we adopt here is slightly different from that used by \citet{Hartlepetal2017} to best fit their DNS results. As they discuss,
the largest spatial scales are not yet in the inertial range and some scale dependence is found even when using the $St_\ell$ scaling. In \citet{Hartlepetal2017}, this dependence appeared in $a_2$ and $b_2$ but we have captured it here in the prefactor $\beta_{\Phi,\textrm{min}}$, which itself varies slowly with scale until some small scale $\ell$ = L/16, below which we assume full scale independence with a constant value.
This is in fair agreement with the scale at which even possibly anisotropic turbulent structures at the forcing scale ``return to isotropy'' \citep{KatoYoshizawa1997}.
For $\beta_{\Phi, \textrm{min}}$ we use the expression
\begin{align}
  \beta_{\Phi, \textrm{min}} = \left\{
     \begin{aligned}
        & \beta_1 \left( \frac{\ell}{\ell_1} \right)^\frac{\log{\beta_2/\beta_1}}{\log{\ell_2/\ell_1}} & \forall  \, \ell \ge L/16 \quad \\
        & \left.\beta_2 \right. & \forall \,  \ell < L/16 \quad
     \label{Eqn:BetaMinParticles}
     \end{aligned}
     \right.
\end{align}
with parameters
\begin{align}
   \beta_1 &= 11, & \beta_2 &= 3, &
    \ell_1 & = L/2, & \ell_2 &= L/16. &
   \label{Eqn:BetaMinParametersParticles}
\end{align}
The parameter $\beta_1$, which represents the value of $\beta_{\Phi, \textrm{min}}$ at the upper scale $\ell_1$, comes from a fit to the DNS data in~\cite{Hartlepetal2017}.

It should be noted that the DNS simulations used for developing this model did not take into account potential backreactions of the particles onto the gas.
However, at large mass loading (large values of $\Phi$) such backreactions do become important; they effectively shut down the turbulent concentration effect and limit how high $\Phi$ can grow through the turbulent concentration mechanism.
\cite{HoganCuzzi2007} found this limit to be approximately $\Phi=100$ or so but more work needs to be done to refine this.
We account for this limit separately when computing planetesimal Initial Mass Functions (Section~\ref{sec:model:integration}).

\subsubsection{Enstrophy Model} \label{sec:model:enstrophy}

Fluid enstrophy (the volume integral of the squared vorticity $\left|\nabla\times \vec{u}\right|^2 = (\partial_i u_j) (\partial_i u_j) - (\partial_i u_j) (\partial_j u_i)$, with $\vec{u}$ being the fluid velocity vector) crucially affects the stability of particle clusters through the threshold description (Section~\ref{sec:model:thresholds}), and needs to be modeled to determine whether or not a particle cluster can collapse to form a planetesimal.
In particular, the thresholds are written in terms of a quantity $S$ which is the enstrophy normalized by the mean enstrophy at any given scale.
~\cite{Hartlepetal2017} analyzed enstrophy multipliers in DNS data and found their width parameter $\beta_S$ to decrease with decreasing spatial scale, reaching an asymptotic value $\beta_S \approx 4$ below $\ell/L \approx 50$.

Extensive atmospheric studies at much higher Reynolds numbers \citep{MeneveauSreenivasan1987,Meneveauetal1990, SreenivasanStolovitsky1994} have shown that multipliers for {\it dissipation} $\epsilon$ are fit by $\beta_\epsilon \approx 3$ from $\ell/L \approx 1/16$ all the way through the entire inertial range.
While these studies did not address enstrophy {\sl per se}, \citet{Hartlepetal2017} note that since enstrophy is at least as intermittent, if not more, as dissipation \citep[see e.g.,][]{Chenetal1997}, $\beta_\epsilon$ should be at least as small as that for dissipation. For the current model, we adopt the speculation by~\cite{Hartlepetal2017} for high Reynolds numbers, but make the conservative assuption that the asymptotic value for enstrophy is the same as for dissipation in the high-$Re$ limit. Specifically, we use the dependence
\begin{align}
  \beta_S = \left\{
     \begin{aligned}
        & \beta_1 \left( \frac{\ell}{\ell_1} \right)^\frac{\log{\beta_2/\beta_1}}{\log{\ell_2/\ell_1}} & \forall  \, \ell \ge L/16 \quad \\
        & \left.\beta_2 \right. & \forall \,  \ell < L/16 \quad
     \label{Eqn:BetaMinFluid}
     \end{aligned}
     \right.
\end{align}
with parameters
\begin{align}
   \beta_1 &= 22.65, & \beta_2 &= 3, &
    \ell_1 & = L/2, & \ell_2 &= L/16. &
   \label{Eqn:BetaMinParametersFluid}
\end{align}
We call the combination of Equations~(\ref{Eqn:BetaModel}-\ref{Eqn:BetaMinParametersFluid}) our ``conservative, speculative model" to distinguish it from the only slightly different form of the functional fit of \citet{Hartlepetal2017} to their specific DNS data, which are at much lower $Re$ than the actual nebula.

There is reason to expect {\it some} variation of even this ``universal" inertial range  function in going to the much larger $Re$ of the solar nebula. The logic is given in \citet[Section~V.B]{Hartlepetal2017}, and is based on properties derived from observations from the Earth's atmosphere at much higher $Re$ than the DNS simulations modeled by \citet{Hartlepetal2017}. These expectations support the slightly smaller asymptotic values of $\beta_2$ used in our ``conservative, speculative model". Briefly summarized, the DNS asymptotic value found by \citet{Hartlepetal2017} for dissipation of turbulent kinetic energy ($\beta_{\epsilon, \textrm{min}}$) is slightly larger than the value found from atmospheric studies at much higher $Re$. Moreover, the DNS asymptotic value of $\beta_{\Phi, \textrm{min}}$ is comparable to the DNS $\beta_{\epsilon, \textrm{min}}$. Also, as noted above, we expect the asymptotic value of $\beta_{S, \textrm{min}}$ for enstrophy to be even smaller than $\beta_{\epsilon, \textrm{min}}$ at high $Re$, {\it and} we expect $\beta_{\Phi, \textrm{min}}$ to be more likely to track enstrophy than dissipation by the nature of the physics involved. To be ``conservative", we merely set $\beta_2=3$, setting the minimum value for {\it both} particle concentration and enstrophy equal to the asymptotic Earth atmosphere value (even though it could arguably be even smaller). These small numerical tweaks to the cascade parameters are intended only as a nod to plausible $Re$-dependence; a better understanding of this $Re$-dependence would be welcome, to put cascade modeling of particle concentration on more quantitatively solid ground.

\subsubsection{Multiplier anticorrelation} \label{sec:model:anticorrelation}

In turbulence, particle concentration is statistically anticorrelated with enstrophy, that is, particles tend to cluster in regions of low vorticity.
Consequently, the multipliers for particle concentration and enstrophy also show this property.
\cite{HoganCuzzi2007} defined a {\sl correlation parameter} $\Gamma$ as the average fraction of subvolumes for which particle and enstrophy multipliers are both larger or equal to 0.5, or both smaller than 0.5.
That is, multipliers $m$ are determined separately for $\Phi$ and $S$ from their respective PDFs, but when it comes to the association of $m$ or $(1-m)$ for both $\Phi$ and $S$ to specific subvolumes (say, left and right), $\Phi$ and $S$ are statistically partitioned in an anticorrelated way with a probability of $1-\Gamma$.
Using several DNS simulations, \cite{HoganCuzzi2007} found a value of $\Gamma \approx 0.3$ which we adopt here.
Unfortunately, \cite{Hartlepetal2017} were not able to independently determine this correlation parameter from the simulation they used to constrain their cascades\footnote{
Due to the finite number of particles in the simulation, multipliers computed from particle tracking data were affected by small particle number statistics.
\cite{Hartlepetal2017} were able to develop a procedure to correct the concentration multipliers for these effects, but not the enstrophy multipliers.
Instead they could only use a limited number of flow snaphots to determine enstrophy multipliers, and therefore could not accurately compute the concentration and enstrophy correlation statistics}.

\subsection{Cascade Simulation} \label{sec:model:simulation}

\begin{figure}
\centering
%
%
%
%
\includegraphics[width=0.85\linewidth]{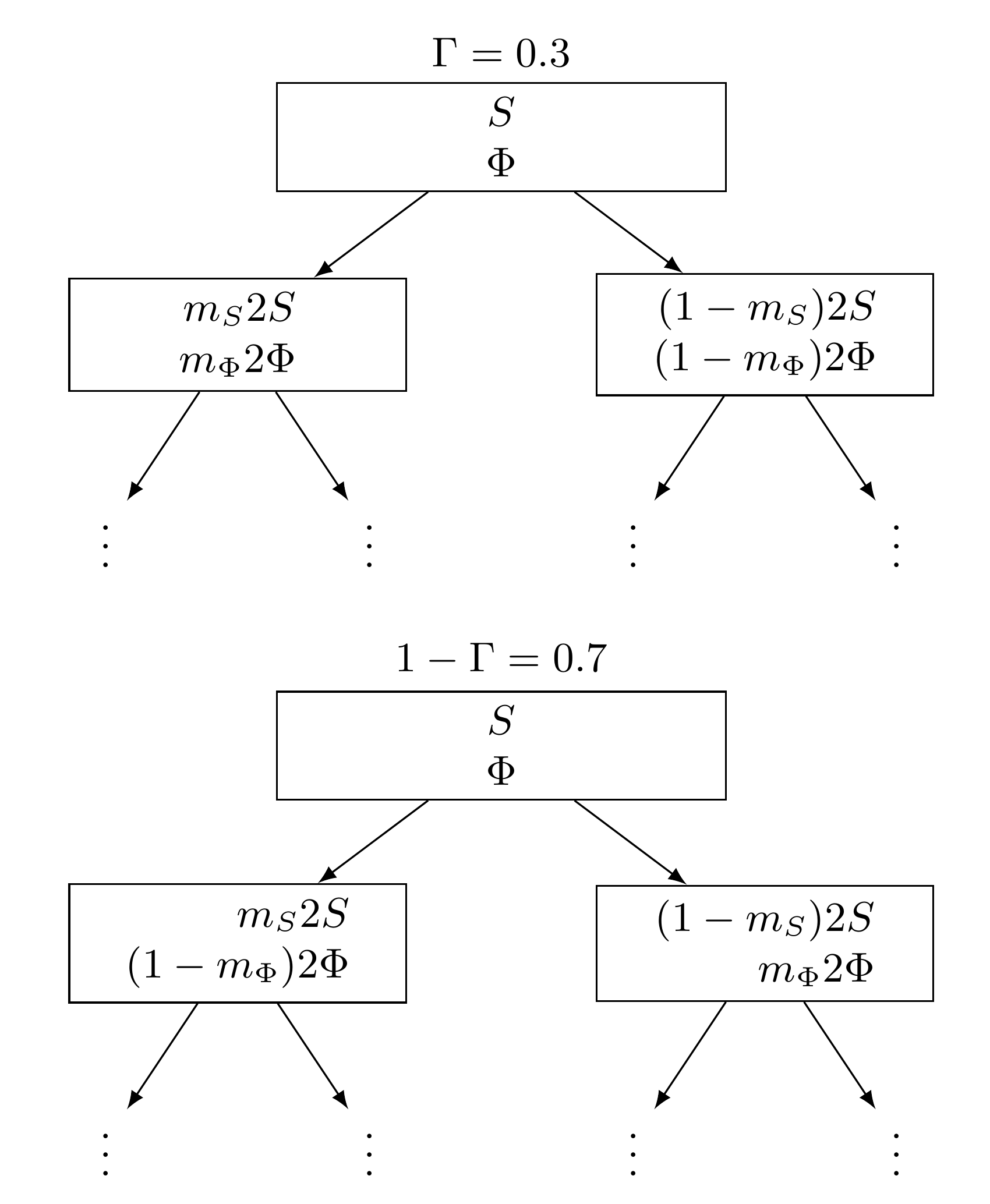}
\caption{Diagram depicting the partion of normalized enstrophy $S$ and mass loading factor $\Phi$ from one cascade level to the next. The multipliers $m_S$ and $m_\Phi$ are assumed to be greater than or equal to 0.5. With a probability of $\Gamma=0.3$, the subvolume with the larger fraction of $S$ also has a larger fraction of $\Phi$ (upper figure), while the reverse is true with a probability of $1-\Gamma=0.7$ (lower figure).}
\label{fig:MultiplierCartoon}
\end{figure}

We use a Monte-Carlo-type simulation to compute the joint probability distribution functions for normalized enstrophy and concentration factor at all relevant spatial scales.
The code is based on~\cite{HoganCuzzi2007} but has been greatly expanded and parallelized.
As in previous work, the code starts at the largest turbulent scale, $L$, where the density of solid particles and enstrophy assume their nominal average values, and works step by step down the cascade until some lower cut-off scale is reached.
At each cascade level $N$ corresponding to a spatial scale of $\ell = 2^{-N/3}L$, random samples $m_\Phi $ and $m_S$ are drawn from the corresponding multiplier distributions for concentration and enstrophy (Sections~\ref{sec:model:concentration} and \ref{sec:model:enstrophy}),
and are used to divide up the concentration and enstrophy values from the previous, larger, scale into subscales.
This process is illustrated in Figure~\ref{fig:MultiplierCartoon}.
Following the anticorrelation rule (Section~\ref{sec:model:anticorrelation}), the subvolume with the larger concentration value gets the larger enstrophy value with a probability of $\Gamma=0.3$,
while the reverse is true with a probability of $1-\Gamma=0.7$.
Each step in the cascade represents a halving of spatial scale along a different orthogonal direction. The effective reduction of scale is therefore $2^{1/3}$.

One such cascade calculation produces a tree of concentration factor and enstrophy values for all the spatial scales considered with the number of concentration and enstrophy values increasing by a factor of 2 at every step.
In the end, we have $2^N$ values each for concentration factor and enstrophy at each level $N$.
By repeating the procedure over and over with newly chosen randomly selected multiplier values, we can accumulate enough samples to compute highly resolved, statistically converegd joint probability distribution functions $P(\Phi,S,\ell)$ at all scales $\ell$.

\begin{figure}
   \vspace{5pt}

   \hspace{-3pt}\includegraphics[width=0.984\linewidth]{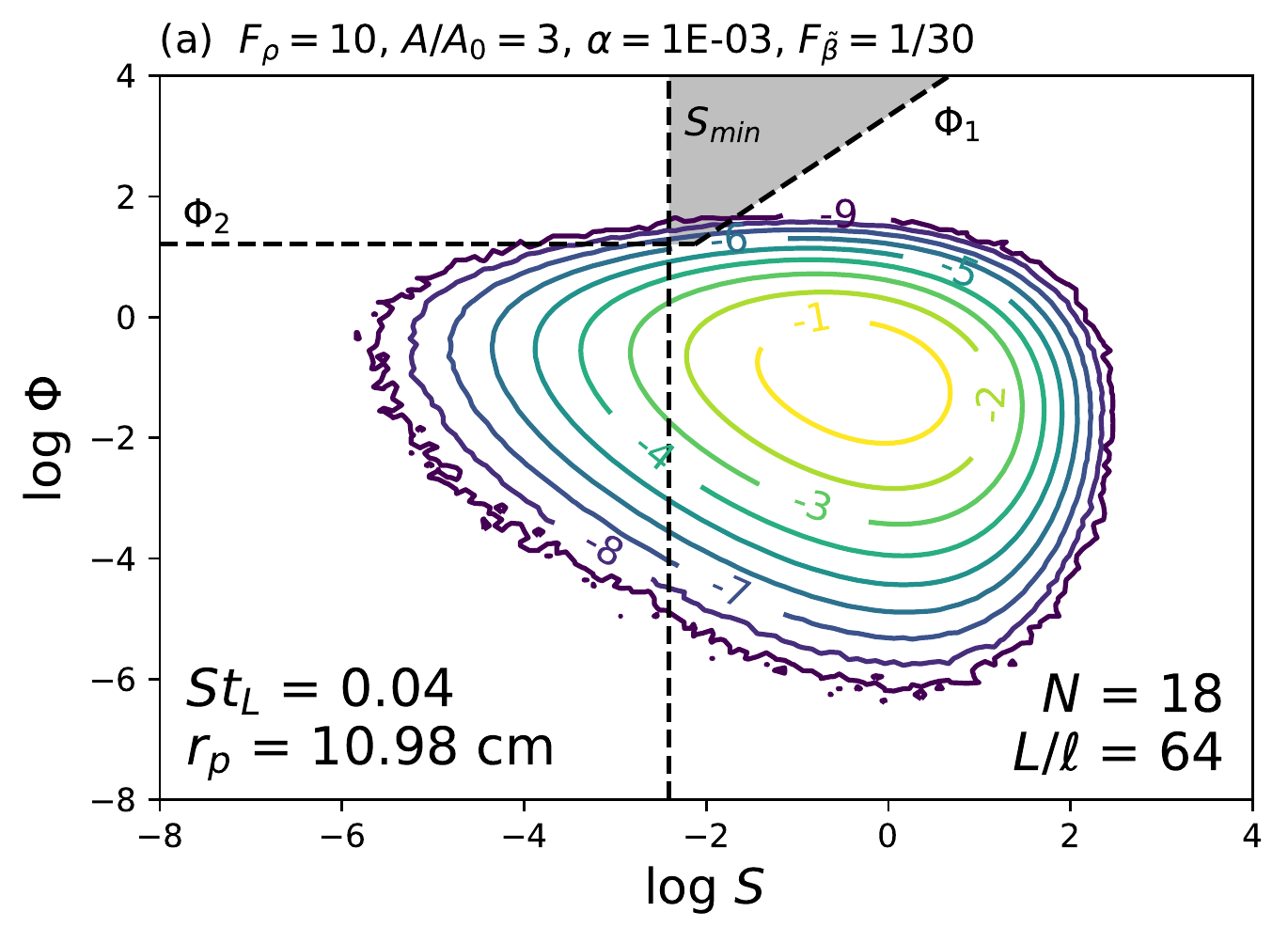}
   \includegraphics[width=\linewidth]{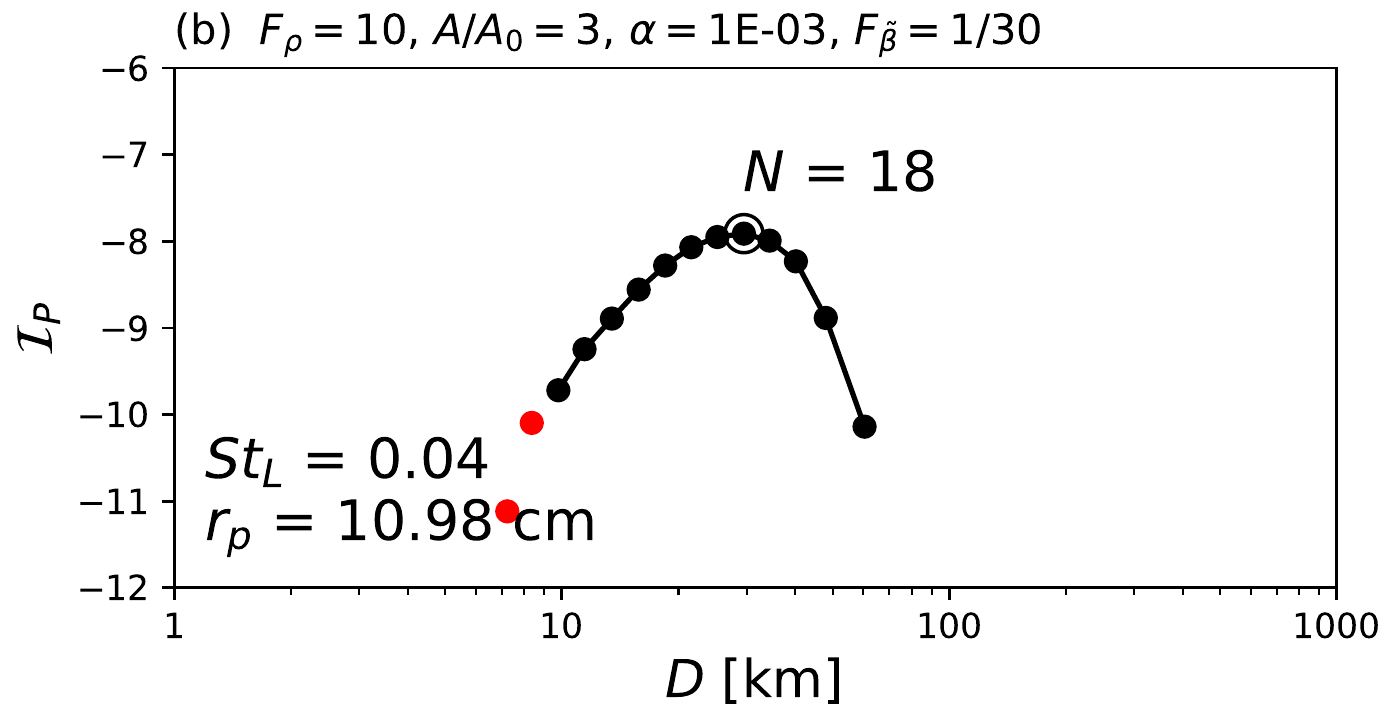}

   \vspace{5pt}

   \caption{
   {\sl(a)} Contours of joint PDF $P(\Phi,S,\ell)$ for mass loading factor $\Phi$ and normalized enstrophy $S$ for Stokes number $St_L = 0.04$ at cascade level $N=18$ and nebula gas parameters as indicated above the plot (see Table~\ref{tab:SymbolsAndParameters} for a description of the symbols). The thresholds $\Phi_1$, $\Phi_2$ and $S_{\rm{min}}$ (Section~\ref{sec:model:thresholds}) are shown by dashed lines. 
   The planetesimal-forming region of the parameter space is shaded in gray. {\sl(b)} The Initial Mass Function (IMF) for this case, computed by integrating the probabilities above the thresholds -- each dot is the result at a single value of $N$. The peak in this particular case (marked with a circle) is reached at cascade level $N = 18$, and corresponds to a planetesimal diameter of 29.5\,km. Black symbols indicate that $\Phi$ is below the mass loading limit while red symbols show where the results are questionable because the mass loading limit of $\Phi_{\rm{limit}}=100$ has been surpassed.
         \label{fig:JointPDFs}}
\end{figure}

Since the number of samples grows exponentially with cascade level, it is easy to generate many samples for the smallest spatial scales but good statistics for the largest scales requires running many such cascades.
In practice, in order to balance the amount of computational work going towards large and small scales, we ``prune'' the cascade tree at some level.
That is, starting at some cascade level $N_{\rm{p}}$ we keep the number of samples constant by randomly selecting only half of the just generated values to be followed further to the next cascade level.
Pruning at a small (shallow) level will, using the same amount of computational time, produce more samples at large scales (small cascade level numbers) {\it vs.} pruning at a larger (deeper) level or not pruning at all.
Storing all samples until the end of the run is usually not feasible since we need a very large number of samples for computing a joint-probability distribution $P(\Phi,S,\ell)$ with good statistics.
Instead, after each new cascade tree has been generated, we bin the samples into $2000\times2000$ logarithmically-spaced concentration and enstrophy bins spanning 20 orders of magnitude in $\Phi$ and $S$ to ensure capturing the entire probability distribution.
An example of such a distribution is shown in Figure~\ref{fig:JointPDFs}a.

\begin{table}
\footnotesize
\centering
\caption{List of cascade simulation runs used in the present work. Cases were run to different maximum cascade levels $N_{\rm max}$, and pruned at different levels $N_{\rm p}$ as explained in the text.} \label{tab:CascadeSimulationRuns}
\begin{tabular}{lccc}
\hline
\hline
& \multicolumn{3}{c}{Number of individual cascade calculations} \\
$St_L$ & $N_{\rm max}=30$ & $N_{\rm max}=30$ & $N_{\rm max}=60$ \\
       & $N_{\rm p}=10$ & $N_{\rm p}=20$ & $N_{\rm p}=20$ \\
\hline
0.16  & $256 \times10^6$ & -- & $50\times10^3$ \\
0.08  & $256 \times10^6$ & -- & $50\times10^3$ \\
0.04  & $256 \times10^6$ & -- & $50\times10^3$ \\
0.02  & $256 \times10^6$ & -- & $50\times10^3$ \\
0.01  & $256 \times10^6$ & -- & $50\times10^3$ \\
0.005 & $256 \times10^6$ & $160\times10^3$ & $50\times10^3$ \\
0.0025 & $256 \times10^6$ & $160\times10^3$ & $50\times10^3$ \\
0.0001 & -- & -- & $50\times10^3$ \\
\hline
\end{tabular}
\end{table}

Table~\ref{tab:CascadeSimulationRuns} lists the simulation runs performed for and used in this study.
We consider seven different Stokes numbers $St_L$ defined using the large eddy lifetime in the nebula.
These correspond to the different particle sizes depending on the specific nebula conditions\footnote{
Stokes number relates to the particle size through the aerodynamic stopping time $t_s$ for which we use an Epstein-to-Stokes transition formula \citep{Podolaketal1988, Estradaetal2016}.
Specifically, we use Equations~70 and 80 of \citet{Estradaetal2016} but with the transition between Epstein and Stokes regimes at the more typical value of $r_p/\lambda_g = 9/4$, where $r_p$ is the particle radius and $\lambda_g$ the gas mean free path for which we use the value for an $H_2$ gas.
The gas dynamic viscosity appears in these equations for which we use a temperature dependent, generalized Sutherland formula \citep{Cloutman2000}, again for an $H_2$ gas.
}, and range from millimeters to decimeters in the inner nebula, and from sub-millimeters to centimeters in the outer nebula.
Since the cascade model is formulated in a non-dimensional fashion, no separate simulations are needed for the different nebula parameters which enter into the analysis through the thresholds (Section~\ref{sec:model:nebula_and_thresholds}) except for the mean mass loading factor which is varied by rescaling the $\Phi$ values.

One final note is that the current cascade model does not directly take into account the back effects of high mass loading ($\Phi\gtrsim100$) onto the flow which causes the turbulent clustering process to stall.
We account for this separately (see Section~\ref{sec:model:integration}).

\subsection{Nebula Model and Thresholds} \label{sec:model:nebula_and_thresholds}

Here we outline the description in \citet{Cuzzietal2010}, to which we refer the reader for details. In a more familiar ``gravitational collapse", like a Jeans instability, dense clumps can collapse on the dynamical timescale $t_{\rm dyn}=(4 G \Phi \rho_g)^{-1/2}$, which for $\Phi\sim 10-100$, is less than an orbit time and swamps other local environmental factors. This sort of collapse is what would occur if the particle stopping times $t_s$ were all longer than $t_{\rm dyn}$. However, for the small particle sizes in question here, it has been shown that gas pressure gradient effects prevent prompt dynamical collapse,  and allow only slow sedimentation of particles toward their mutual center on much longer timescales of $100-1000$ orbits \citep{Sekiya1983, Cuzzietal2008, ShariffCuzzi2015} Thus, in order for a clump of small particles to survive long enough to sediment slowly and gently into a ``sandpile'' planetesimal, it has to have the right properties to resist various disruptive mechanisms for many orbits.

\citet{Cuzzietal2010} proposed three simple, physics-based thresholds to constrain the part of the density-enstrophy phase space that contains such planetesimal-forming clumps, and we adopt them here. The thresholds are applied to the cascade PDFs as shown in Figure~\ref{fig:JointPDFs}, and filter out nearly all clumps except for the densest ones at any lengthscale. This dense subset represents only a small volume fraction of the disk, and a small mass fraction of the particles. It is this statistical filter that gives this planetesimal formation process its characteristic gradual formation of planetesimals over time, instead of immediate transformation of most solids into planetesimals all across some unstable region as in other scenarios. Essentially, the same three thresholds were used by \citet{Chambers2010}.

\subsubsection{Baseline Nebula Model} \label{sec:model:nebula}

The thresholds are easily implemented for an arbitrary location in the nebula (in this paper, we select 3\,AU and 25\,AU), using a simple radial powerlaw model of nebula gas surface mass density and temperature:
\begin{eqnarray}\nonumber
  \sigma(a)= \sigma(a_o)(a/a_o)^{-p}, \\
  T(a) = T(a_o) (a/a_o)^{-q}. \label{eqn:sigma_and_T}
\end{eqnarray}
From this follow the gas density scale height $H(a) = c(a)/\Omega(a)$, gas density $\rho_g(a) = \sigma(a)/2H(a)$ and pressure gradient\footnote{\citet{Cuzzietal2010} called the pressure gradient $\beta$ but in this paper we denote it as $\tilde\beta$ to avoid confusion with the $\beta$ values associated with the cascade models.} $\tilde\beta(a) = H(a)^2/a^2$.
The latter drives the headwind that makes particles drift inwards and can disrupt strengthless clumps.
Sound speed and orbital frequency are given by their usual expressions $c(a)=\sqrt{\gamma k_B T(a) / m_{H_2}}$ and $\Omega(a) = \sqrt{M_\odot G } a^{-3/2}$ with adiabatic index $\gamma=1.4$, molecular mass of hydrogen gas $m_{H_2}=3.35\times10^{-24}$\,g, solar mass $M_\odot$, and gravitational constant $G$.
The mean local density in solids is given by $\left< \rho_p(a) \right> = A \rho_g(a) H(a)/h_d(a)$ where $A$ is the global abundance of solids compared to gas, and sets the global mean of the concentration factor $\Phi = \rho_p/\rho_g$.
The additional factor of $H(a)/h_d(a)$ arises due to particle settling towards the midplane.
Settling is resisted by turbulent diffusion and this interplay results in a settled particle scale height of
\begin{equation}
h_d = H/{\sqrt{1 + St_L/\alpha}}
\label{eqn:h_d}
\end{equation}
\citep[their Equation~101]{YoudinLithwick2007, Carballidoetal2011, Estradaetal2016}.

Our reference nebula parameters are as follows: $\sigma(a_o)$ = 1700\,g\,cm$^{-2}$, $T(a_o)=300$\,K (both at $a_o$ = 1\,AU), exponents $p = 3/2$, $q=1/2$, and a canonical abundance of solids of $A_o = 0.01$.
Here, we have adopted the \citet{Hayashi1981} Minimum Mass Solar Nebula (MMSN) as a convenient baseline, but consider various enhancement factors (see below) for gas density {\it and} solids-to-gas ratio, motivated by more sophisticated nebula models \citep[e.g.,][and others]{Estradaetal2016, Deschetal2017}.
After all, a MMSN makes the unrealistic assumption that all of the initially available solids are  transformed into planets with no losses into the sun.
Also, nebula surface mass densities decrease with time and T-Tauri-stage estimates from millimeter-wavelength observations may be underestimates of conditions at early planetesimal formation stages \citep{Andrews2015}.

\subsubsection{Model Variations} \label{sec:model:variations}

Since there is considerable uncertainty in the physical properties of the protoplanetary solar nebula \citep{Cuzzietal2010}, we consider a wide range of nebula models through various enhancement factors relative to our baseline MMSN.
We consider enhancements in local gas density through a factor $F_\rho$ which is varied between $0.3$ and $10$ (in the inner nebula) or $30$ (in the outer nebula), and enhancements in the solids-to-gas ratio, $A/A_o$, which is varied between $1$ and $10$ except for the $F_\rho=0.3$ case where we consider values up to $30$.
Note that our calculations consider only one Stokes number (or particle size) at a time, but in reality there is likely a distribution of particle sizes present.
Since particles with disparate Stokes numbers cluster in different regions and so can be considered separately, it might be best to think of $A/A_o$ not as the overall solids-to-gas ratio but rather the mass fraction of particles with (or near) a particular Stokes number, which may be only a fraction of the total solids in the nebula.
Some combinations of the above parameters, in particular simulaneously high values of $F_\rho$ and $A/A_o$, correspond to unrealistic amounts of solids in the planetesimal formation regions, and we therefore exclude them from our analysis.
Specifically, we only consider the cases for which the total solid mass in the
asteroid-forming region ($2-4$\,AU) is between $5$ and $250$ Earth masses, $M_\oplus$.
Put another way, our nebula surface (solid) densities at $3$\,AU range from $3.5$ to $180$\,g\,cm$^{-2}$.
For comparison, the MMSN with a canonical solids-to-gas ratio of $A_o=0.01$ has a surface mass density for solids of $3.3$\,g\,cm$^{-2}$.
Similarly, for the outer nebula between $16$ and $30$\,AU, where trans-Neptunian objects (TNOs) are presumed to have formed, we only consider models with total mass of solids between $10$ and $2000$\,$M_\oplus$, or put another way, the surface density of solids at $25$\,AU is between $0.14$ and $26$\,g\,cm$^{-2}$ where the lower value corresponds to the MMSN.

In addition to varying gas and solid abundances, we also consider reduced values of the pressure gradient through a scale factor $F_{\tilde\beta}$ emulating the peloton-like effect where clumps of particles are shielded from the mean pressure gradient by virtue of being embedded in larger structures.
How strong this effect is is poorly understood and so we consider a wide range of values for $F_{\tilde\beta}$ from $1$ to $1/100$.

\subsubsection{Thresholds} \label{sec:model:thresholds}

{\it $\Phi_1$: Local rotation and gravitational binding:} This threshold is basically a generalization of the traditional gravitational binding criterion to allow for variable rotation rates of dense clumps in eddies of different sizes. It assumes the clump has the rotation frequency of an eddy of the same lengthscale, based on a Kolmogorov energy spectrum, and requires it to have sufficient mass to be bound at that frequency. This can be expressed by requiring the local gravitational timescale $t_G=t_{\rm dyn}=(4 G \Phi \rho_g)^{-1/2}$ be shorter than the local eddy time $1/\omega(\ell)$, where in the inertial range the mean eddy frequency $\left<\omega(\ell)\right>=\Omega(L/\ell)^{2/3}$ with $\Omega$ the orbital frequency and $L=H\alpha^{1/2}$ being the large eddy size. Since $\Omega$ can be rewritten in terms of the solar mass and distance to the sun, the threshold can be expressed for arbitrary distance $a$ in terms of the reference distance $a_o$ and the gas density $\rho_{go} \equiv \rho_{g}(a_o)$. We use the cascade relation $\ell = 2^{-N/3}L$ to express $(L/\ell)^{2/3} = 2^{2N/9}$. To allow $\Phi_1$ to be represented on a $(S,\Phi)$ plot, we use the definition of normalized enstrophy $S = \omega^2(\ell)/\left<\omega^2(\ell)\right>$
\footnote{In this context, $\omega^2(\ell)$ is understood to be the {\sl coarse-grained} enstrophy at some spatial scale $\ell$, and $\left<\omega^2(\ell)\right>$ is its mean.
This mean is scale-dependent.
In subtle ways, this is different from \citet{Hartlepetal2017} who derived enstrophy multipliers by binning the {\sl fine-grained} enstrophy.
The mean over some scale $\ell$ of the fine-grained vorticity is not scale dependent.
However, the multipliers themself do not know anything about the mean at any given scale, they only describe how a quantity partitions from one scale to another.
In effect, these subtleties only require that the cascade calculation of enstrophy at any spatial scale $\ell$ has this normalized definition.
}.
We end up with \citep[as in][Equation~6]{Cuzzietal2010}
\begin{equation}
\Phi_1(S,a) = 2^{4N/9} K_0 S \left({a \over a_o}\right)^{p-3/2},
\end{equation}
where $K_0 \equiv 3 M_{\odot}/ 4 \pi \rho_{go} a_o^3$. The threshold $\Phi_1(S,a)$ appears as a sloping line in Figure~\ref{fig:JointPDFs}.

{\it $S_{\rm{min}}$: Global nebula shear:} $S_{\rm{min}}$ is closely related to $\Phi_1$, but  captures the fact that on long timescales, a dense particle clump of finite extent cannot avoid the systematic Keplerian radial (tidal) shear of the nebula, which acts like $\omega$ in $\Phi_1$. That is, regardless of the statistical likelihood that there {\it will} be local patches of low local vorticity in homogeneous, isotropic turbulence (see the PDF in Figure~\ref{fig:JointPDFs}), $S_{\rm{min}}$ imposes a minimum local vorticity given by the global orbital shear rate $\Omega$ and is given by:
\begin{equation}
   S_{\rm{min}} = \frac{\Omega^2}{\left<\omega^2(\ell)\right>} = \frac{\Omega^2}{2^{2N/9}\Omega^2} = 2^{-4N/9}
   \label{Eqn:S_min}
\end{equation}
where, as above, $N$ and $\ell$ are the cascade level and corresponding spatial scale.

{\it $\Phi_2$: Ram pressure and the gravitational Weber number:} The very dense, particle-rich clumps envisioned here will become decoupled from the gas and start to behave like individual bodies, tending towards Keplerian motion. If they are formed at high nebula altitude, they will settle towards the midplane, and even when lying in the midplane they experience a headwind from the pressure-supported gas. Because they have no intrinsic strength, they are susceptible to disruption by the ram pressure or the associated vortex instabilities associated with these headwinds. A similar situation is found with raindrops, in which surface tension provides the force to resist these disruptive effects as defined by the so-called Weber number $We$. \citet{Cuzzietal2008} argued that self-gravity of these strengthless clumps plays a comparable role, derived a so-called gravitational Weber number $We_G$, and supported the argument with numerical simulations that indicated a critical value $We^*_G\sim 1$.
From it they derived a second threshold $\Phi_2$, that is independent of $S$ but depends on lengthscale $\ell$ (or cascade level $N$) and, importantly, the pressure gradient parameter $\tilde\beta$. Ultimately this threshold becomes \citep{Cuzzietal2008,Cuzzietal2010}:
\begin{equation}
\Phi_2(a) = 2^{N/3}\left({  a_o \tilde\beta(a_o)\over H(a_o) } \right)
                      \left({ 2 K_0 \over \alpha We_G^* } \right)^{1/2}
                      \left({ a\over a_o} \right)^{(p-3/2)/2}.
                         \label{eqn:phi2}
\end{equation}

These simple thresholds ($\Phi_1, \Phi_2$, and $S_{\rm min}$) are subject to some uncertainty, of course. For example, in a more refined analysis, \citet{Sekiya1983} obtained a result for $\Phi_1$ differing by a factor of $10/3$ from ours.
Also, there is a range of uncertainty for the critical Weber number -- \citet{Cuzzietal2008,Cuzzietal2010} argued that is should be somewhere between 1 and 10 -- while we use $We^*_G=1$.
Larger values ``relax'' the $\Phi_2$ criterion, so that a larger part of the parameter phase-space can form planetesimals (Figure~\ref{fig:JointPDFs}(a)).
In Sections~\ref{sec:innernebula} and \ref{sec:outernebula}, we will explore the effects on the planetesimal IMF of relaxing the nominal thresholds. Two other possible choices of physically-based thresholds, erosion of the clump by shear in the surrounding fluid, and diffusion of the clump by eddies on comparable or smaller lengthscales, were discussed and shown to be less restrictive than our selected thresholds  \citep{Cuzzietal2010, ShariffCuzzi2015}. Clearly, even our chosen three are simplified, but we believe they capture the essence of the problem while we await advances in numerical capabilities.

\subsection{Integration} \label{sec:model:integration}

By integrating the part of the PDF lying above the thresholds, we can compute the expected distribution of planetesimal sizes and the rate at which they form.
Specifically, for each particle Stokes number $St_L$ and cascade level $N$ (and corresponding scale $\ell$) we consider the integrals:
\begin{eqnarray}
  {\cal I}_P(l) = \int_{\ge (\Phi_1, \Phi, S_{\rm min})} P(\Phi, S) d\log\Phi d\log S \\
  \overline{\Phi}(l) = \frac{1}{{\cal I}_P}\int_{\ge (\Phi_1, \Phi, S_{\rm min})} \Phi P(\Phi, S) d\log\Phi d\log S.
\end{eqnarray}
Integral ${\cal I}_P$ is the total probability that a given clump of size $\ell$ is stable against distruption and will eventually form a planetesimal,
and $\overline{\Phi}$ is the mean mass loading factor of such stable clumps.
We can use $\overline{\Phi}$ to derive the average mass of solids in the clump by multiplying it by the gas density and clump volume, that is $M = \overline{\Phi} \rho_g \ell^3$.
For simplicity, we assume that the clump will form a single, spherical planetesimal \footnote{Here, we do not consider subsequent evolution of clumps such as a possible bifurcation due to rotational fission as suggested by \citet{Nesvornyetal2010}.}.
Its diameter is obtained from $M = \rho_s 4/3 \pi (D/2)^3$ which gives
\begin{equation}
  D = \left( \frac{6 M}{\pi \rho_s} \right)^{1/3},
  \label{eqn:D}
\end{equation}
where $\rho_s$ is the mass density of the final planetesimal which we here assume to be 2\,g\,cm$^{-3}$.
Plotting ${\cal I}_P$ against $D$ for all clump sizes yields the Initial Mass Function (IMF) of formed planetesimals except for a normalization factor.
An example of such a size distribution is shown in Figure~\ref{fig:JointPDFs}(b).
An important feature of these distributions in our formation scenario is that they have a distinct peak -- a preferred planetesimal size $D_{\rm peak}$.
We denote the probability at that scale as ${\cal I}_{P,{\rm peak}}$ and the mean mass loading factor as $\overline{\Phi}_{\rm peak}$.
The mass rate at which such planetesimals form is then obtained by multiplying this probability by the volume of the formation region and the mass density of solids in the stable clumps, and dividing by a formation time scale $T_{\rm pa}$ (discussed below):
\begin{equation}
    \dot{M}_{\rm pa} = 2\pi (a_2^2 - a_1^2) h \rho_g {\cal I}_{P,{\rm peak}} \overline{\Phi}_{\rm peak} / T_{\rm pa},
    \label{eqn:Mdotpa}
\end{equation}
where $a_1$ and $a_2$ define the nebula region, and $h$ represents the thickness of the layer participating in the formation process.
Due to settling, the density scale height of solids, $h_d = H/{\sqrt{1 + St_L/\alpha}}$, is less than the pressure scale height \citep{Dubrulleetal1995, Estradaetal2016}.
On the other hand, the disruptive effect of the settling velocity excludes clumps formed higher than $H\sqrt{\tilde\beta}$ above the nebula midplane from participating in the formation process.
Therefore, the participating layer height $h$ is the minimum of these two heights.

The choice of timescale $T_{\rm pa}$ in Equation~\ref{eqn:Mdotpa} represents a significant difference between this work and both \citet{Cuzzietal2010} and \citet{Chambers2010}, who themselves adopted different values.
 \citet{Chambers2010} and  \citet{2012LPI....43.2536C} argued that the timescale $T_{\rm pa}$ is not the time it takes for a clump to sediment, which can be many orbital times, because once a clump is bound it is not relevant how long it actually takes to sediment. Differing from  \citet{Chambers2010}, \citet{2012LPI....43.2536C} argued that the formation rate is the creation rate of a new, independent set of clumps that {\it are} bound and destined to inexorably sediment into planetesimals. This latter criterion is in turn the time for a new, independent set of eddies and clumps to form that offer the right conditions for planetesimal formation. This is the timescale on which physically and statistically independent realizations of the particle and fluid velocity and density fields are manifested in the turbulent nebula, that is, $T_{\rm pa} \sim \tau_L$, where the large eddy lifetime $\tau_L$ is close to the orbit period. Studies of the velocity autocorrelation functions in turbulence routinely show that $\tau_L$ is the time within which the flow loses memory of prior states. Moreover, numerical simulations of particles in turbulence show particle clustering statistics to asymptote on about the same timescale \citep{ReutschMaxey1992, Cencinietal2006}. Structures on {\it smaller} scales can come and go on shorter timescales of course \citep[for instance, chose the timescale corresponding to the eddy with the spatial scale of the clump]{Chambers2010}, but we believe that our selection of $\tau_L$ is more conservative in the spirit of $\dot{M}_{\rm pa}$, and more appropriate in the sense of wiping the entire fluid slate clean, guaranteeing an independent manifestation of the particle field. We will follow this latter choice of $T_{\rm pa} = \tau_L$.

Lastly, we need to discuss what happens at large mass-loading factors.
When the solids-to-gas ratio is high, there is backreaction of the particle motion onto the gas which limits how strongly turbulence is able to concentrate particles.
\citet{HoganCuzzi2007} have shown this limits $\Phi$ to values around 100 or so.
Our new cascade model, despite its many improvements compared to previous works, does not explicitly take this effect into account.
Instead, we disregard {\it a posteriori} results with $\overline{\Phi}$ larger than this mass-loading limit.

\section{Results} \label{sec:results}

\subsection{Inner nebula -- Asteroid belt} \label{sec:innernebula}

\begin{figure*}
   \centering
   \vspace{10pt}

   \begin{minipage}{0.49\linewidth}

   \includegraphics[width=\linewidth]{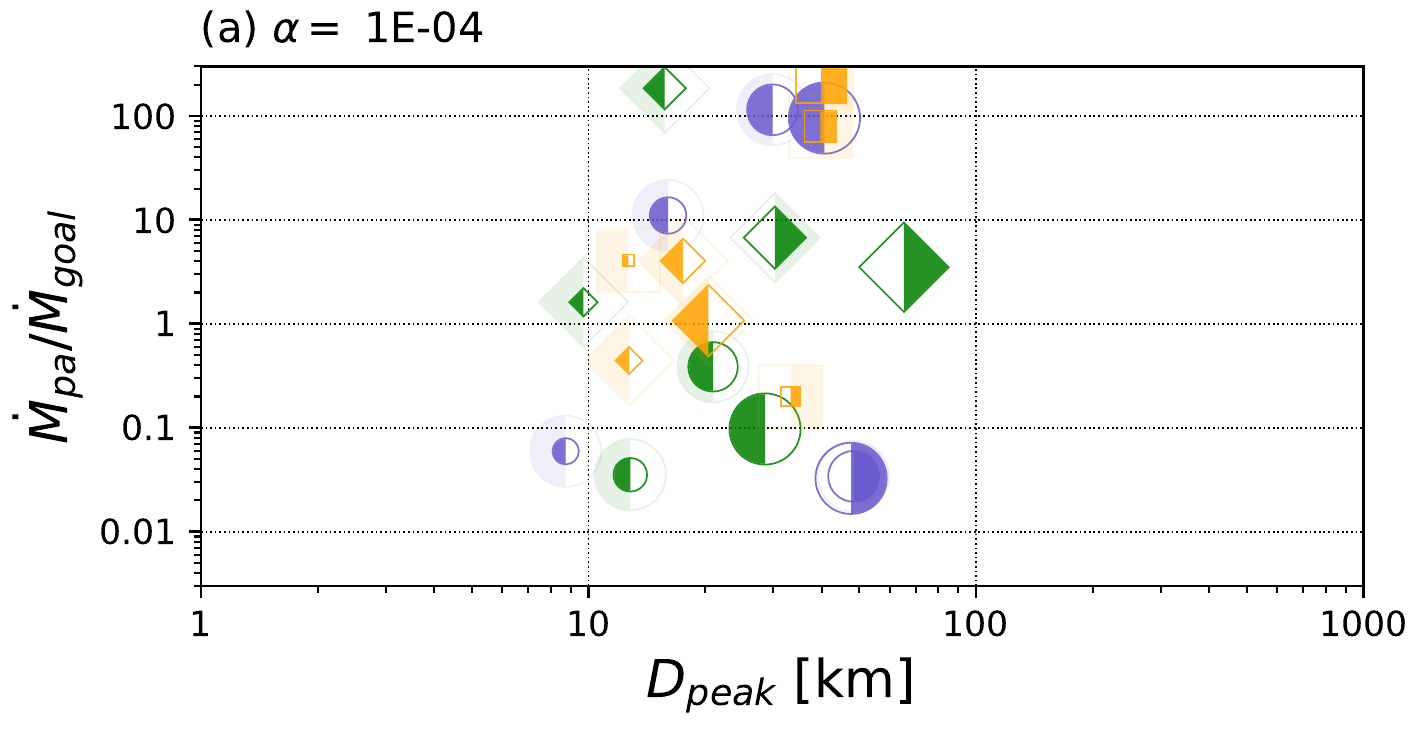}

   \vspace{-24pt}

   \includegraphics[width=\linewidth]{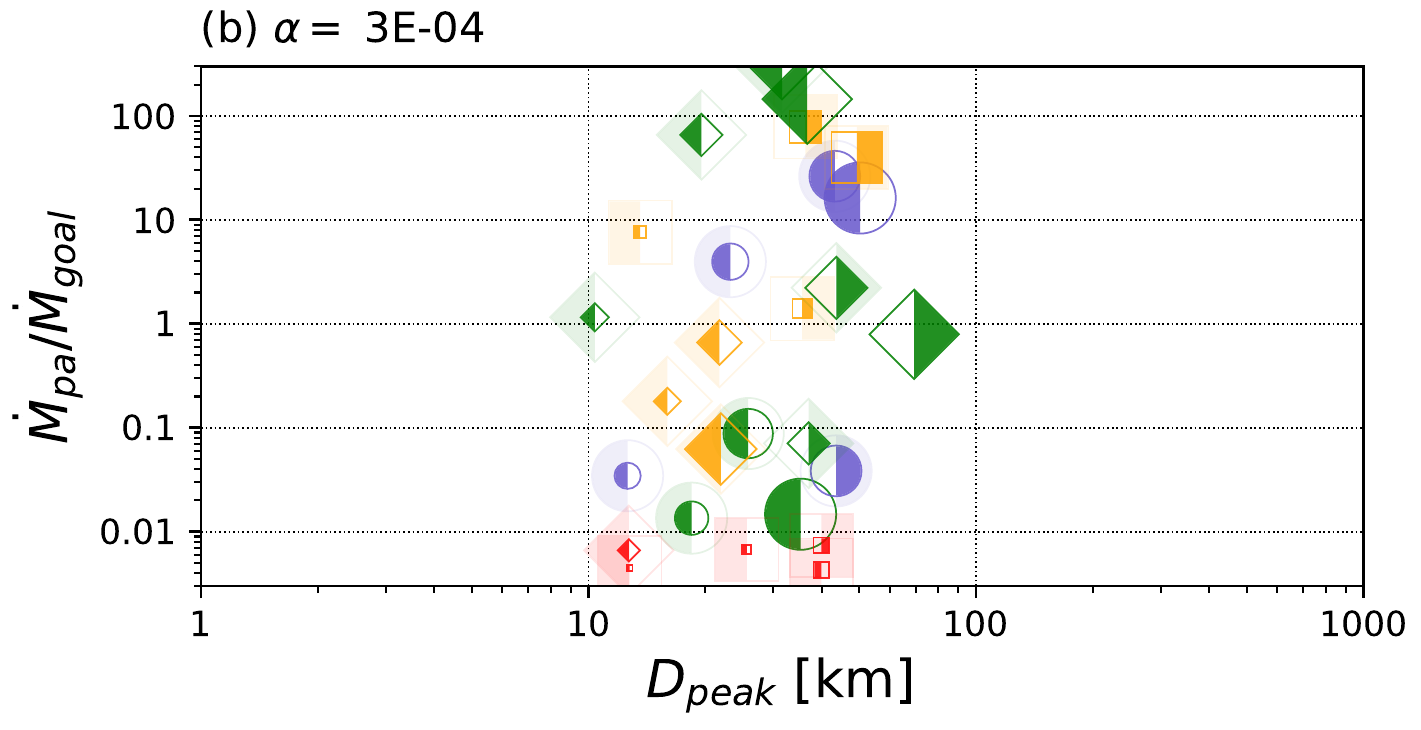}

   \end{minipage}
   \begin{minipage}{0.49\linewidth}

   \includegraphics[width=\linewidth]{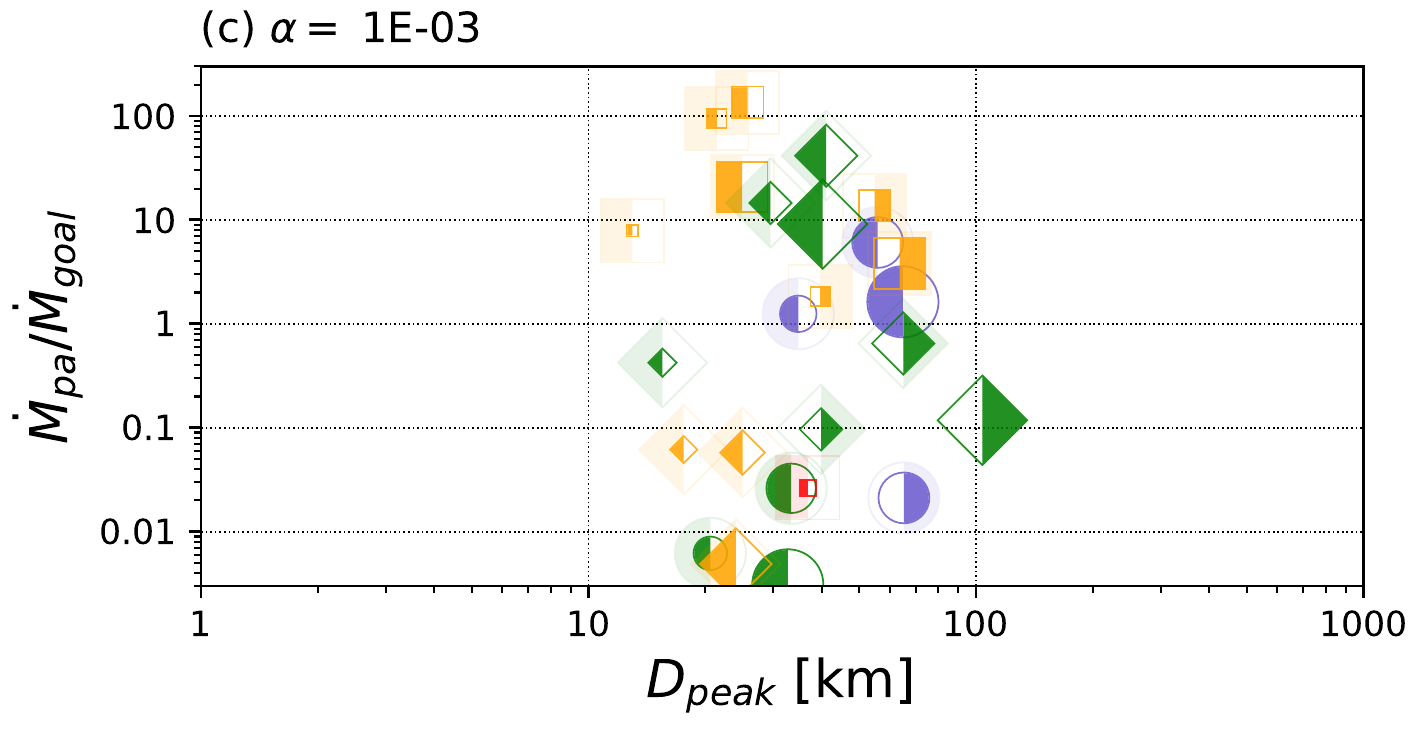}

   \vspace{-24pt}

   \includegraphics[width=\linewidth]{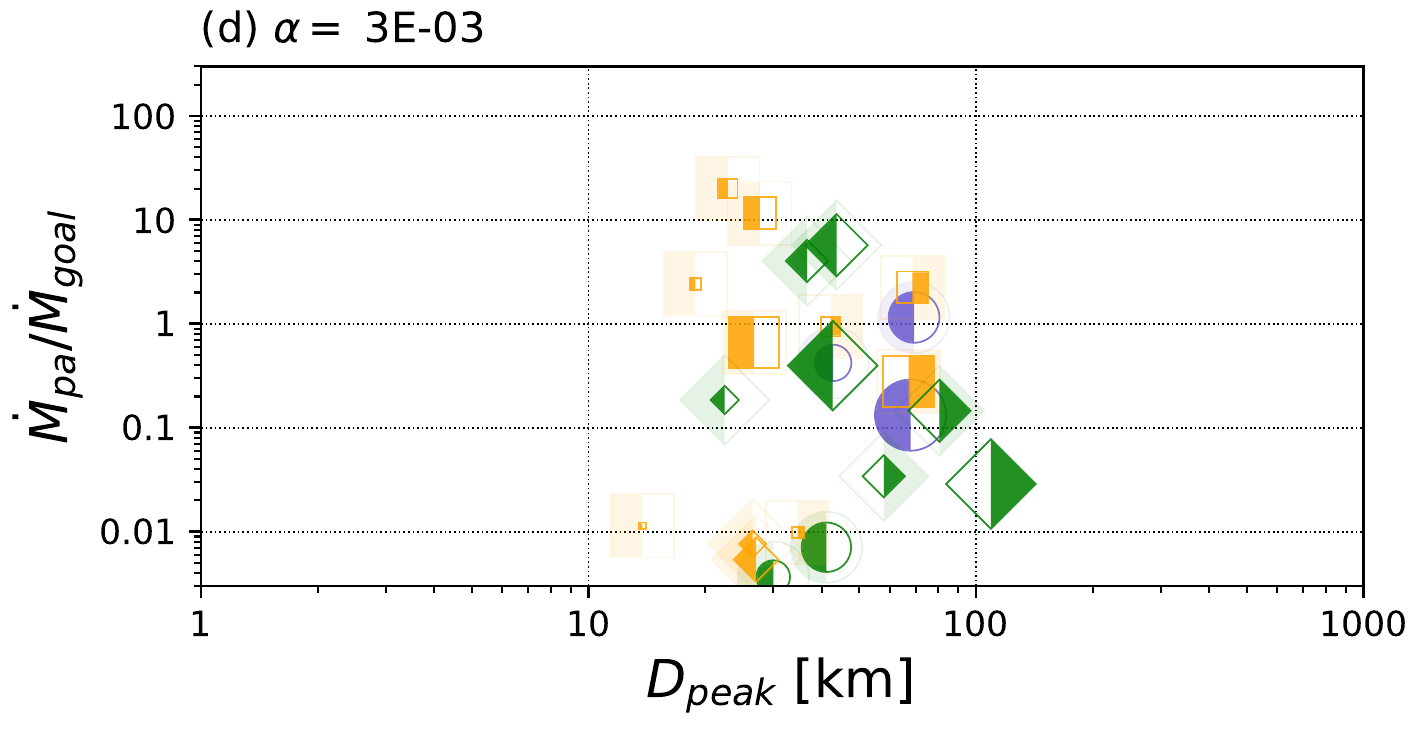}

   \end{minipage}

   \vspace{-5pt}

   \includegraphics[width=0.44\linewidth]{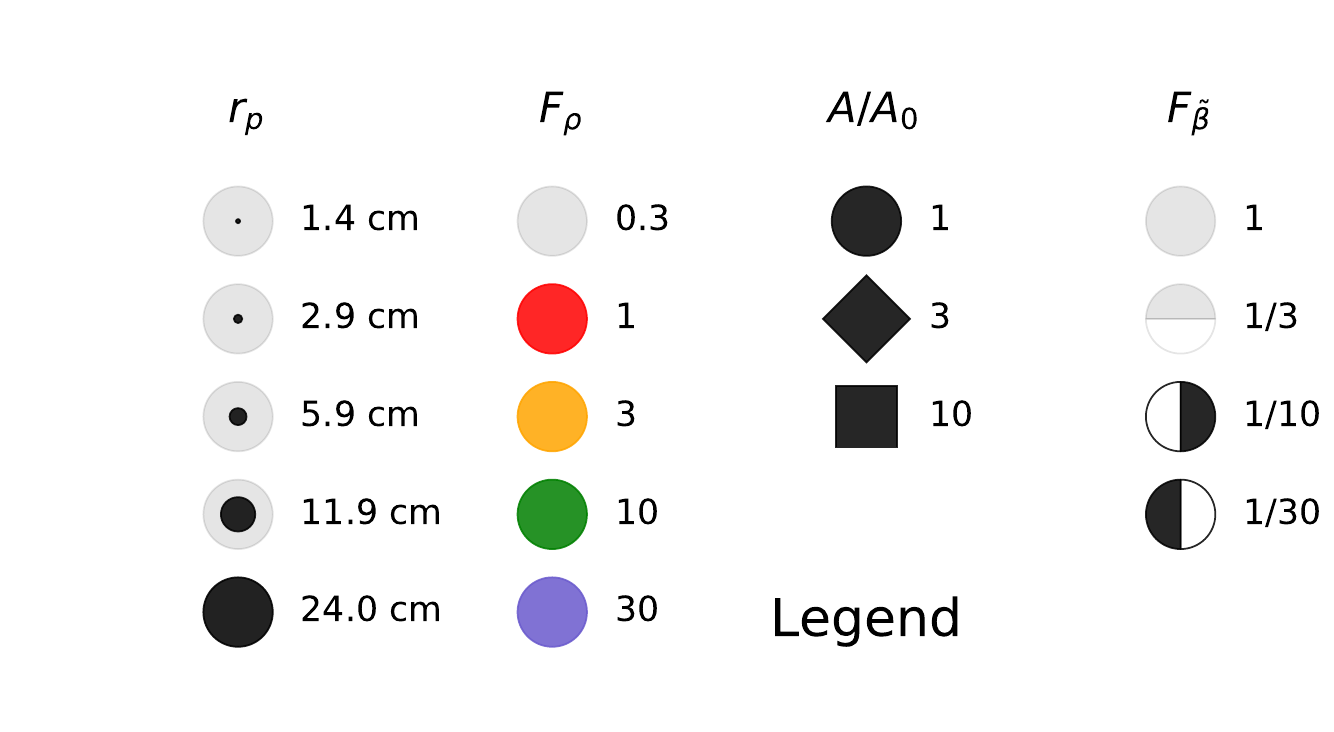}

   \vspace{0pt}

   \caption{
    Model results for the inner nebula, showing the peak of the planetesimal IMF, $D_{\rm{peak}}$, and the rate of formation, $\dot{M}_{\rm{pa}}$, relative to the expected rate, $\dot{M}_{\rm{goal}}$, for a wide range of nebula parameters and particle sizes.
    Thresholds use their nominal, non-``relaxed" values.
    The panels show results for different values of $\alpha$, while (within each panel) colors, symbol shapes and fill styles denote gas density enhancement factor, $F_\rho$, solids enhancement factor, $A/A_o$, and the scale factor for the headwind parameter, $F_{\tilde\beta}$.
    The size of the symbols scales with the particle radius $r_p$.
    The smallest and largest particle sizes that resulted in planetesimals are shown in the legend, together with some representative values in between.
    The legend symbols for $F_{\rho}$, $A/A_o$ and $F_{\tilde\beta}$ show all parameter values that were computed and included in the figure if they produced planetesimals in the range.
    Those parameters that did not produce results in the plot range have been greyed out.
         \label{fig:InnerNebulaPlanetesimals}}
\end{figure*}

\begin{figure*}
   \centering
   \vspace{10pt}

   \begin{minipage}{0.49\linewidth}

   \includegraphics[width=\linewidth]{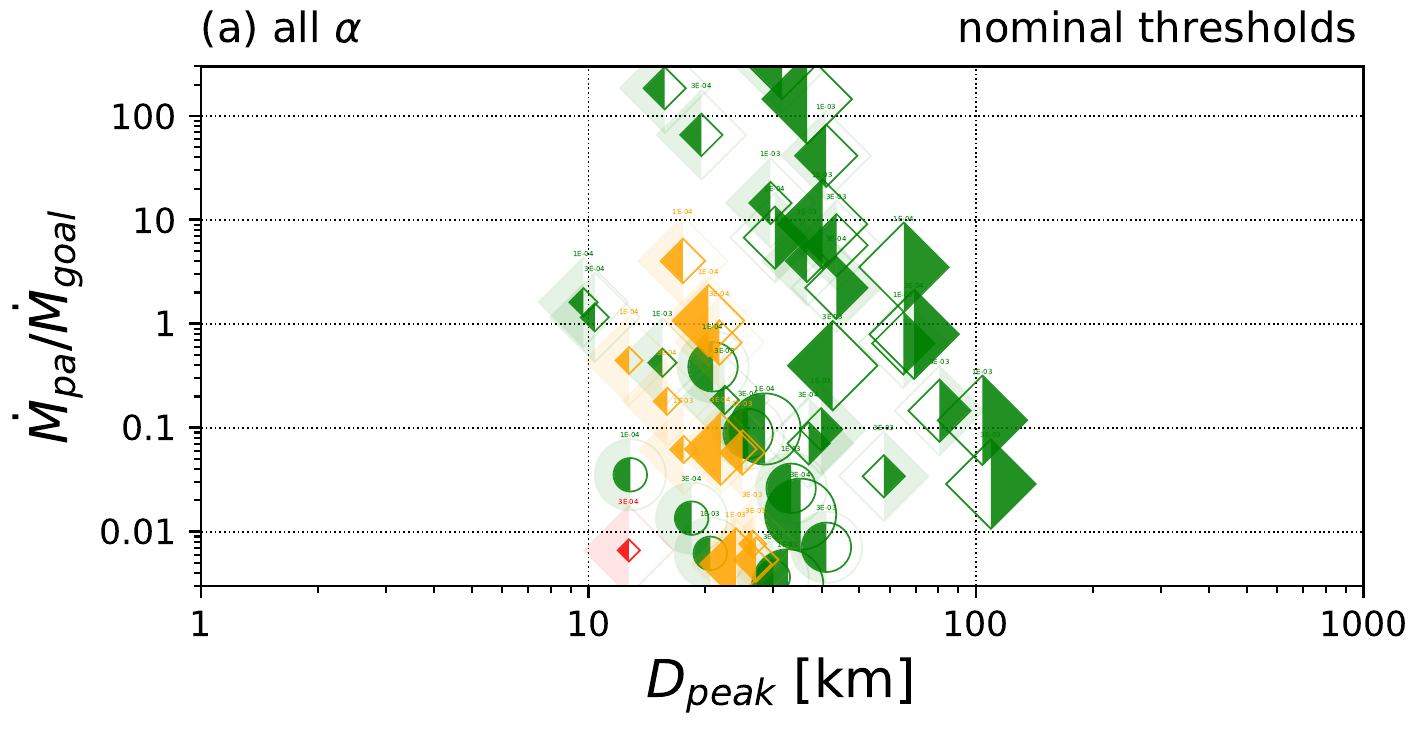}

   \vspace{-24pt}

   \includegraphics[width=\linewidth]{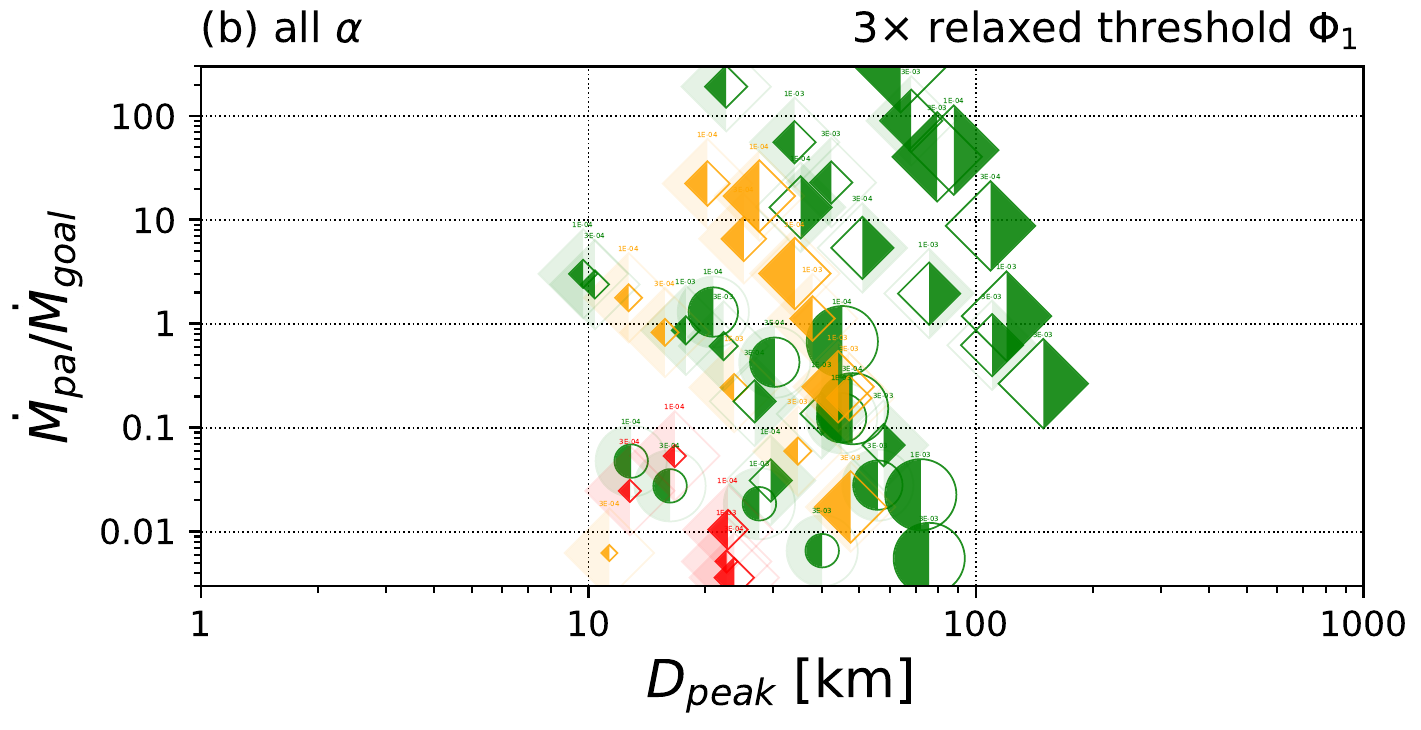}

   \end{minipage}
   \begin{minipage}{0.49\linewidth}

   \includegraphics[width=\linewidth]{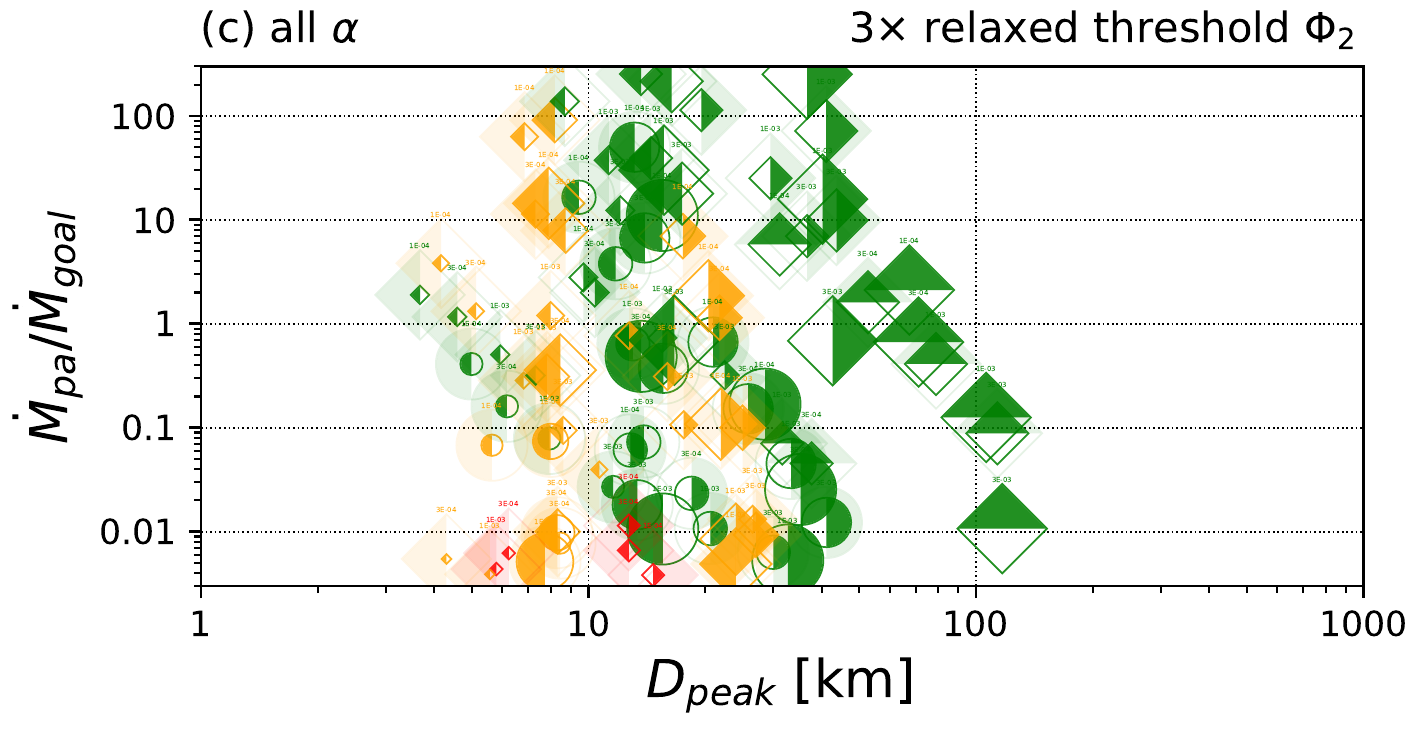}

   \vspace{-24pt}

   \includegraphics[width=\linewidth]{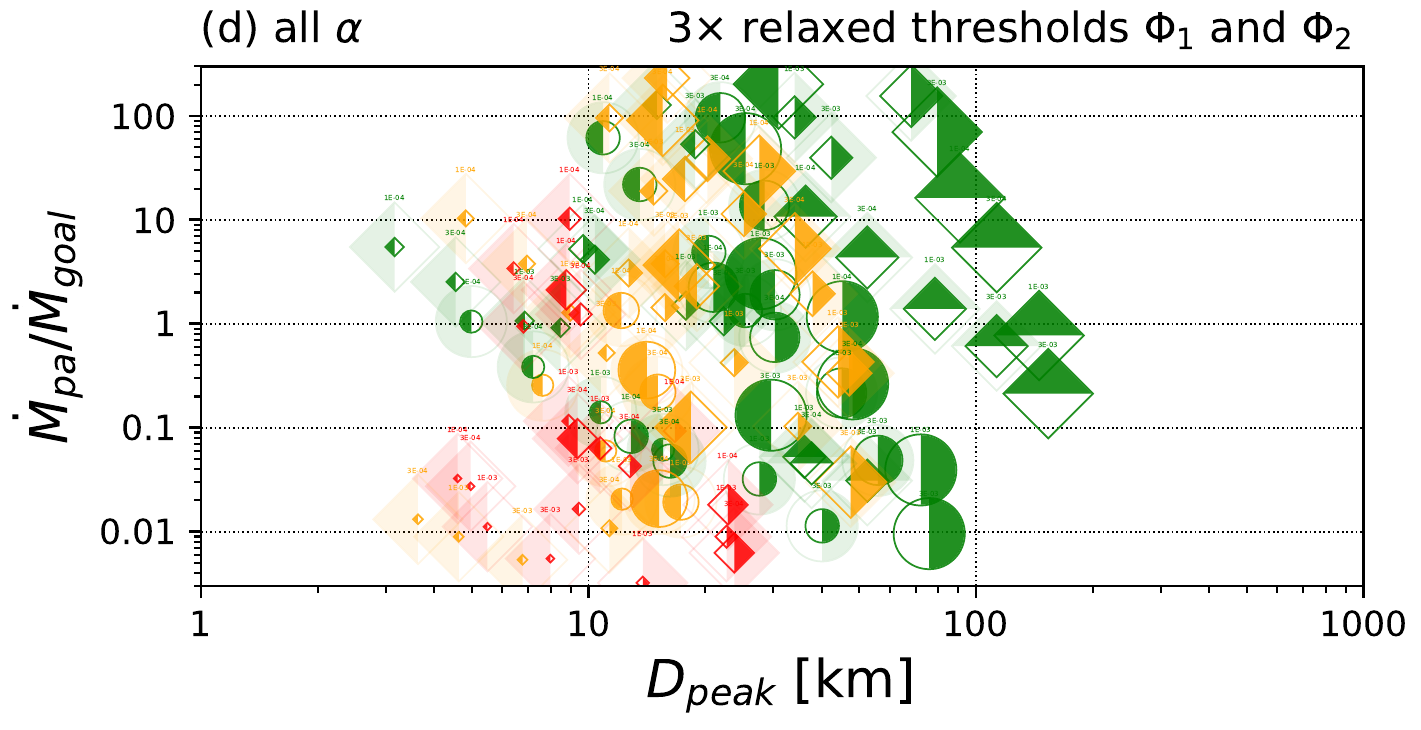}

   \end{minipage}

   \vspace{0pt}

   \includegraphics[width=0.44\linewidth]{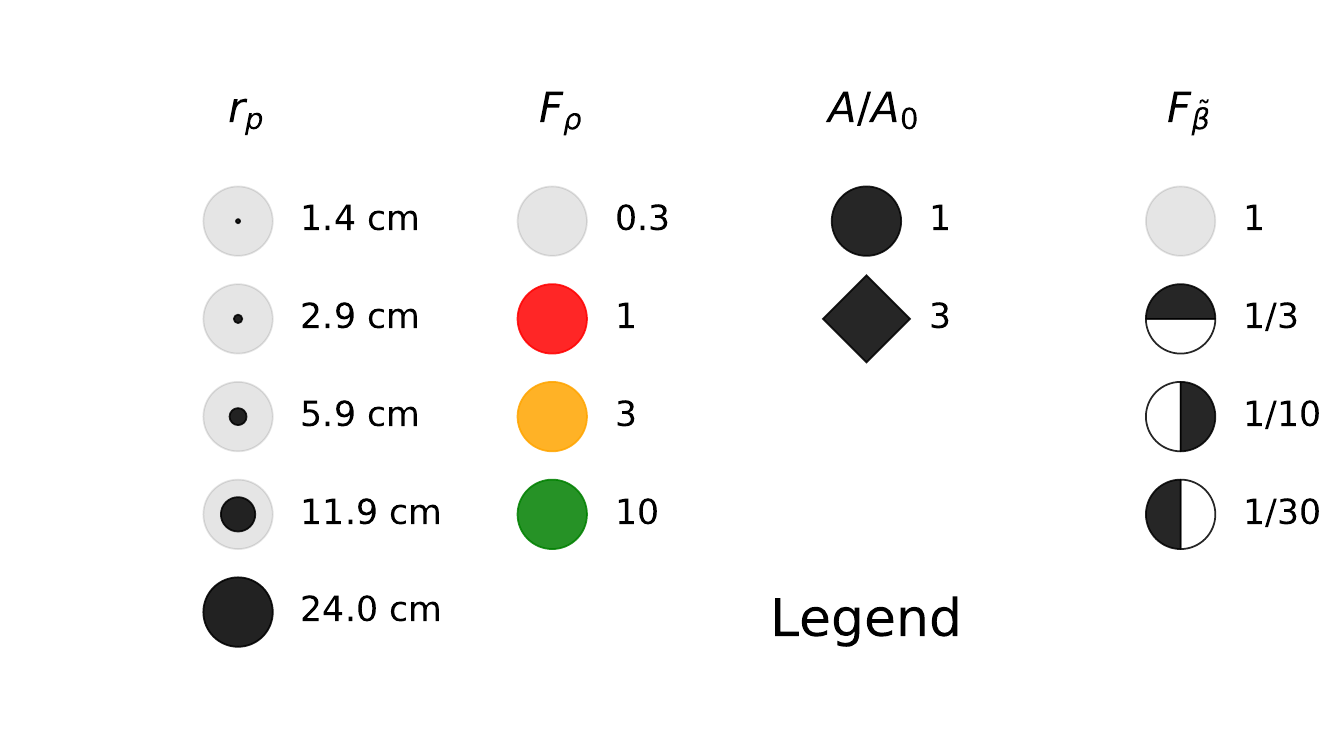}

   \vspace{0pt}

   \caption{Inner nebula results as in Figure~\ref{fig:InnerNebulaPlanetesimals}, but here all $\alpha$ values have been plotted together and the panels vary the thresholds $(\Phi_1, \Phi_2)$.
   As before, colors, symbol shapes and fill styles denote gas density enhancement factor, $F_\rho$, solids enhancement factor, $A/A_o$, and the scale factor for the headwind parameter, $F_{\tilde\beta}$.
   However, the range of values for $F_\rho$ and $A/A_o$ has been restricted compared to Figure~\ref{fig:InnerNebulaPlanetesimals}.
   Plots like these may seem ``busy'' but make it easy to detect general trends such as minimum and maximum size of planetesimals, and dependence of planetsimal size on particle size and $\alpha$.
   Details are more easily explored under magnification, in particular the value of $\alpha$ which is printed in small font above each datapoint, or in Figures~\ref{fig:InnerNebulaPlanetesimals} and~\ref{fig:InnerNebulaStLDependence}.
   \label{fig:InnerNebulaPlanetesimalsAllAlpha}}
\end{figure*}

\begin{figure*}[p]
   \centering

   \vspace{3pt}

   \begin{minipage}{0.49\linewidth}

     \includegraphics[width=\linewidth]{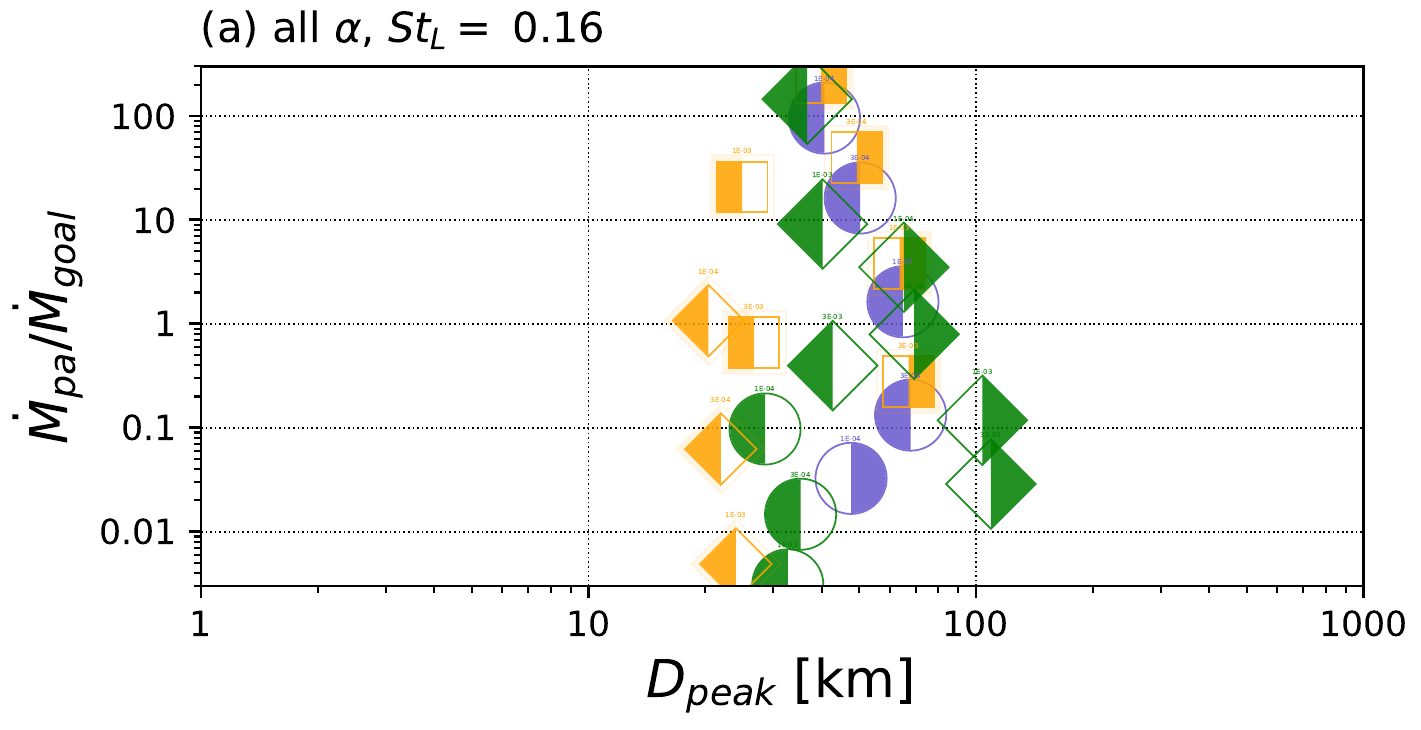}

     \vspace{-24pt}

     \includegraphics[width=\linewidth]{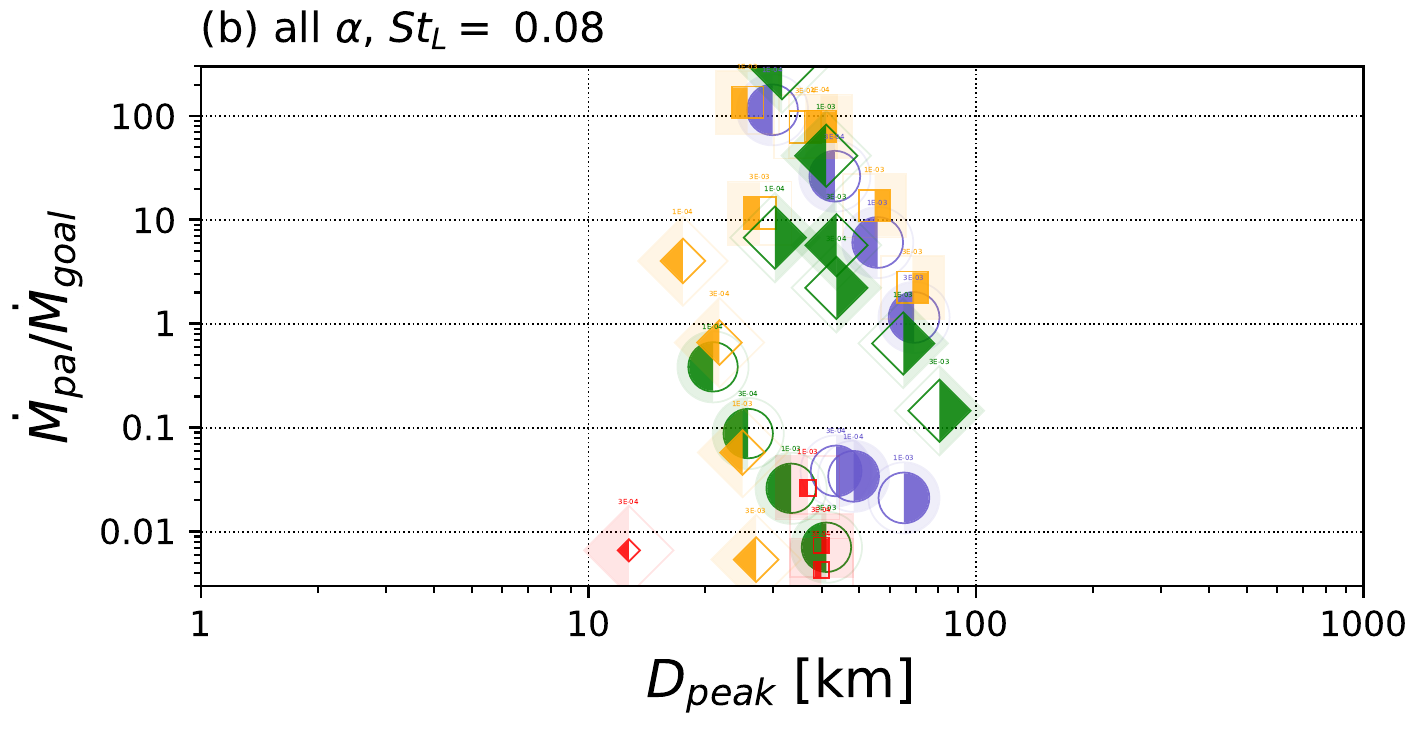}

     \vspace{-24pt}

     \includegraphics[width=\linewidth]{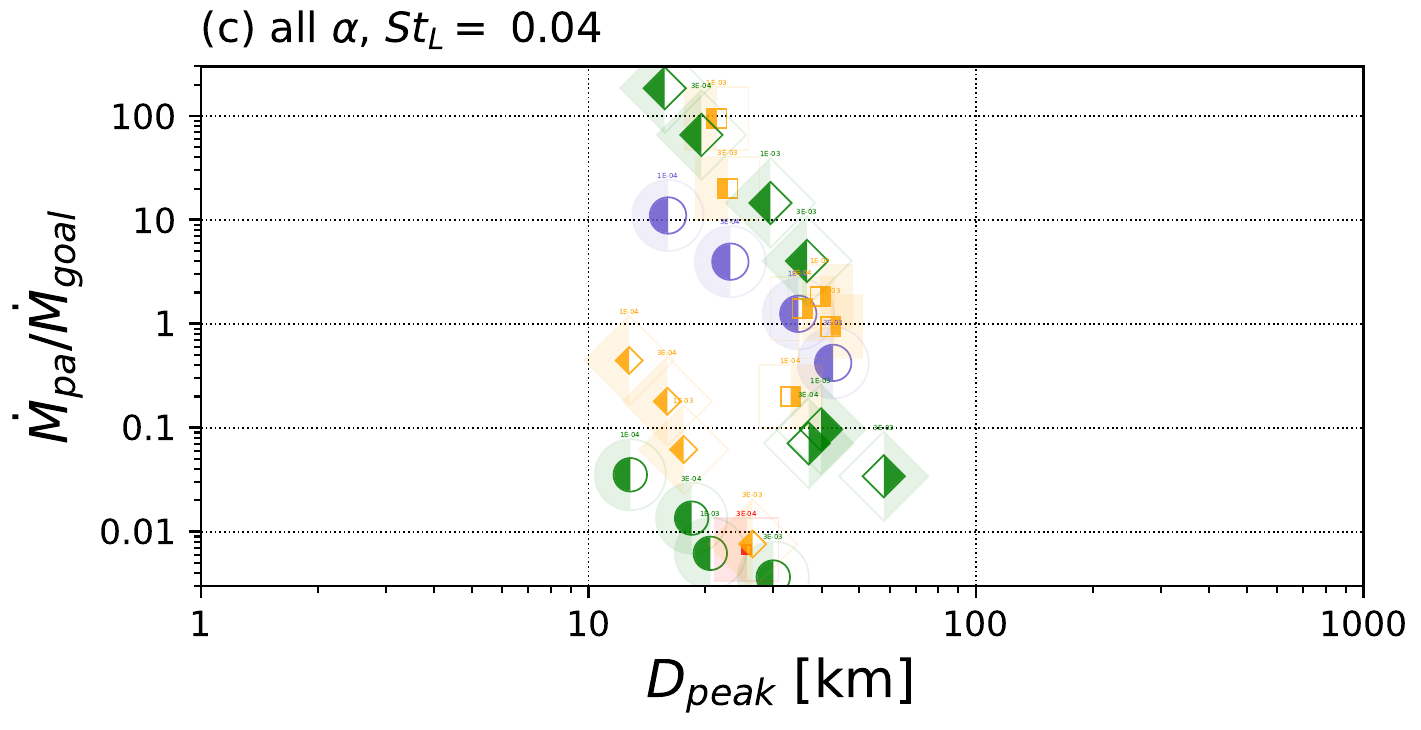}

     \vspace{-24pt}

     \includegraphics[width=\linewidth]{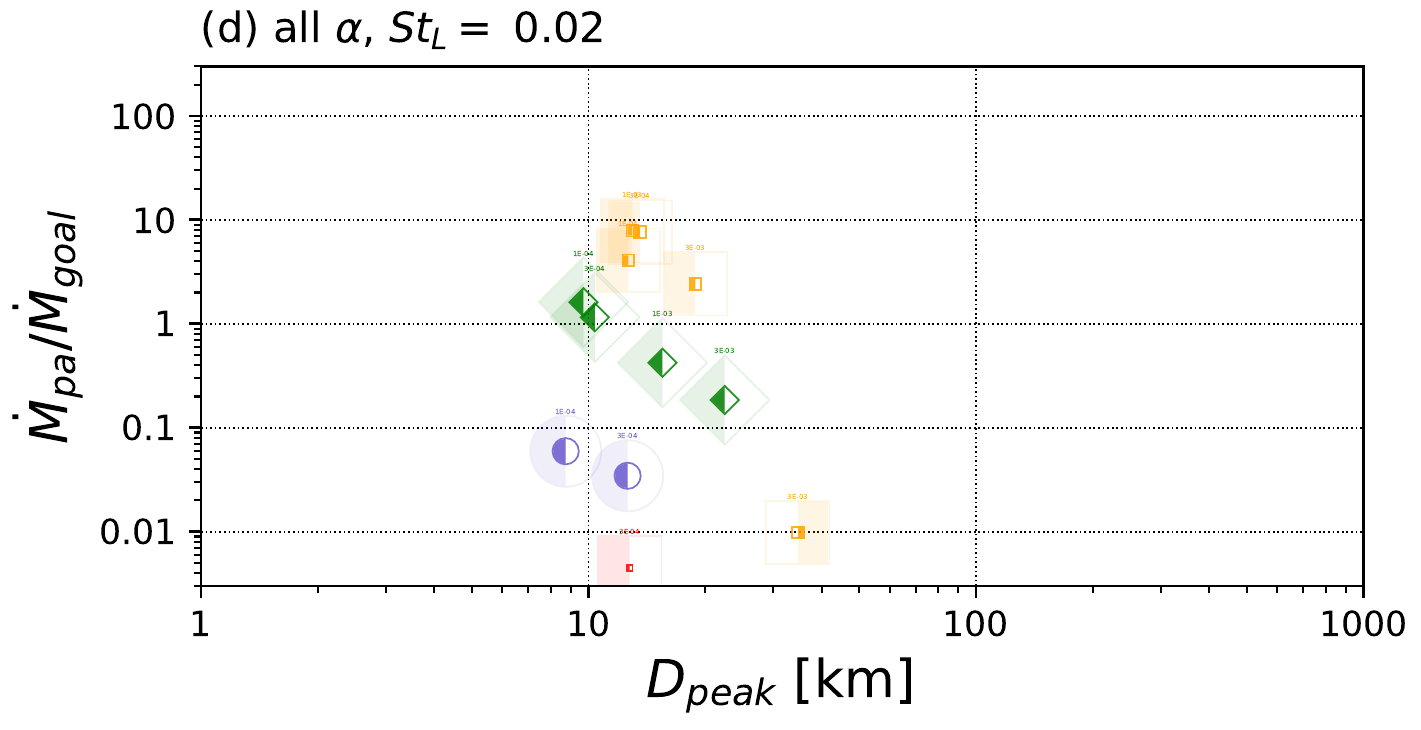}

   \end{minipage}
   \begin{minipage}{0.49\linewidth}

     \vspace{18pt}

     \includegraphics[width=\linewidth]{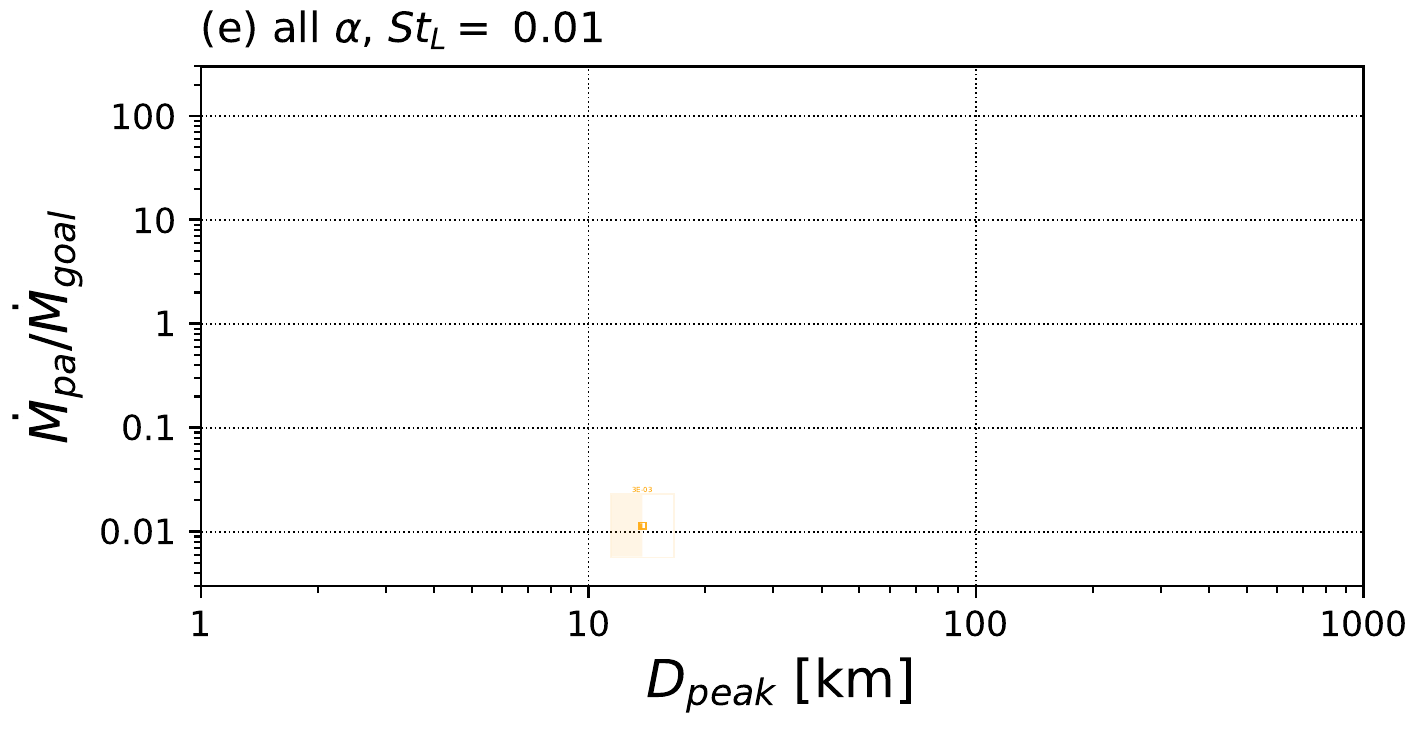}

     \vspace{-24pt}

     \includegraphics[width=\linewidth]{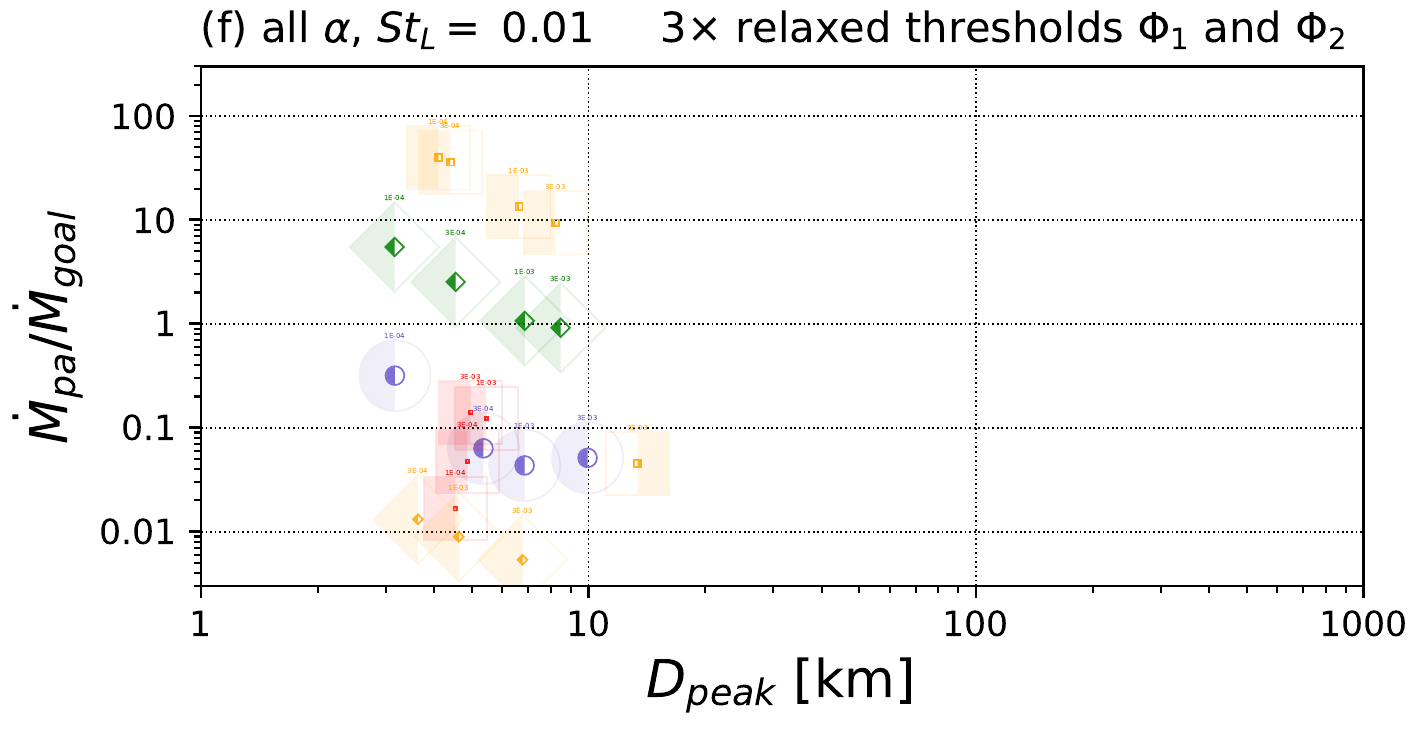}

     \vspace{-24pt}

     \includegraphics[width=\linewidth]{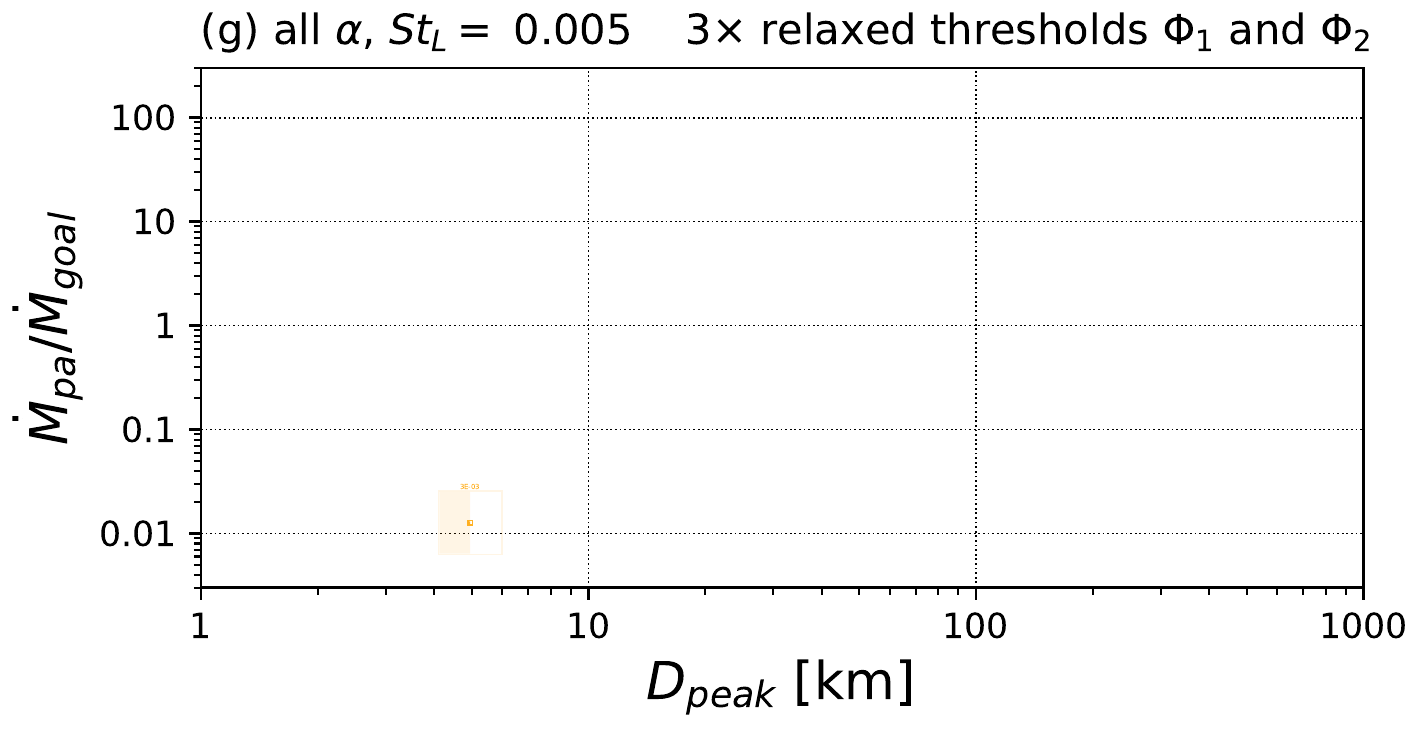}

     \vspace{-3pt}

     \includegraphics[width=0.916\linewidth]{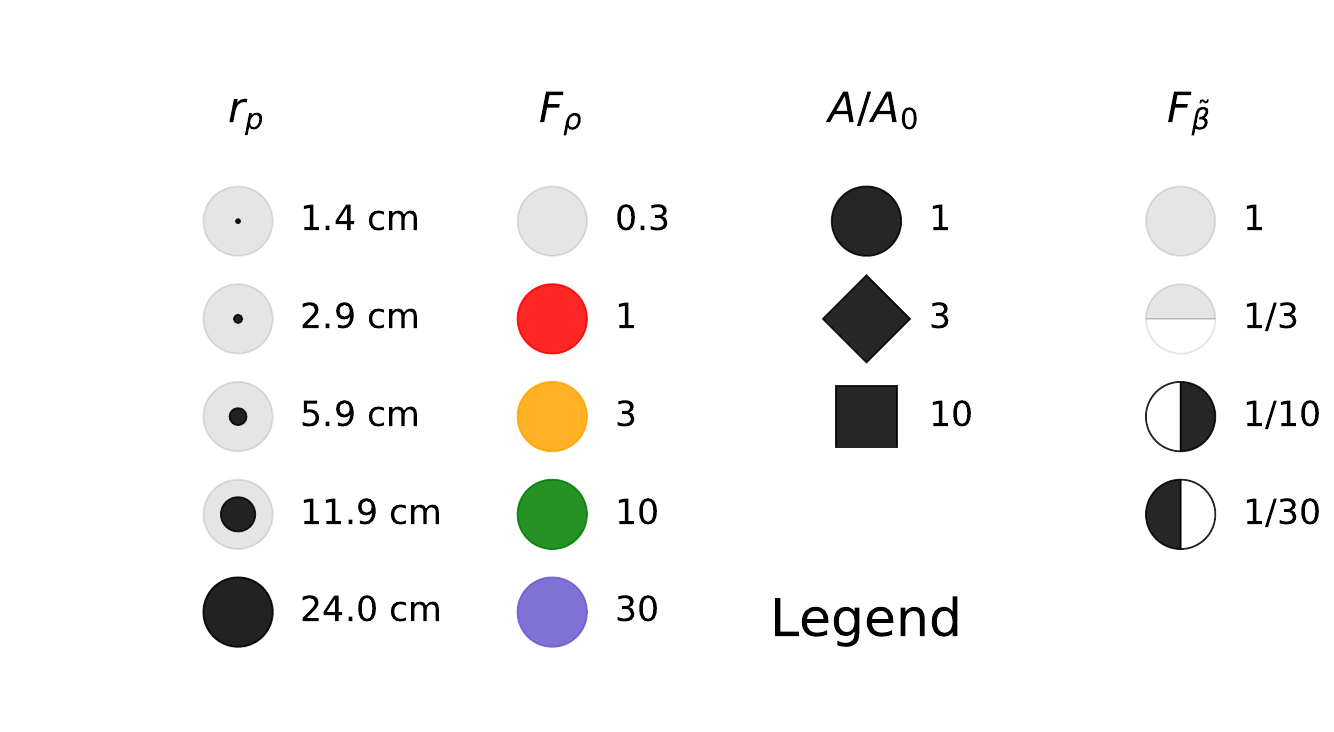}

   \end{minipage}

   \vspace{3pt}

   \caption{Inner nebula results for each $St_L$ shown in separate panels for nominal thresholds (panels {\sl (a)--(e)\/}) and relaxed thresholds (panels {\sl (f)} and {\sl (g)\/}).
   Again, colors, symbol shapes and fill styles denote factors for gas density enhancement, solids enhancement and pressure gradient parameter, respectively.
   The value of $\alpha$ for each case is printed above each datapoint in small font best seen under magnification.
   The smallest planetesimal-producing Stokes numbers for these cases are 0.01 for the nominal thresholds, and 0.005 for the relaxed thresholds.
   \label{fig:InnerNebulaStLDependence}}
\end{figure*}

The results of the present model for the formation of planetesimals in the inner nebula, specifically their sizes and production rates, are shown in Figures~\ref{fig:InnerNebulaPlanetesimals}--\ref{fig:InnerNebulaStLDependence}.
The rate of planetesimal production, $\dot{M}_{\rm{pa}}$, by our process of primary accretion is normalized by the estimated {\it required} rate, $\dot{M}_{\rm{goal}} \equiv M_{\rm{goal}} / T_{\rm{neb}}$.
As in \citet[their section 3.3.1.]{Cuzzietal2010}, we assume that a mass of $M_{\rm{goal}}=2M_\oplus$ was turned into planetesimals, in the $2-4$\,AU region, within the lifetime of the nebula ($T_{\rm{neb}} = 2$\,Myrs).
Clearly, these numbers are uncertain.
Given these and other uncertainties in the study (including the cascade model itself and the threshold treatment), one can probably consider values of $\dot{M}_{\rm{pa}}/\dot{M}_{\rm{goal}}$ two orders of magnitude around unity as plausible.

The results of Figures~\ref{fig:InnerNebulaPlanetesimals}--\ref{fig:InnerNebulaStLDependence} show that larger particles generally make larger planetesimals, and that particles of at least cm size are needed to form planetesimals.
This is the major difference from the \citet{Cuzzietal2010} models that allowed even chondrule size particles to form planetsimals directly, because of their incorrect cascade model.
We will discuss implications of these new results in Section~\ref{sec:conclusions}.

Figure~\ref{fig:InnerNebulaPlanetesimals} also explores how the intensity of turbulence affects the formation process by plotting results separately for different values of $\alpha$.
The $\alpha$-dependence is not very strong; larger values of $\alpha$ do yield slightly larger planetesimals but at lower production rates.

Changing the uncertain threshold limits is explored in Figure~\ref{fig:InnerNebulaPlanetesimalsAllAlpha}.
Relaxing, that is, lowering them by factors of three does affect the minimum and maximum size of possible planetesimals but does not change the smallest size of particles able to form planetesimals.

Particles of a given size will have different Stokes numbers $St_L$ under different nebula conditions,  and it is really the Stokes number that is relevant for the turbulent concentration effect.
The dependence on $St_L$ is investigated in Figure~\ref{fig:InnerNebulaStLDependence} which shows that
large Stokes numbers more easily lead to formation of large planetesimals.
This is due to the fact that large Stokes numbers reach large concentrations at larger scales than do smaller $St_L$, and therefore more mass is available in a clump.
For nominal thresholds, we find that few to no planetesimals can form for $St_L \lesssim 0.01$.
Relaxing thresholds only reduces the minimum required $St_L$ slightly, and the planetesimals that do form for small $St_L$ are rather small ($\lesssim 10$\,km).

Another parameter in our model that can be explored is the pressure parameter  $\tilde\beta$, which we vary through the scale factor $F_{\tilde\beta}$ relative to the nominal $\tilde\beta$ value (see Section~\ref{sec:model:nebula}).
Although evident in any of the figures, it is most easily seen in Figure~\ref{fig:InnerNebulaPlanetesimals} that smaller $\tilde\beta$ values produce smaller planetesimals but with a higher rate of formation (symbols move up and left in the plot).
Reducing $\tilde\beta$ has the same effect as relaxing the $\Phi_2$ threshold (see Equation~\ref{eqn:phi2}) in that it allows clumps of less strongly concentrated particles (which occur more often) to form planetesimals.

\subsection{Outer nebula -- TNOs} \label{sec:outernebula}

\begin{figure*}
   \vspace{10pt}
   \centering

   \begin{minipage}{0.49\linewidth}

   \includegraphics[width=\linewidth]{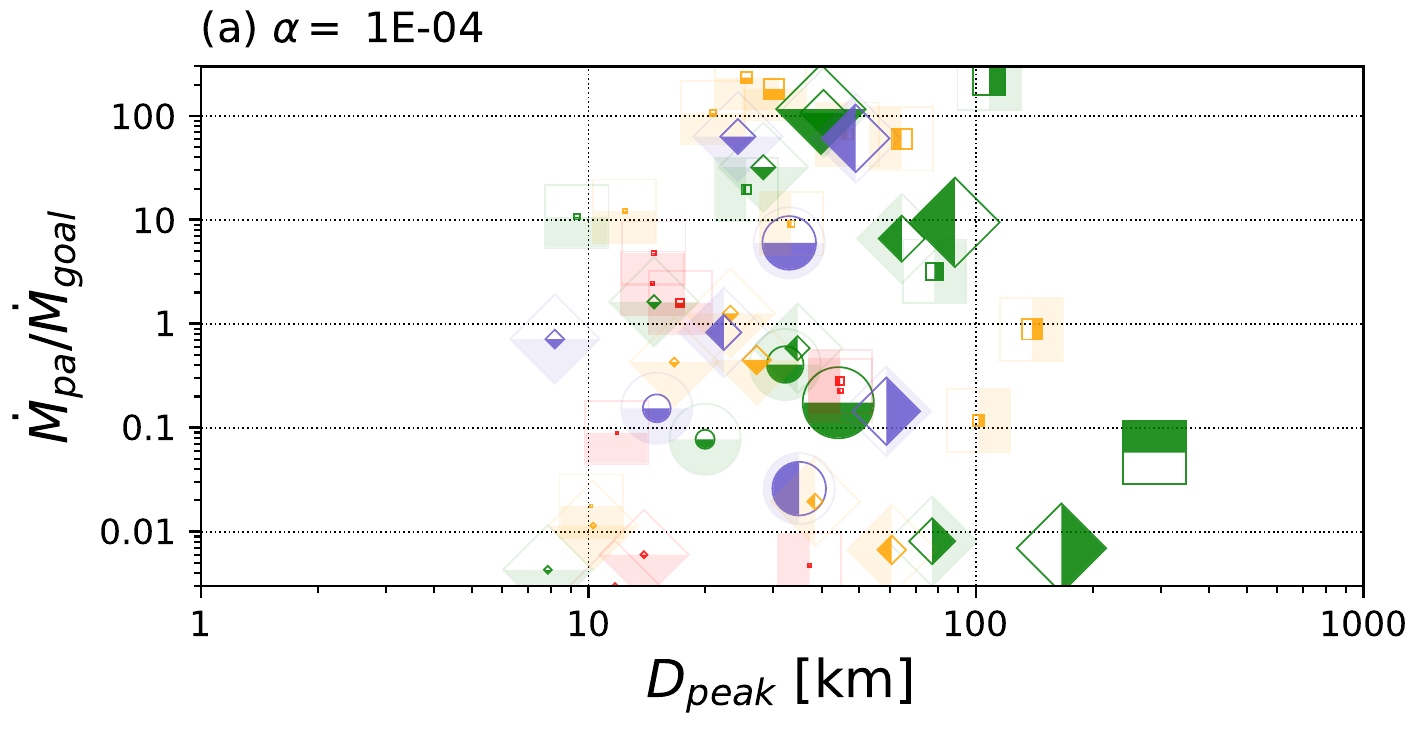}

   \vspace{-24pt}

   \includegraphics[width=\linewidth]{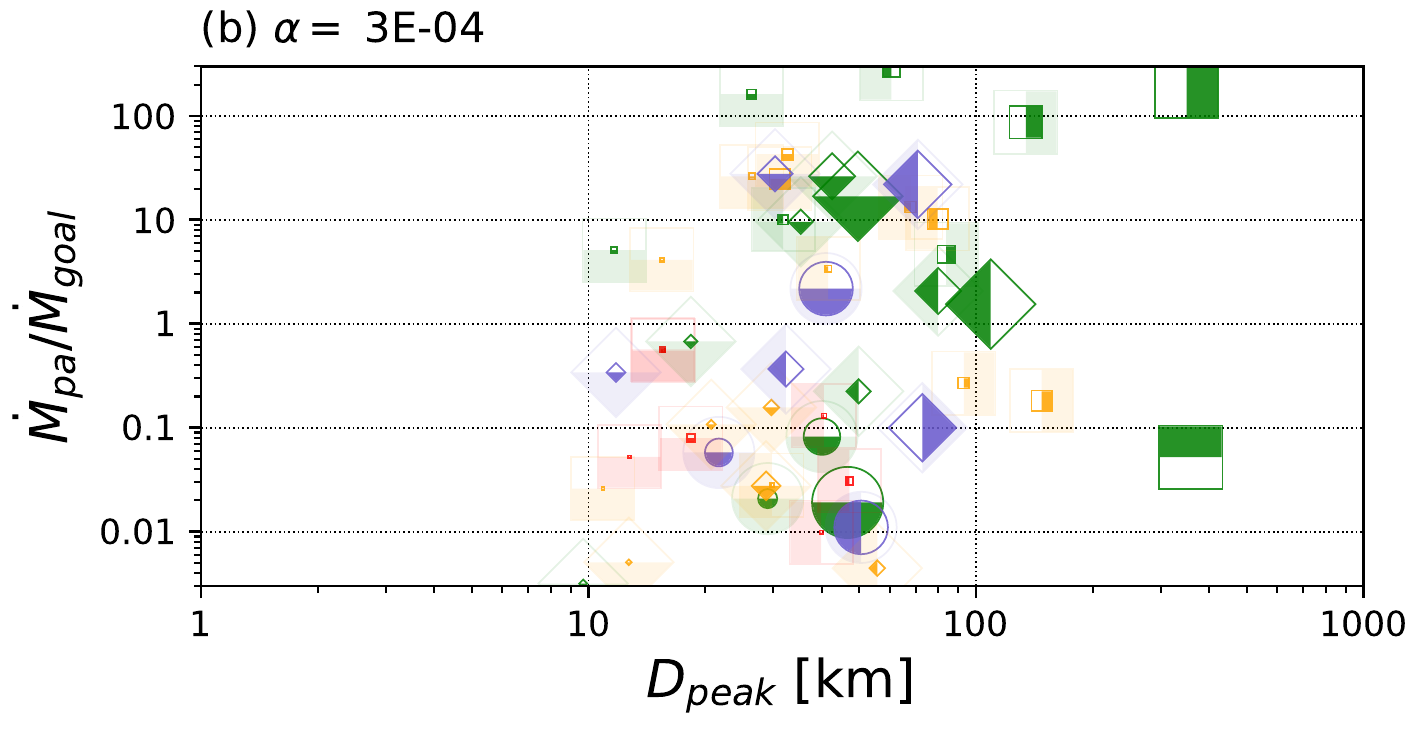}

   \end{minipage}
   \begin{minipage}{0.49\linewidth}

   \includegraphics[width=\linewidth]{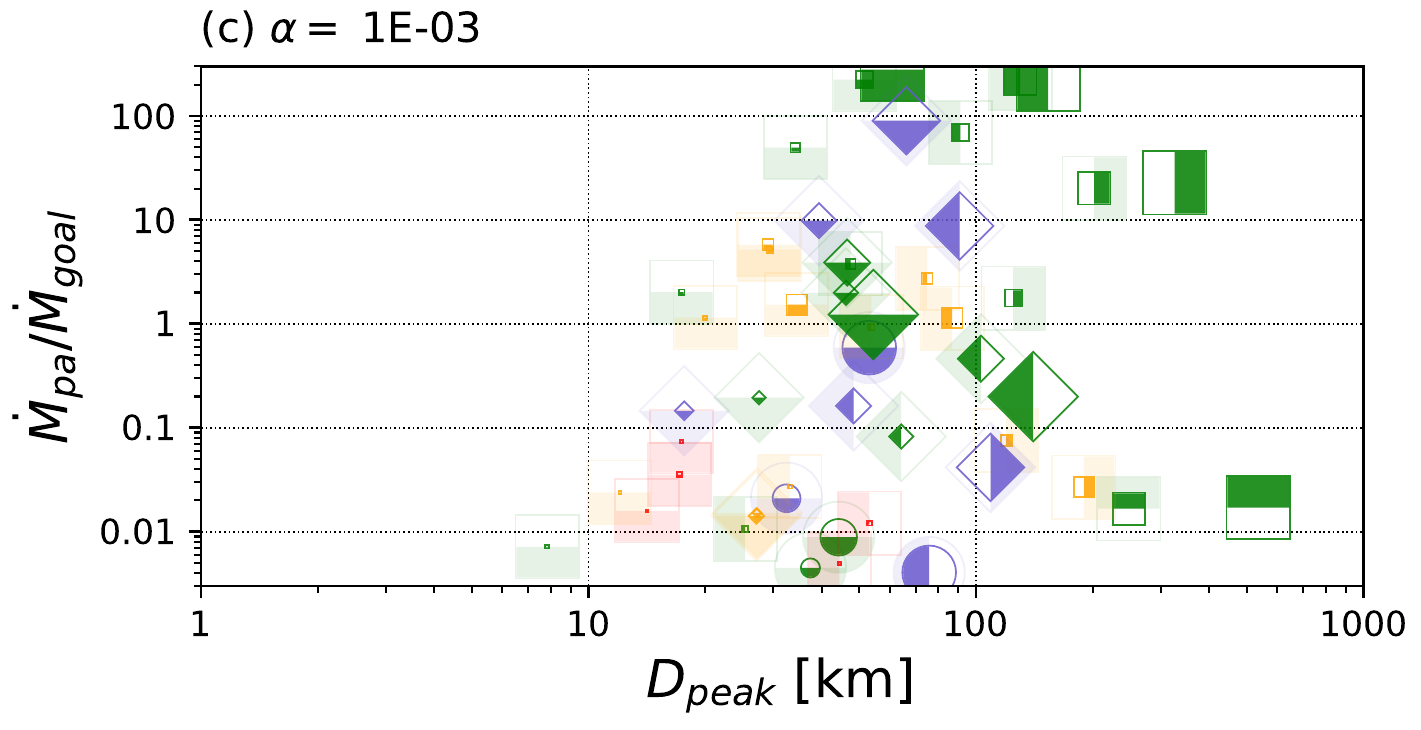}

   \vspace{-24pt}

   \includegraphics[width=\linewidth]{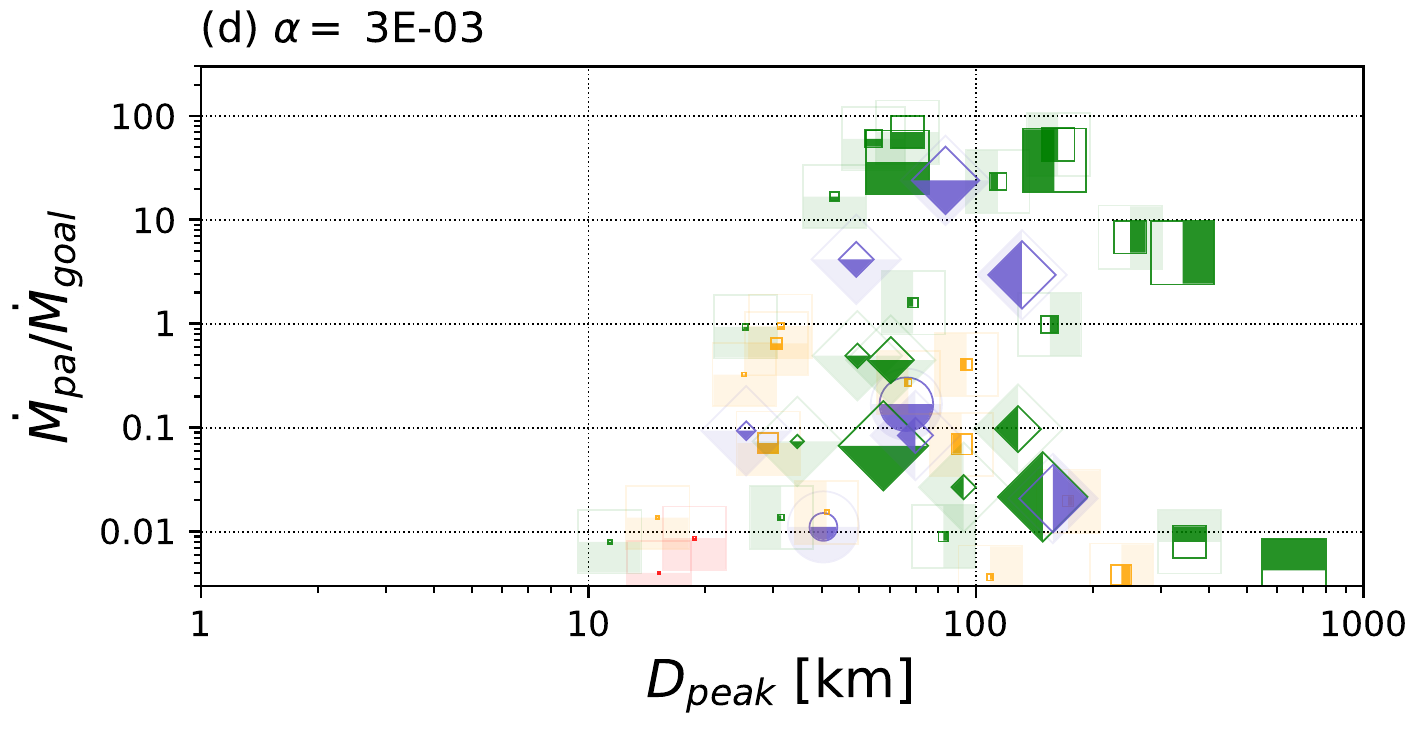}

   \end{minipage}

   \vspace{0pt}

   \includegraphics[width=0.44\linewidth]{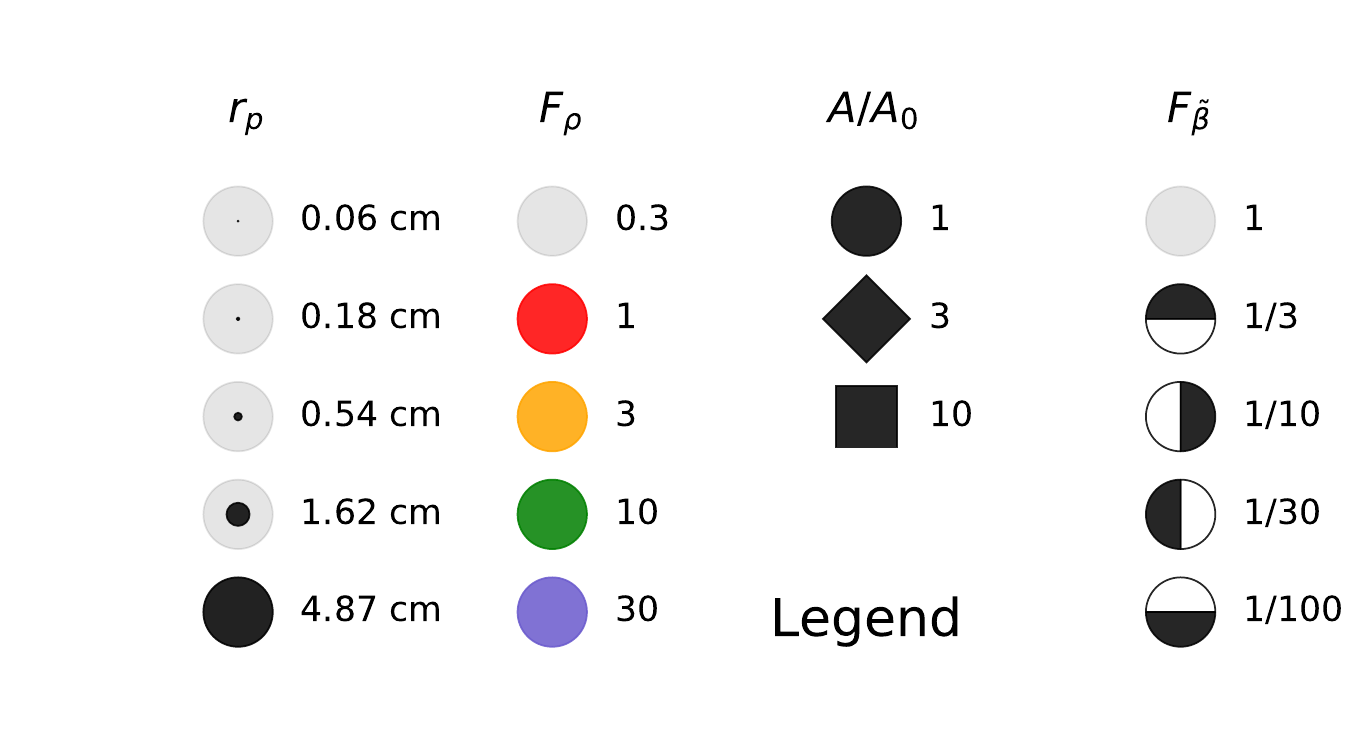}

   \vspace{0pt}

   \caption{
   Model results for the outer nebula region.
   Similar to Figure~\ref{fig:InnerNebulaPlanetesimals}, the figure shows the peak of the planetesimal IMF, $D_{\rm{peak}}$, and the rate of formation, $\dot{M}_{\rm{pa}}$, relative to the excepted rate, $\dot{M}_{\rm{goal}}$, for the nominal range of possible nebula parameters and particle sizes. The panels {\sl(a)} through {\sl(d)} show results for different values of $\alpha$, while within each panel colors, symbol shapes and fill styles denote gas density enhancement factor, $F_\rho$, solids enhancement factor, $A/A_o$, and headwind parameter scale factor, $F_{\tilde\beta}$. The size of the symbols scales with the particle size. Results for particles larger than 5\,cm have been omitted.
   \label{fig:OuterNebulaPlanetesimals}}
\end{figure*}

\begin{figure*}[p]
   \centering

   \vspace{3pt}

   \begin{minipage}{0.49\linewidth}

     \includegraphics[width=\linewidth]{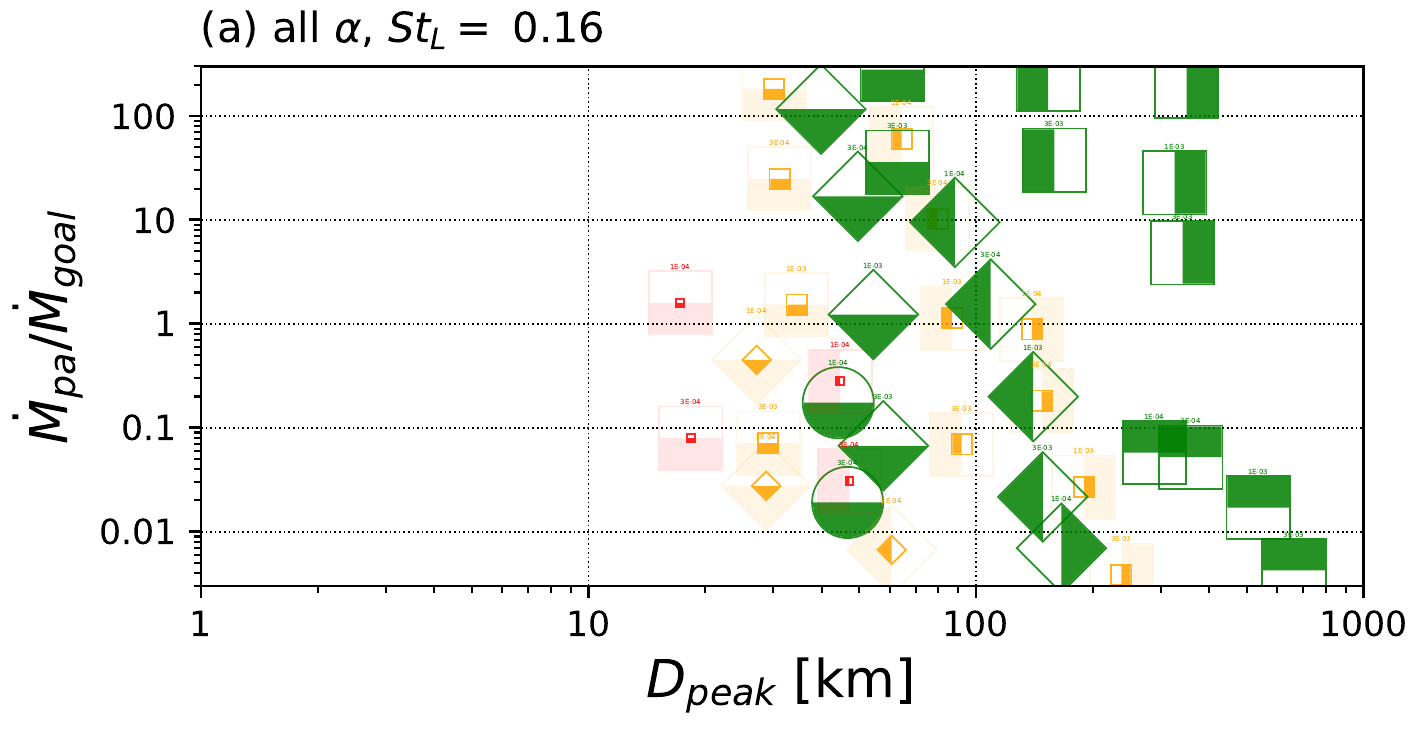}

     \vspace{-24pt}

     \includegraphics[width=\linewidth]{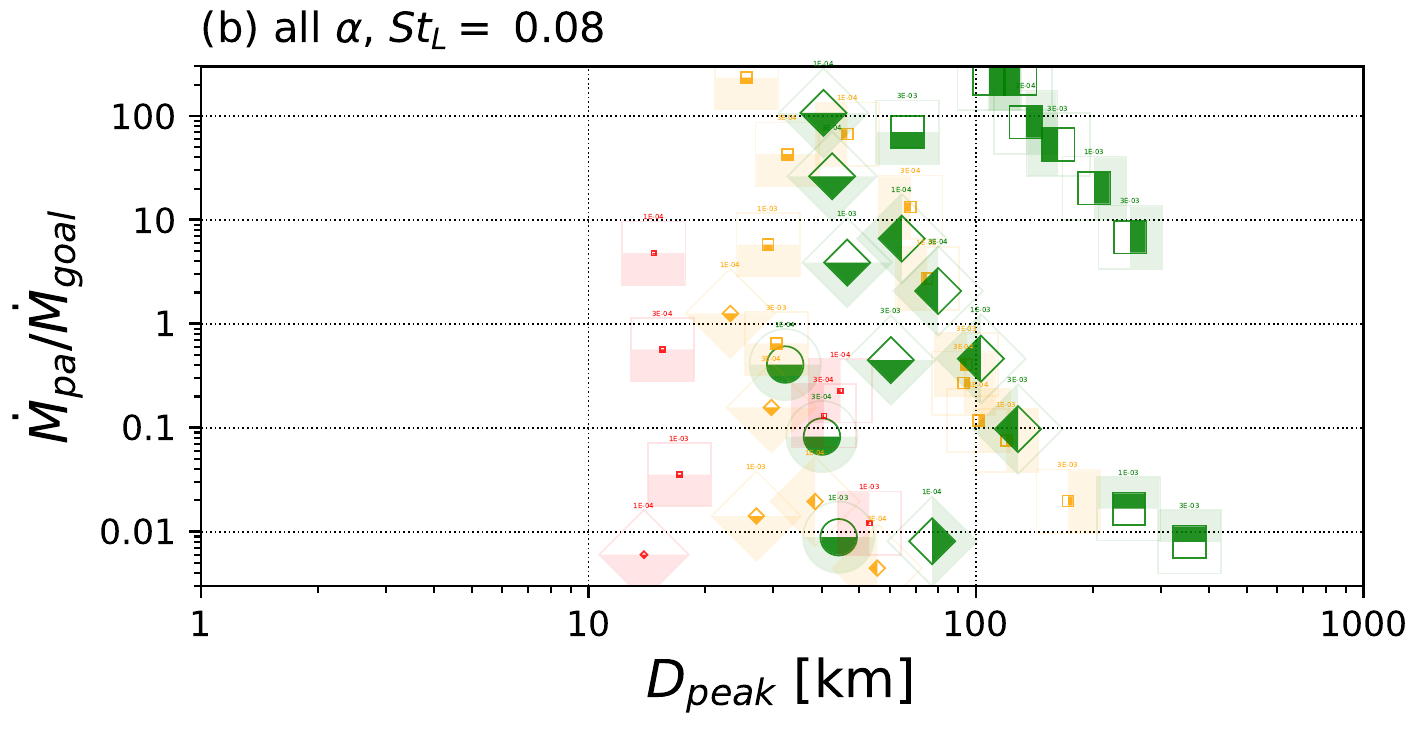}

     \vspace{-24pt}

     \includegraphics[width=\linewidth]{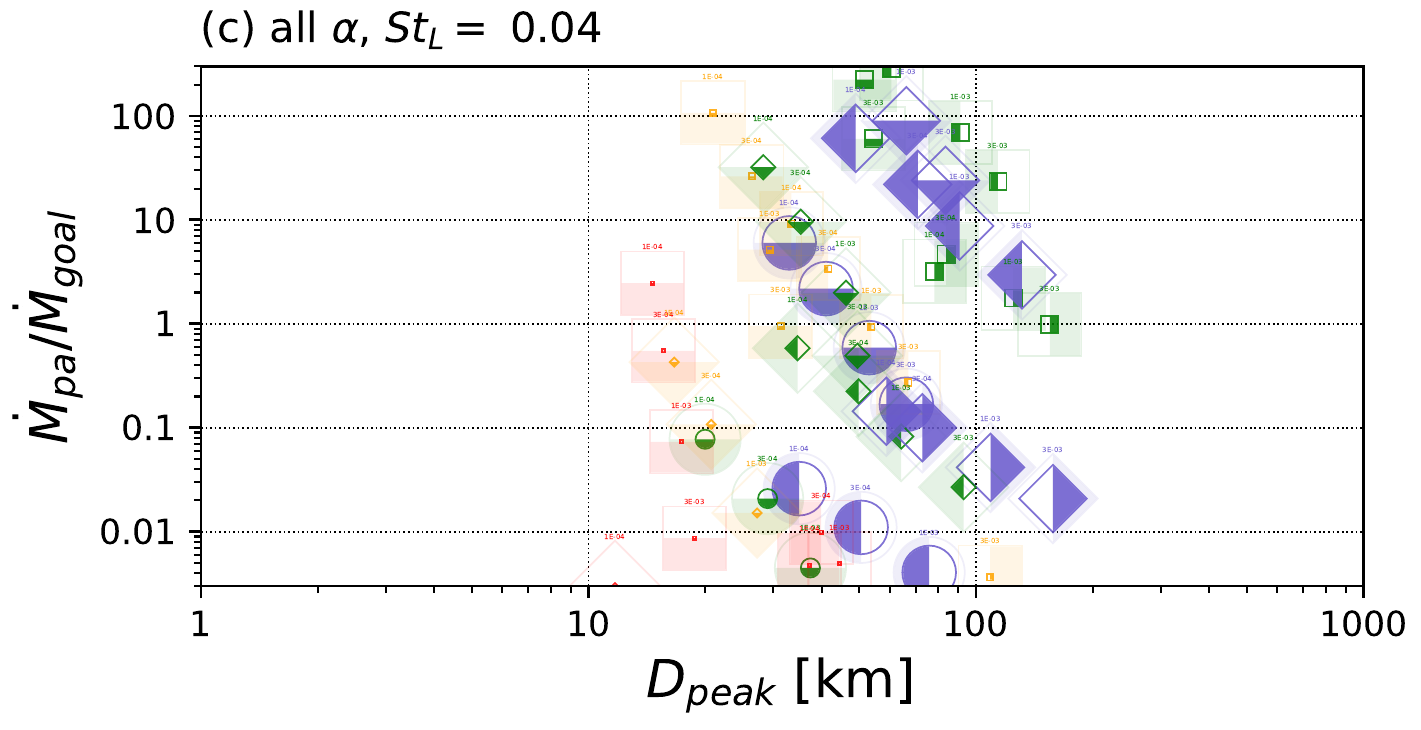}

     \vspace{-24pt}

     \includegraphics[width=\linewidth]{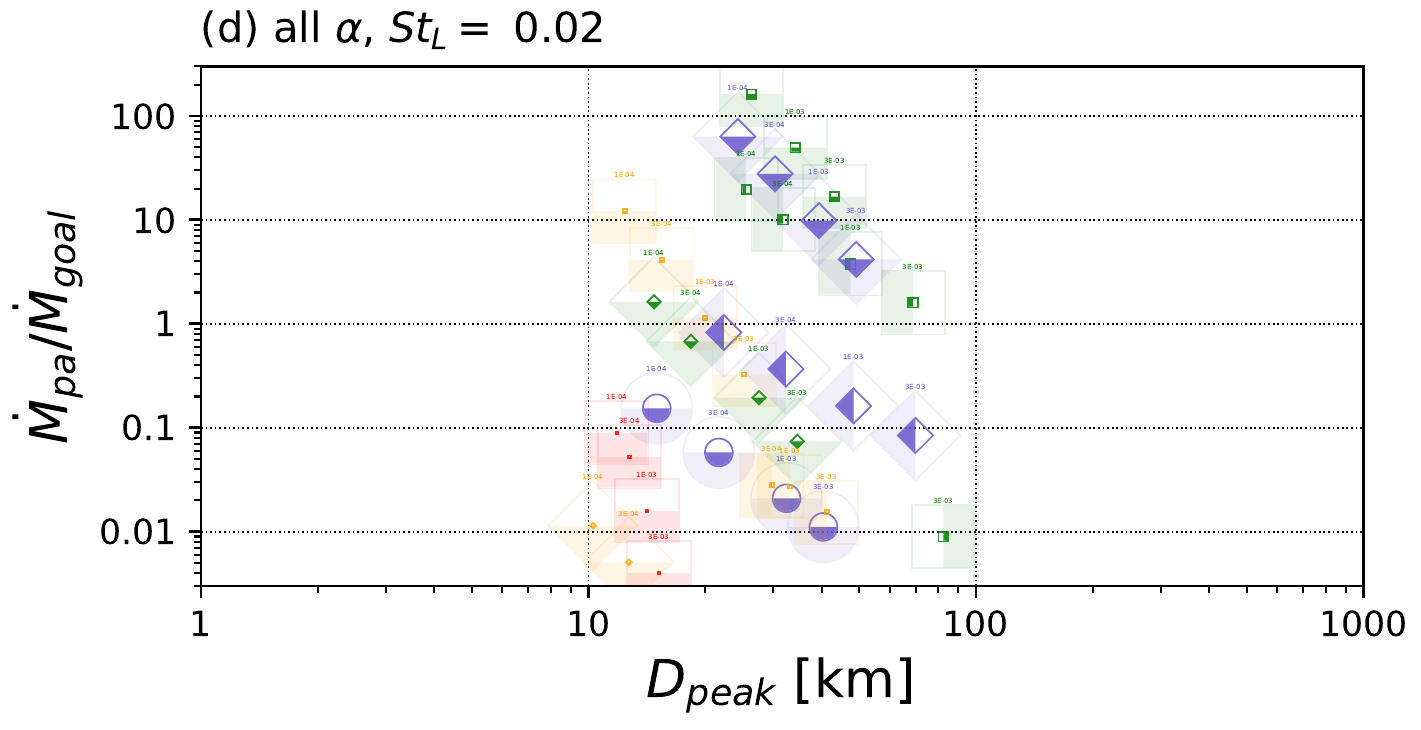}

   \end{minipage}
   \begin{minipage}{0.49\linewidth}

     \includegraphics[width=\linewidth]{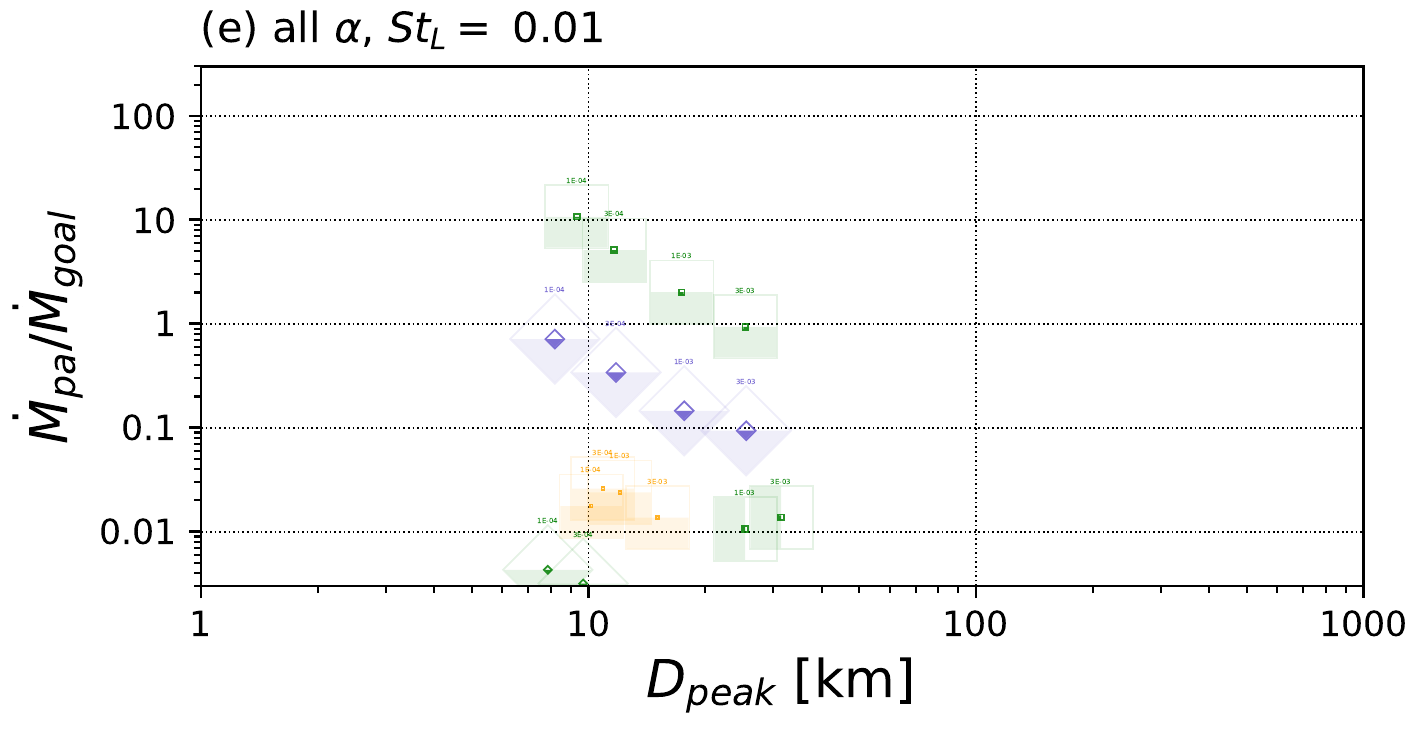}

     \vspace{-24pt}

     \includegraphics[width=\linewidth]{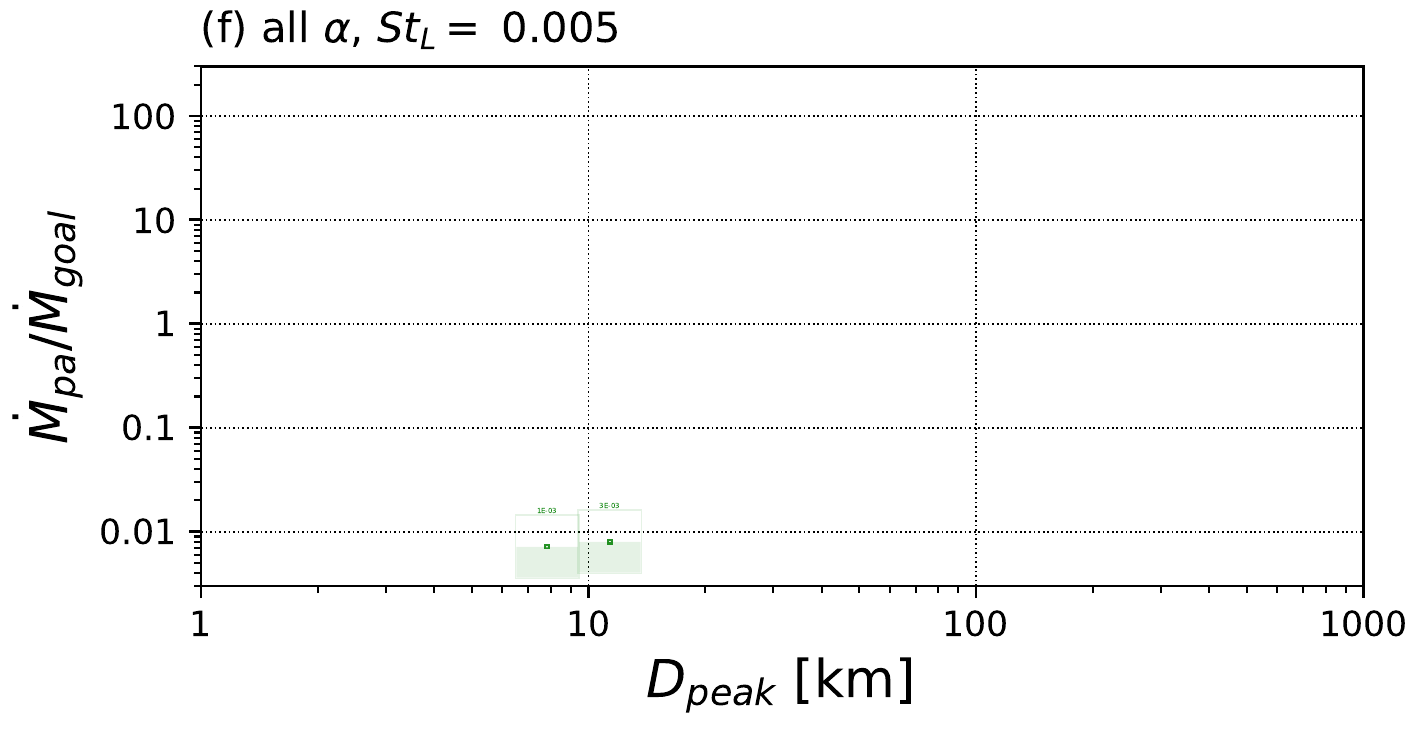}

     \vspace{-24pt}

     \includegraphics[width=\linewidth]{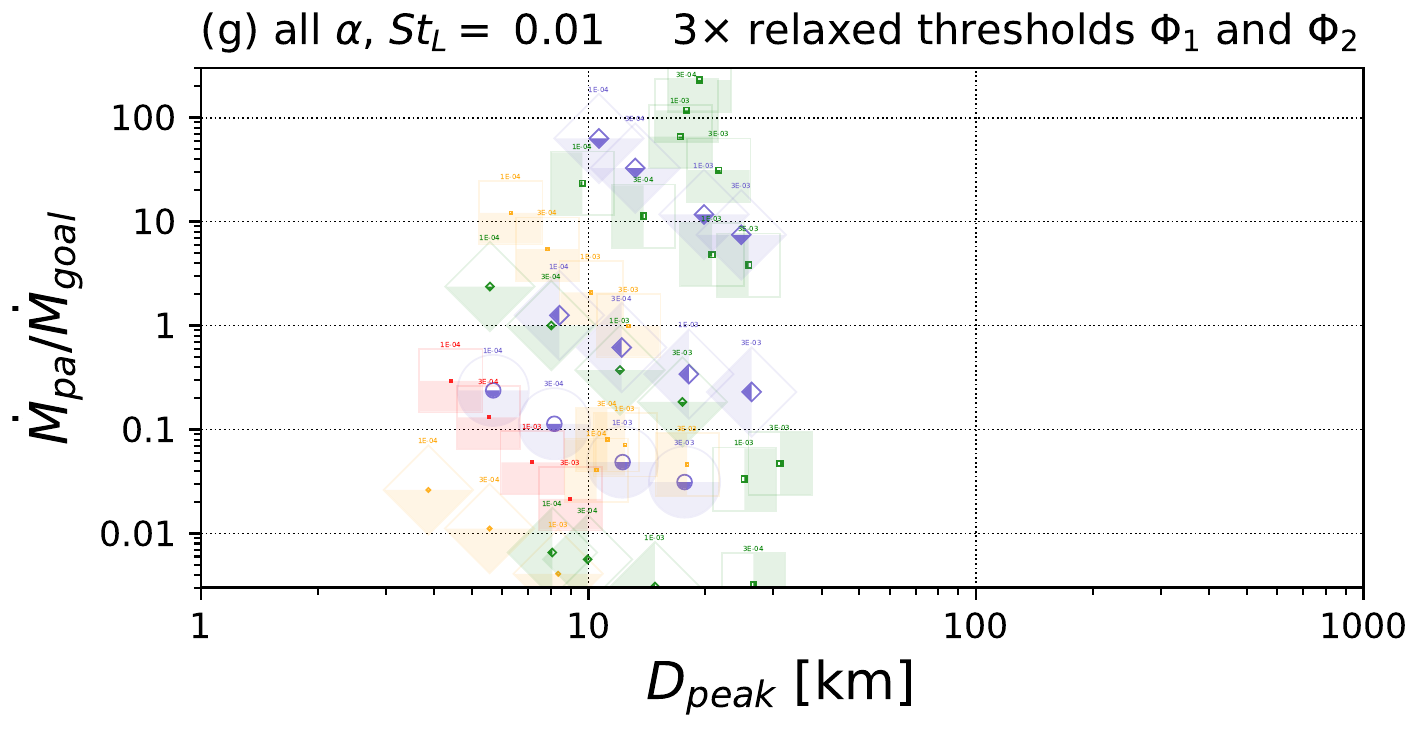}

     \vspace{-24pt}

     \includegraphics[width=\linewidth]{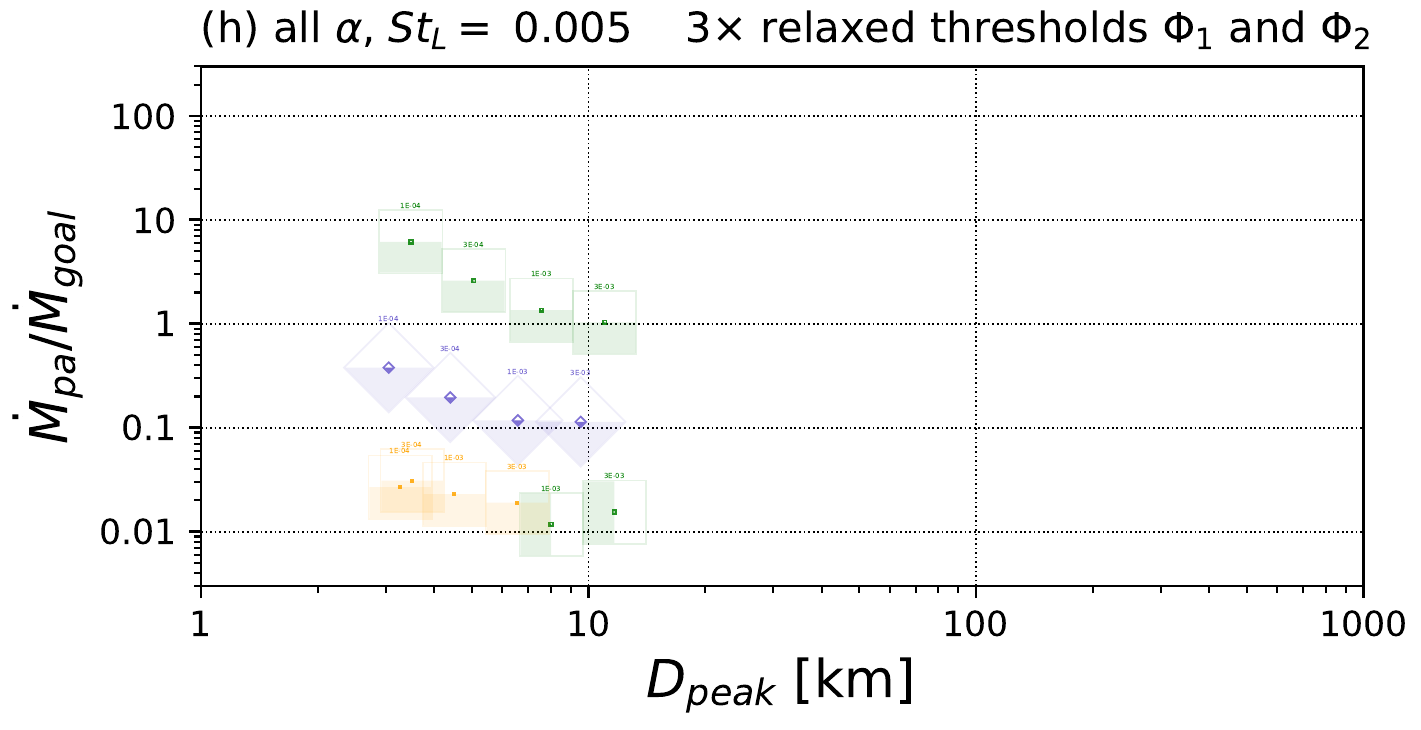}

   \end{minipage}

   \vspace{-3pt}

   \includegraphics[width=0.449\linewidth]{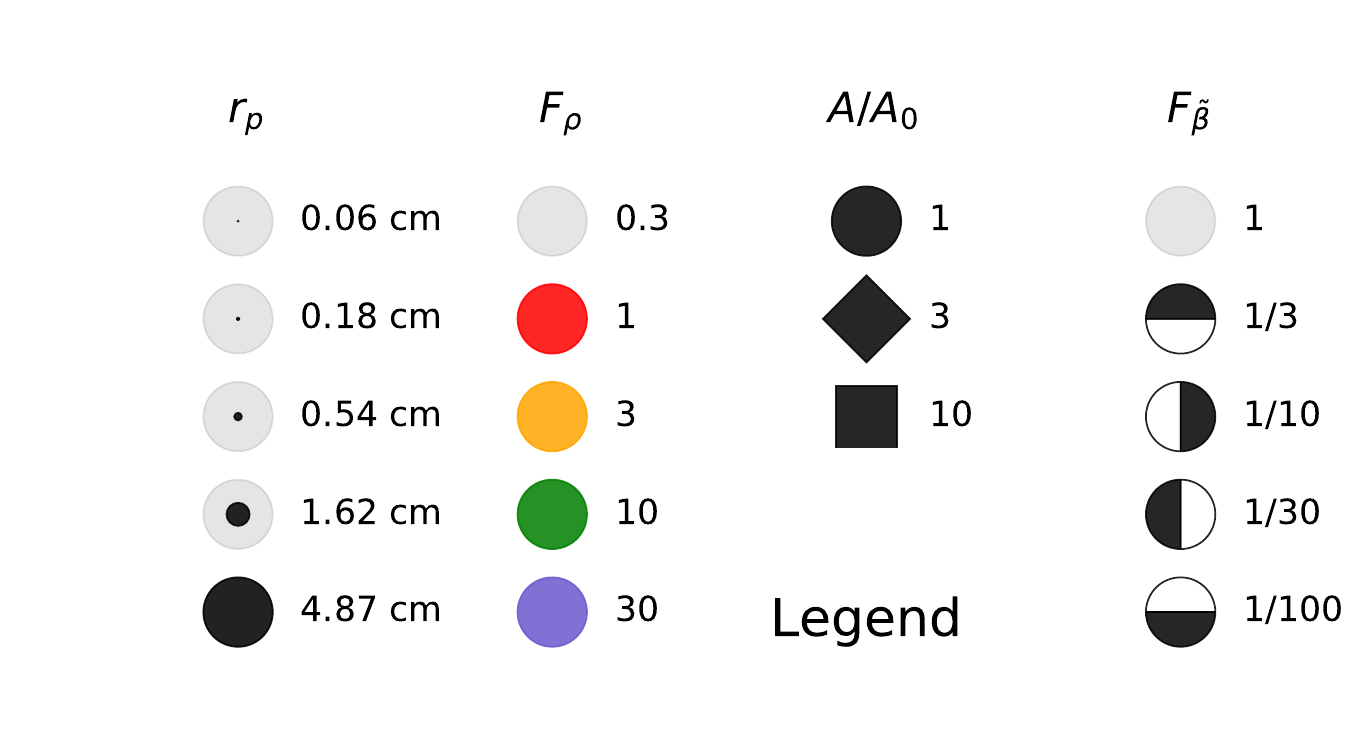}

   \vspace{3pt}

   \caption{Model results for the outer nebula where similar to Figure~\ref{fig:InnerNebulaStLDependence} Stokes numbers are plotted in separate panels for nominal thresholds (panels {\sl (a)--(f)\/}) and relaxed thresholds (panels {\sl (g), (h)\/}).
   Again, colors, symbol shapes and fill styles denote factors for gas density enhancement, solids enhancement and pressure gradient parameter, respectively.
   The value of $\alpha$ for each case is printed above each datapoint in small font best seen under magnification.
   Particle sizes above 5\,cm have been omitted for legibility.
   The smallest planetesimal-producing Stokes number is $St_L=0.005$, both for nominal and relaxed thresholds.
   \label{fig:OuterNebulaStLDependence}}
\end{figure*}

\begin{figure*}
   \centering

   \vspace{3pt}

   \begin{minipage}{0.49\linewidth}

     \includegraphics[width=\linewidth]{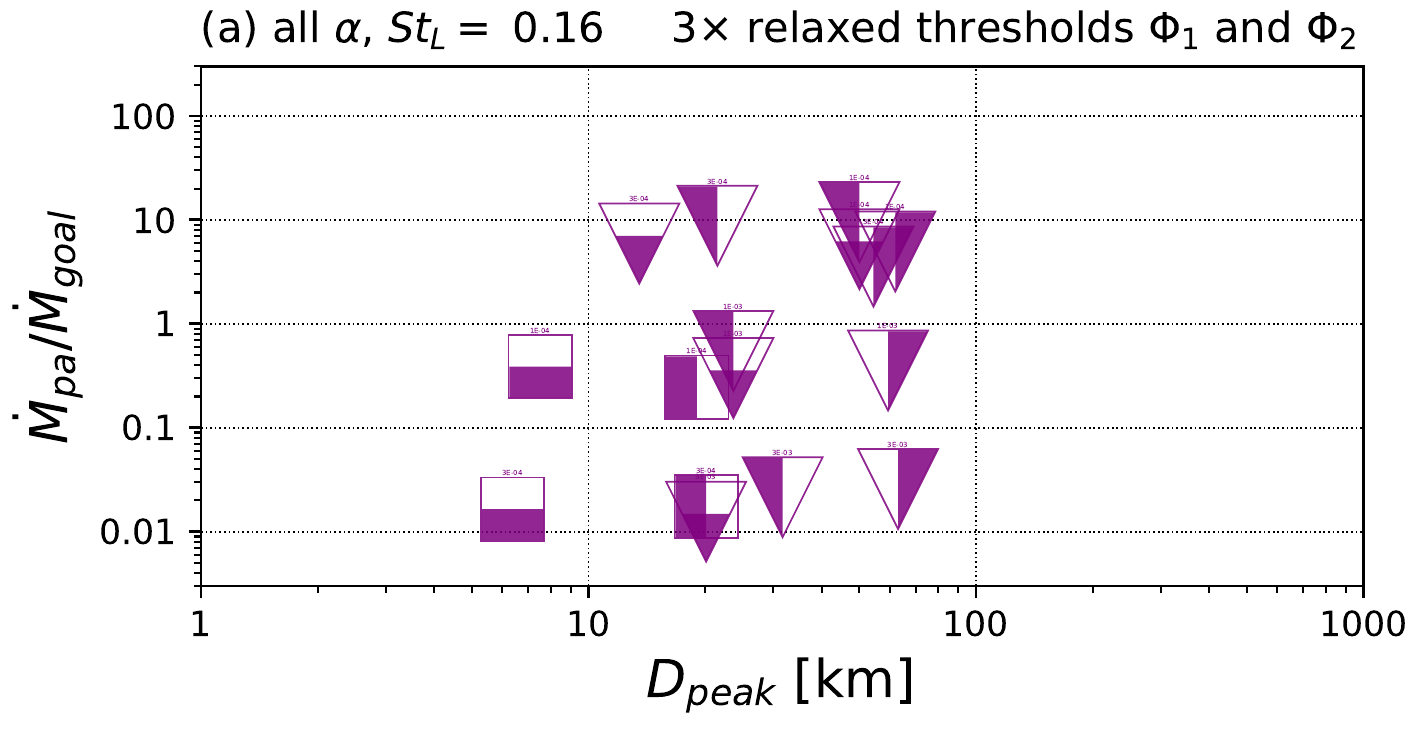}

     \vspace{-24pt}

     \includegraphics[width=\linewidth]{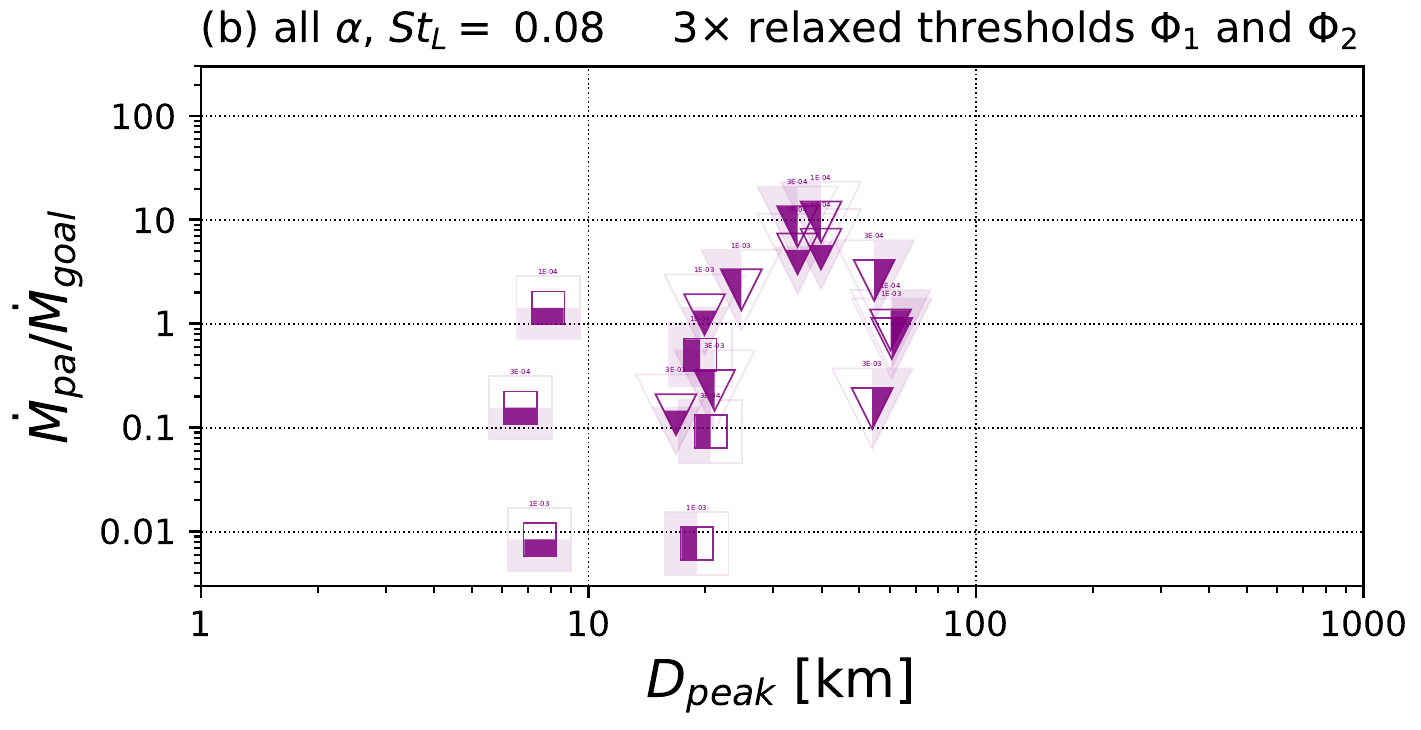}

     \vspace{-24pt}

     \includegraphics[width=\linewidth]{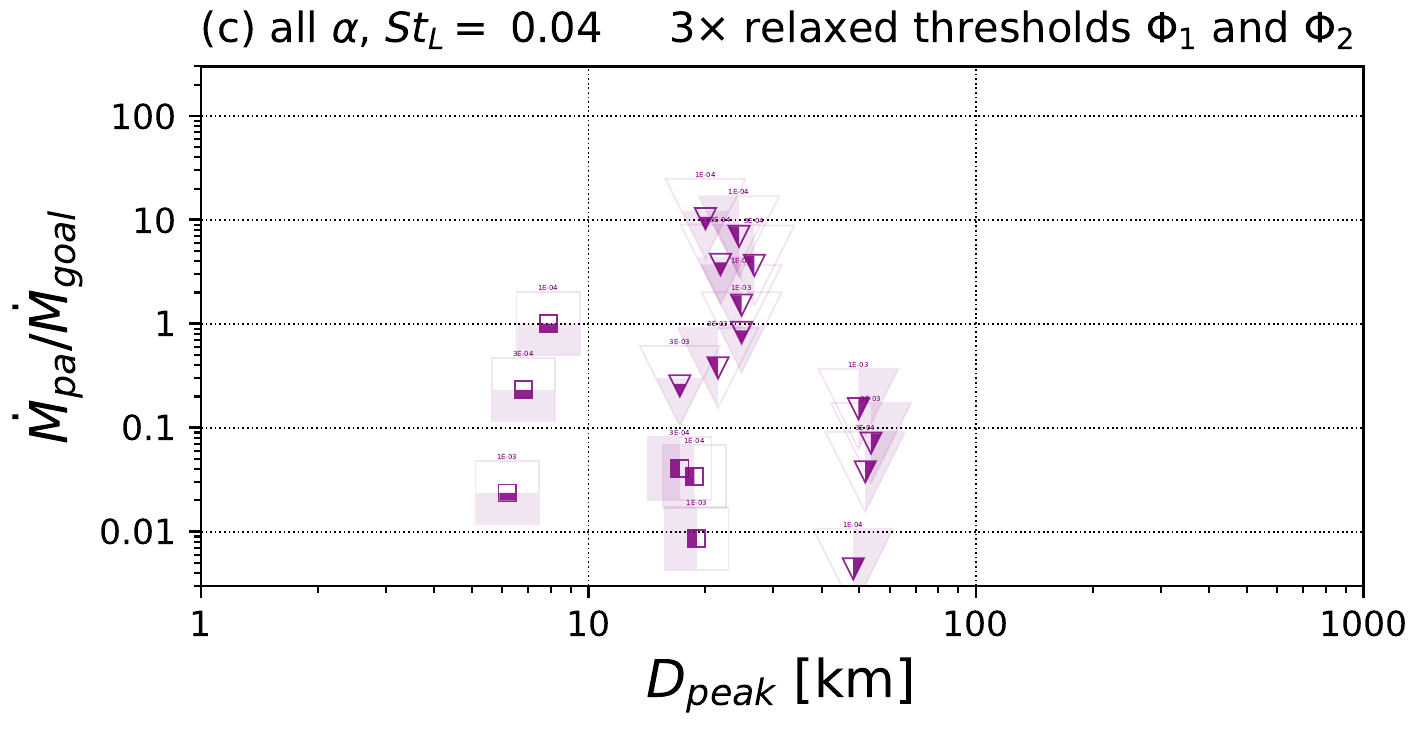}

   \end{minipage}
   \begin{minipage}{0.49\linewidth}

     \vspace{16pt}

     \includegraphics[width=\linewidth]{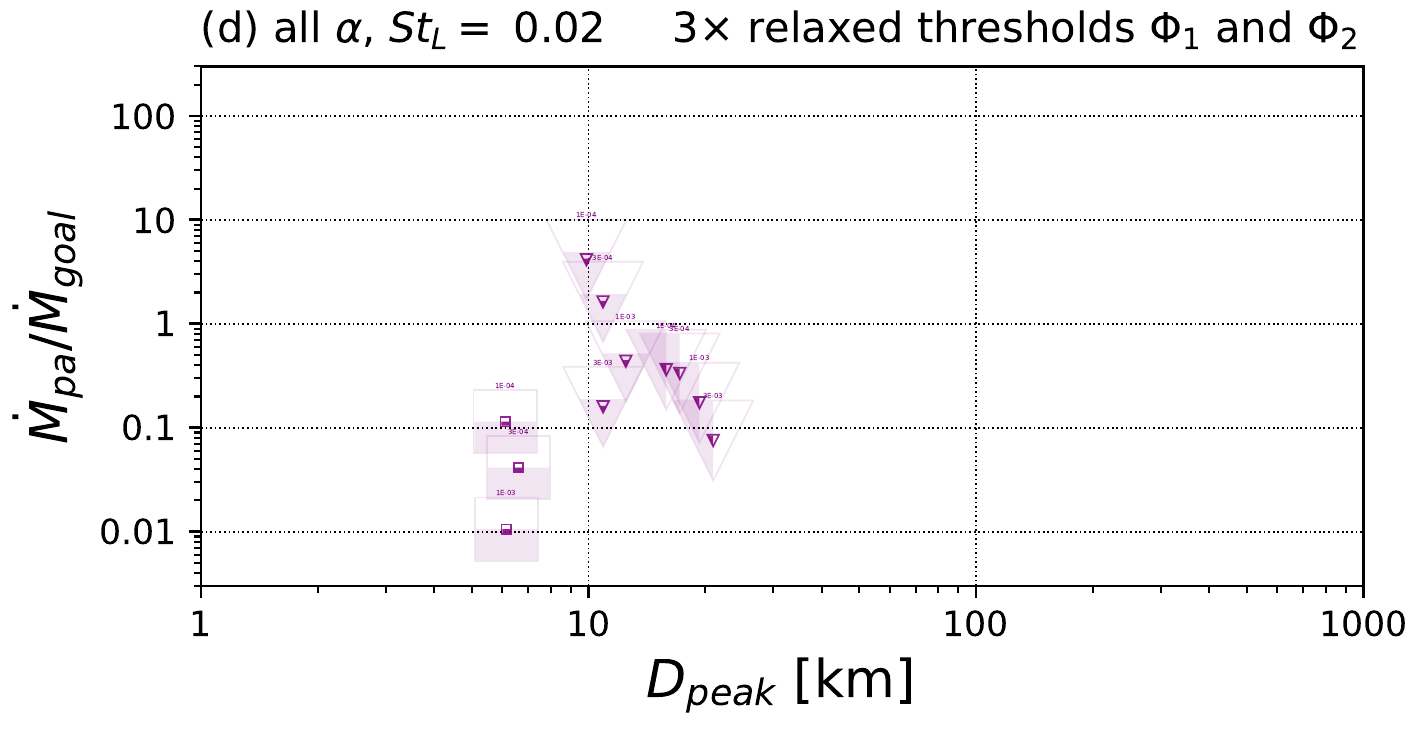}

     \vspace{-24pt}

     \includegraphics[width=\linewidth]{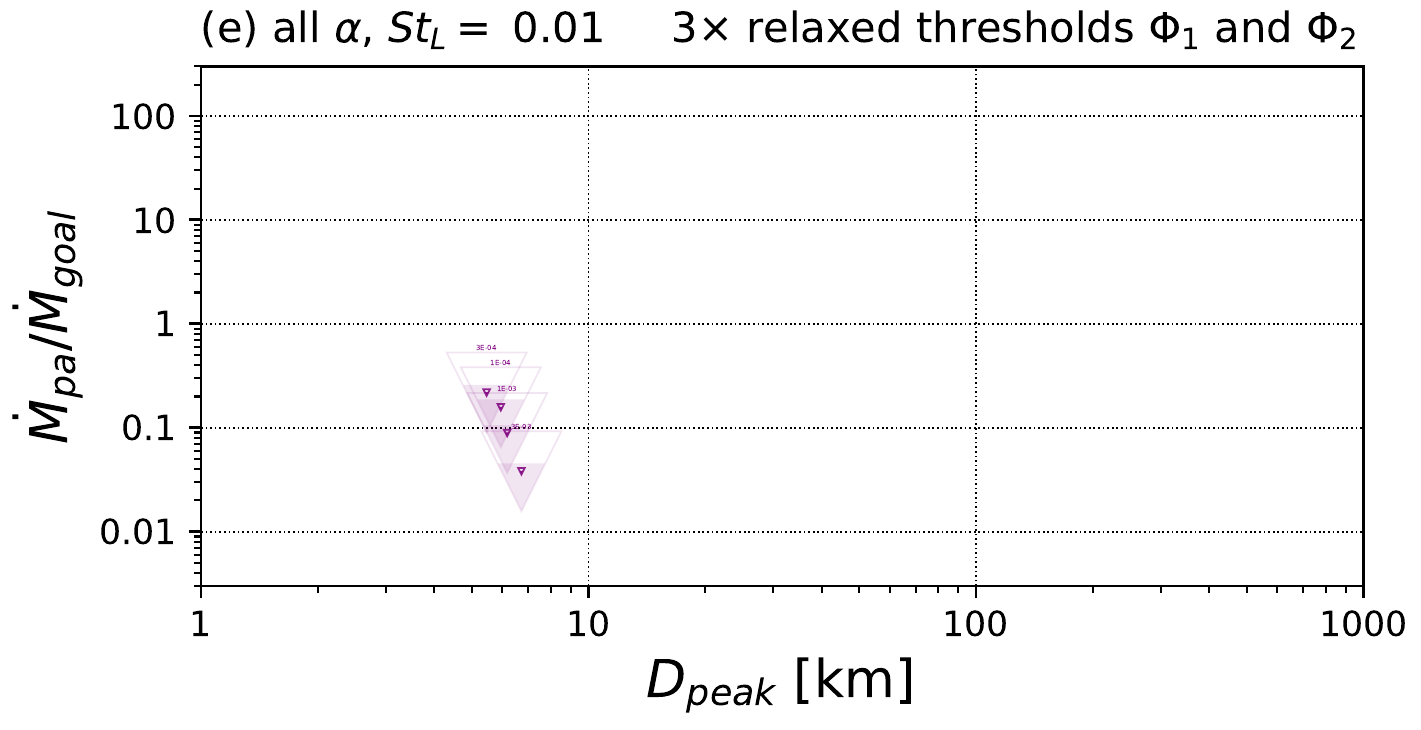}

     \vspace{-3pt}

     \includegraphics[width=0.916\linewidth]{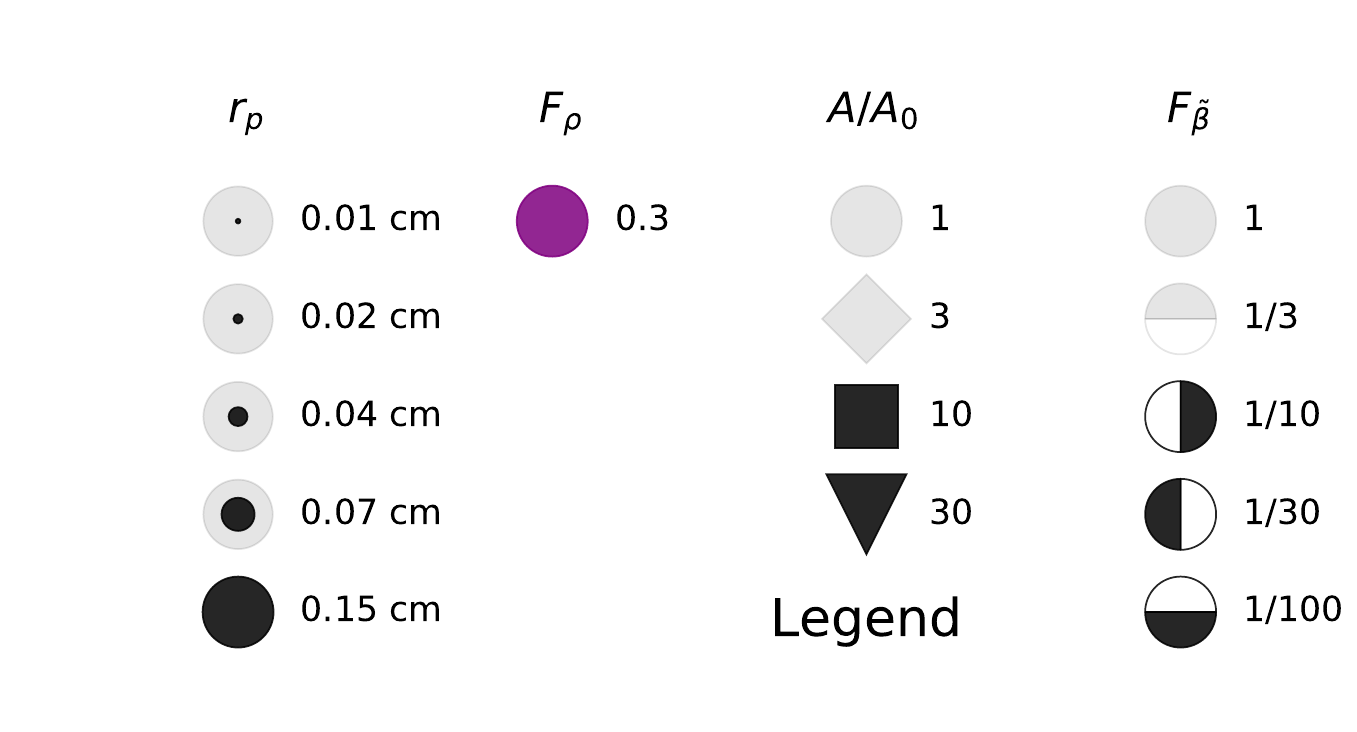}

   \end{minipage}

   \vspace{3pt}

   \caption{Outer nebula results for a gas density {\sl lower} than the MMSN ($F_\rho=0.3$), for an extended range of solids-to-gas ratios ($A/A_o \in [1,30]$) and relaxed thresholds.
   Each panel shows results for different Stokes numbers, and where $St_L=0.01$ is the smallest Stokes number that produced planetesimals in the plot range.
   As before, symbol shapes and fill styles denote factors for solids enhancement and pressure gradient parameter, respectively, and the value of $\alpha$ for each case is printed above each datapoint in small font best seen under magnification.
   The symbol size again scales with the particle size as indicated in the legend, but note that the size range is very different from the other outer nebula figures (Figures~\ref{fig:OuterNebulaPlanetesimals} and~\ref{fig:OuterNebulaStLDependence}).
   \label{fig:OuterNebulaStLDependenceF03}}
\end{figure*}

The model results for the outer nebula are presented in Figures~\ref{fig:OuterNebulaPlanetesimals} and \ref{fig:OuterNebulaStLDependence} in similar fashion to the corresponding asteroid-belt results (Figures~\ref{fig:InnerNebulaPlanetesimals} and \ref{fig:InnerNebulaStLDependence}).
Again, we normalize the rate of planetesimal production, $\dot{M}_{\rm{pa}}$, to the expected rate, $\dot{M}_{\rm{goal}} \equiv M_{\rm{goal}} / T_{\rm{neb}}$, for which we here assume that a mass of $M_{\rm{goal}}=40~M_\oplus$ was turned into planetesimals within the lifetime of the nebula, $T_{\rm{neb}} = 2~\rm{Myrs}$.

The general result for the dependence on $\alpha$ and $\tilde\beta$ seen in the inner nebula holds here as well, that is, larger $\alpha$ produce larger planetesimals but at smaller rate, while reducing $\tilde\beta$ allows smaller planetesimals at higher rate to form.
The smallest Stokes number that produced planetsimals with formation rates similar to the expected values is $St_L=0.005$, both for nominal and relaxed thresholds.

By looking at the outer nebula figures, many more planetesimals of large size ($\gtrsim$100\,km) can be seen compared to the inner nebula.
This, however, is mostly due to the range of nebula parameters we consider in the outer nebula {\it vs} the inner nebula.
The very large planetsimals form mostly for nebula conditions for which gas density enhancement and enhancement of solids-to-gas ratio is simultaneously large, e.g., $F_\rho=10$ {\it and} $A/A_o=10$.
Such parameter combinations correspond to disks with large amounts of total solids.
Since we restrict the {\it total} mass in the asteroid-forming region to no more than 250 Earth masses (see Section~\ref{sec:model:variations}), such combinations were disregarded in the inner nebula.
Conditions in the outer nebula during planetesimal formation are even more uncertain than in the inner nebula, and we here allow a total mass of available solids up to 2000 Earth masses. As did \citet{Cuzzietal2010}, we find primary accretion of planetesimals to be rather inefficient at least for this process, in which case the MMSN concept is not a realistic starting condition.

However, even with gas densities that likely were {\sl larger} than a MMSN initially, the gas density did get smaller later in the lifetime of the disk, and it may be of interest to see if there are still parameter combinations that allow for planetesimal formation in such environments.
For our nominal thresholds, we did not find any planetesimals with reasonable formation rates, but for the relaxed thresholds formation in the outer nebula seems possible at large solids-to-gas ratios.
Figure~\ref{fig:OuterNebulaStLDependenceF03} presents those results.
Note that due to the small gas density, the Stokes numbers we considered correspond to fairly small particles, sub-mm to 1.5\,mm in radius.
Still, sizable planetesimals can form under such conditions if $A/A_o$ is 10 or 30. Such solids-to-gas ratios do not seem too unreasonable in a scenario where a large fraction of gas was lost during disk evolution, thus increasing $A/A_o$ over time from initial values not too far from cosmic abundances.

\subsection{Limits on planetesimal formation explained} \label{sec:sizelimits}

\begin{figure}
   \centering
   \begin{minipage}{\reducedlinewidth}
   \includegraphics[width=\linewidth]{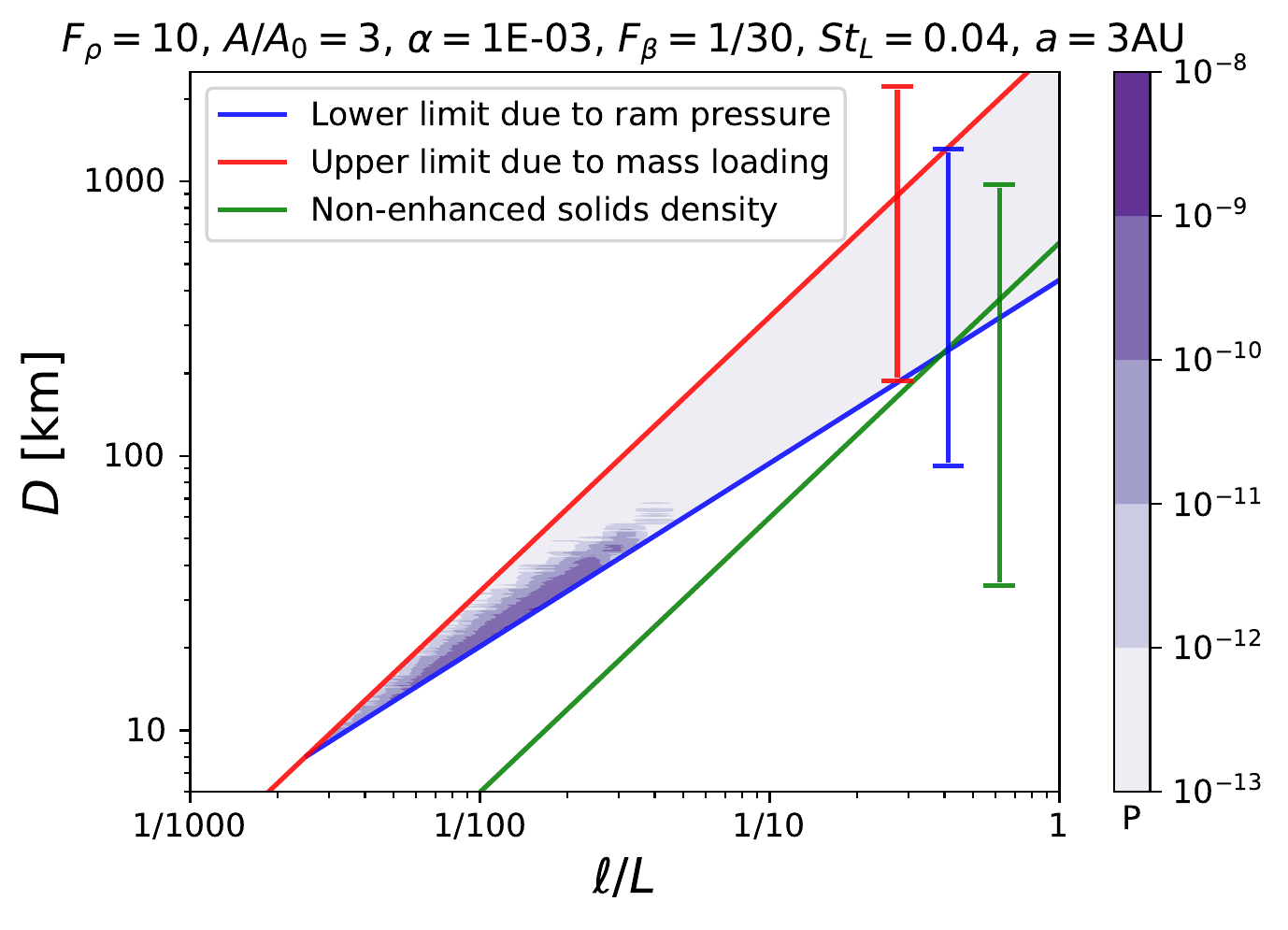}
   \end{minipage}
   \caption{Diameter, $D$, of a solid planetesimal forming at 3\,AU by gravitational sedimentation of a bound, dense region of size $\ell$. The blue line shows the minimum size for the clump to be stable against ram pressure. The red line shows the maximum size possible when turbulent clustering reaches the mass-loading limit, and the green line shows the resulting size assuming turbulent clustering is ineffective and the solids density remains at its nominal background value (except for the effect of settling). The lines are for a specific set of particle and nebula parameters, indicated above the plot. Vertical bars in the same colors indicate the range of possible outcomes (all having the same slopes as the plotted lines) across a range of plausible nebula parameters and Stokes numbers: $F_{\rho} \in [1,30]$, $A/A_o \in [1,10]$, $\alpha \in [10^{-4}, 3\times10^{-3}]$ (all of which affect the red and green curves), $St_L \in [0.0025,0.16]$ (which only effects the green curve due to particle settling), and $F_{\tilde\beta} \in [1/100, 1]$ (which only affects the ram pressure limit in blue). The actual solids concentrations reached depend on the amount of turbulent clustering at any given scale and must lie between the red and green curves.
   Meanwhile, only those values above the blue curve correspond to clumps which are stable against disruption by ram pressure and may in fact collapse into planetesimals.
   Purple contours show the actual probability, from our numerical cascade, for the specific parameter set shown, that a clump of size $\ell$ has the right conditions to form a planetesimal of diameter $D$.
   \label{fig:PlanetesimalSizeLimit}}
\end{figure}

From the cascade model results shown in Figures~\ref{fig:InnerNebulaPlanetesimals}--\ref{fig:OuterNebulaStLDependenceF03} it is apparent that there are both lower and upper size limits beyond which planetesimals do not form.
We here offer some physical explanation for the general result that primary accretion by this process leads to bodies mostly within the 10 to few 100\,km diameter range.
We will also discuss why no planetesimals form for Stokes numbers $St_L \lessapprox 0.005$.

\subsubsection{Lower size limit} \label{sec:lowersizelimit}

The lower size limit is in general set by the threshold $\Phi_2$ (Equation~\ref{eqn:phi2}), as can be seen in Figure~\ref{fig:InnerNebulaPlanetesimalsAllAlpha} where the size of the smallest planetesimals changes when $\Phi_2$ is varied but other thresholds are kept unchanged. The $\Phi_2$ threshold is related to the ram pressure, and is also a function of $\ell$ (see Section~\ref{sec:model:thresholds}) --
a clump needs to have a density large enough, for any given size, to resist disruption by the ram pressure between the gas and particle clump for the time it takes for gravitational sedimentation to produce a compact object.
This minimum clump density sets the lower limit on the size of resulting planetesimals for any clump size.
Recall that the diameter of a planetesimal is given by the expression $D(\ell) = \ell ( \rho_p / \rho_s)^{1/3}$ (Equation~\ref{eqn:D}) where due to the ram pressure limit $\rho_p \ge \Phi_2 \rho_g$.
This minimum diameter is shown as a function of the scale $\ell$ of the clump from which the planetesimal formed, by the blue line in Figure~\ref{fig:PlanetesimalSizeLimit}. The smallest {\it value} of this minimum diameter is found when the {\it minimum} size (blue) curve crosses the {\it maximum} size (red) curve in Figure~\ref{fig:PlanetesimalSizeLimit}, which is discussed below.
The nominal thresholds used here and a range of plausible nebula parameters lead to a minimum planetesimal diameter around $10$\,km.

This estimated lower limit on planetesimal size is consistent with the results of our cascade model (Figures~\ref{fig:InnerNebulaPlanetesimals} and \ref{fig:InnerNebulaPlanetesimalsAllAlpha}). For the ``relaxed" thresholds (Figure ~\ref{fig:InnerNebulaPlanetesimalsAllAlpha}), the minimum size from the cascade results is approximately 3\,km, also consistent with the relaxed $\Phi_2$.

\subsubsection{Upper size limits} \label{sec:uppersizelimit}

For assessing the maximum possible size of planetesimals in the framework of our model, we can look at two limiting cases.
At the largest scales in the flow near $\ell=L$, turbulent clustering is not able to generate large variations in the particle density and therefore the density will be close to the local mean value (see Section~\ref{sec:model:nebula}).
As scales get smaller, turbulent clustering will generate flow regions with larger particle densities, but there is a ``mass-loading limit'' at $\Phi = \rho_p/\rho_g \sim \Phi_{\rm{limit}} = 100$ at which turbulent clustering saturates. The maximum (pre-collapse) mass density in solids then is
\begin{equation}
  \rho_{p,\rm{max}} = \Phi_{\rm{limit}} \rho_g.
  \label{eqn:ParticleDensityMaxClustering}
\end{equation}
The resulting planetesimal diameters for these nominal and maximum {\it turbulent clustering densities} described above are shown as functions of clump size in green and red lines respectively in Figure~\ref{fig:PlanetesimalSizeLimit}, along with the minimum allowed size given the ram pressure limit (Section~\ref{sec:lowersizelimit}; blue line).
In practice, however, the maximum turbulent densities (red curve) are irrelevant at large scales since turbulence is inefficient there and densities will stay near their nominal values (green curve) corresponding to maximum possible planetsimal sizes, depending on nebula conditions, of up to a few 100\,km consistent with the results of Section~\ref{sec:innernebula}.
Figure~\ref{fig:PlanetesimalSizeLimit} also shows, for a specific set of nebula parameters, the actual probability from our cascade model that a clump can form a planetesimals of a given size.
As can be seen, by nature of the mechanism, the probabilities at a given scale $\ell$ are always larger towards the non-enhanced densities (green curve), and the maximum (red curve) is rarely reached.

\subsubsection{Limiting Stokes number} \label{sec:stokesnumberlimit}

Figure~\ref{fig:PlanetesimalSizeLimit} can also be used to gain understanding of why planetesimals fail to form, or have very small formation probability, below some limiting Stokes number.
The point where the curves for the minimum and maximum planetesimal size due to ram pressure and mass loading limits (blue and red curves in Figure~\ref{fig:PlanetesimalSizeLimit}) intersect marks the smallest scale of clump that can form a planetesimal.
The two curves, and therefore the intersection point, only depend on nebula conditions, and are independent of $St_L$.
The turbulent clustering process of course {\it is} Stokes number dependent.
It is most effective at scales where the particle stopping time is comparable to the eddy time scale, which itself gets smaller at smaller spatial scales.
In effect, large Stokes number particles experience significant clustering already at large spatial scales while small Stokes numbers start to be effected only at small scales.
In effect, Stokes numbers $St_L \lessapprox 0.005$ do not experice enough clustering above that intersection (for reasonable nebula conditions) and therefore fail to produce planetesimals.

\subsection{Comments on general trends}\label{sec:trends}

Given the presence of $1-10$\,cm radius particles, probably aggregates containing many individual chondrules in the inner nebula, we can reach some general conclusions about the sensitivity of IMF modal size on nebula properties. Planetesimal diameters increase slightly with $\alpha$, probably because the concentration cascade can start at a larger energy containing eddy scale $L=H\alpha^{1/2}$, while formation rates more strongly decrease, perhaps because there are fewer volume elements of large size to participate.  Not surprisingly, a smaller headwind, as determined by $\tilde\beta$, makes planetesimal formation easier in the sense that we need smaller overdensities to form planetesimals. This will increase the formation rate although the average size of planetesimals goes down. This size decrease, in turn, is because the lower size limit (Section~\ref{sec:lowersizelimit}) decreases with smaller $\tilde\beta$ and the parameter space between the lower and upper size limits increases (see Figure~\ref{fig:PlanetesimalSizeLimit}). However, planetesimals that {\it do} form at larger $\tilde\beta$ are usually larger (perhaps because of raising the minimum size limit). Decreasing $\tilde\beta$ by adopting $F_{\tilde\beta}<1$ is our approximation for the ``peloton effect" that the current simple model cannot incorporate, but is an essential aspect of the so-called ``streaming instability" or resonant drag instability. Models of this process initially show longitudinally extended arcs or streams of high density, in which the leading ends diminish the headwind experienced by material towards their trailing ends much like a peloton of cyclists. Equations 2.12 and 2.14 in \citet{Nakagawaetal1986} can be combined with our expression for the particle layer thickness, to show that $F_{\tilde\beta}\sim 0.5-0.1$ is within the reach of a settled background particle layer with local solids abundance between $1-10$ times cosmic as may be produced by radial drift, for instance \citep{Estradaetal2016}. Even smaller values of $F_{\tilde\beta}$ might be expected, within dense but still not fully unstable zones arising from SI-like collective effects (Umurhan et al 2019). Of course, this treatment of the uncertain headwind is only a crude approximation, and true global models are really needed.

\subsection{``Typical" particle concentrations}\label{sec:concentrations}

\begin{figure*}[p]
   \centering
   \begin{minipage}{0.47\linewidth}

   \includegraphics[width=\linewidth]{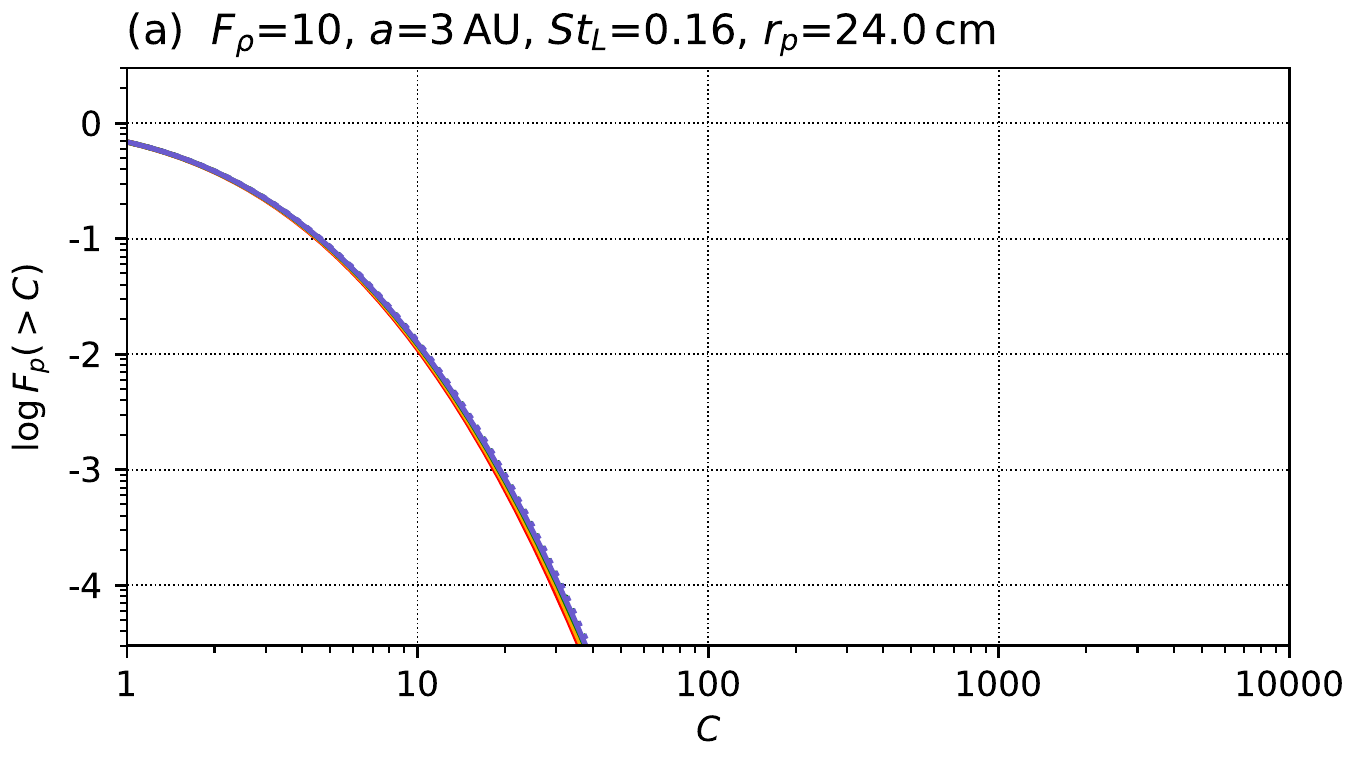}

   \vspace{-18pt}

   \includegraphics[width=\linewidth]{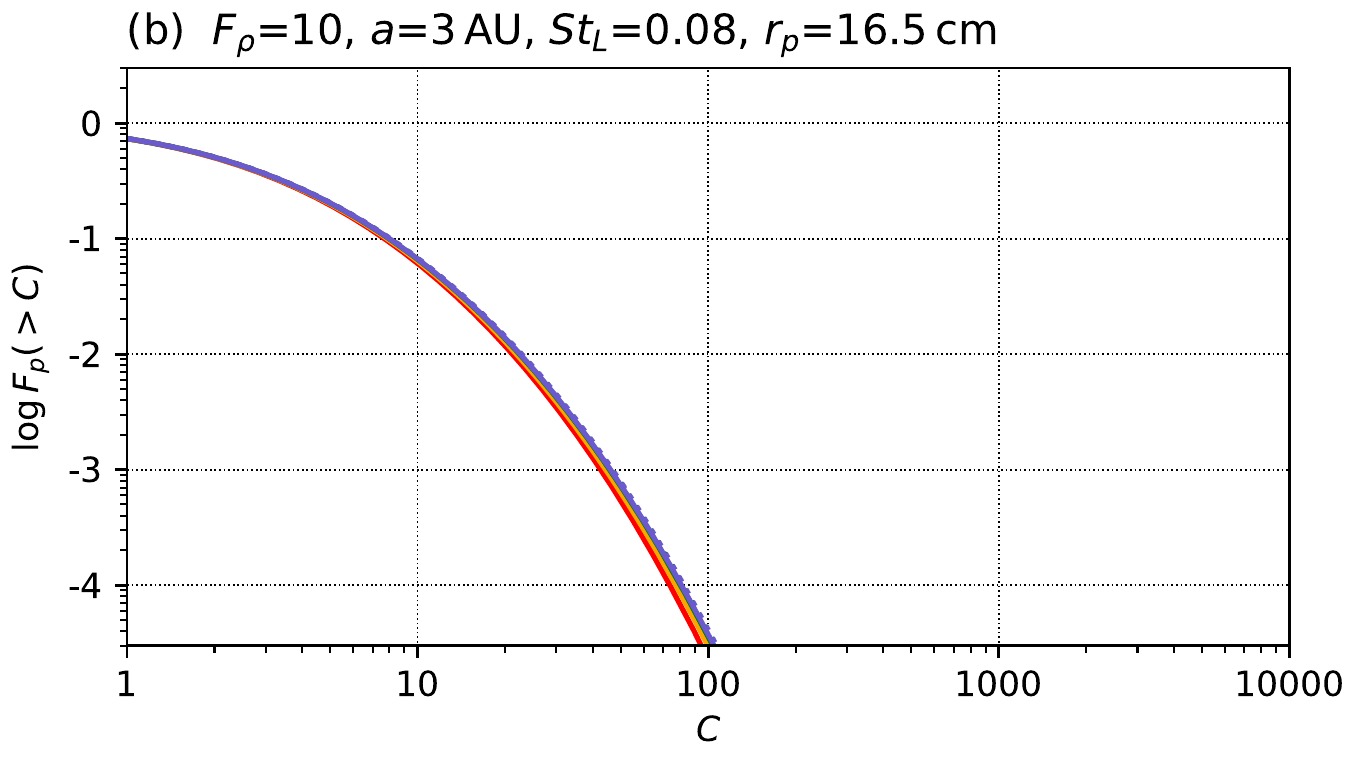}

   \vspace{-18pt}

   \includegraphics[width=\linewidth]{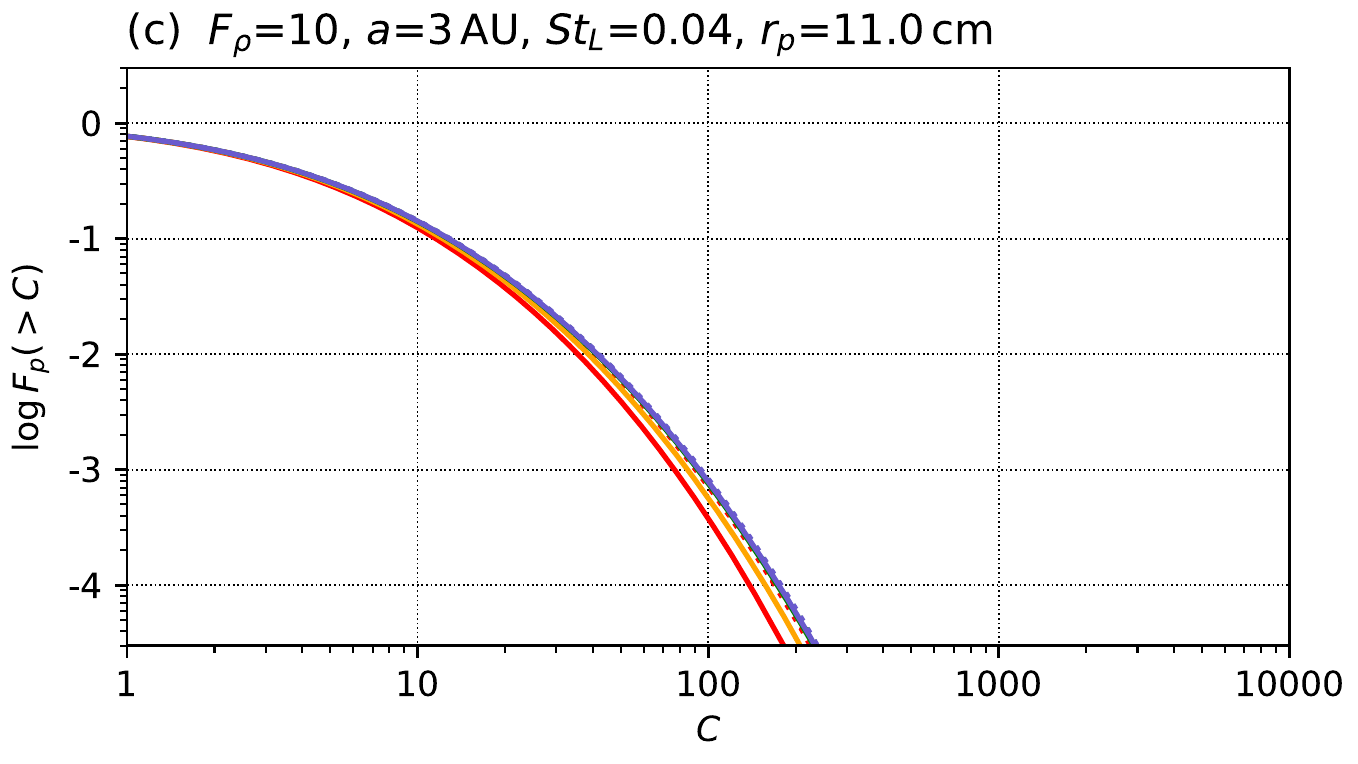}

   \vspace{-18pt}

   \includegraphics[width=\linewidth]{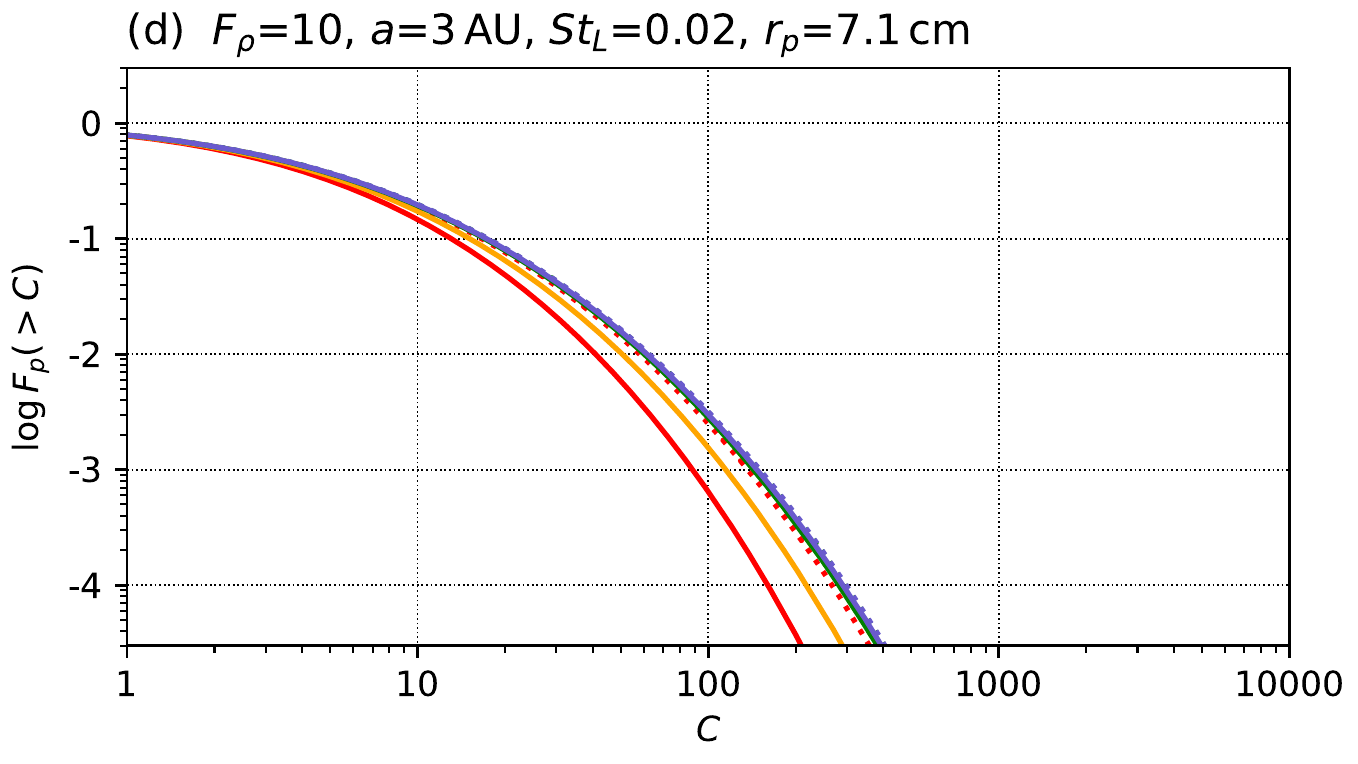}

   \end{minipage}
   \begin{minipage}{0.47\linewidth}

   \includegraphics[width=\linewidth]{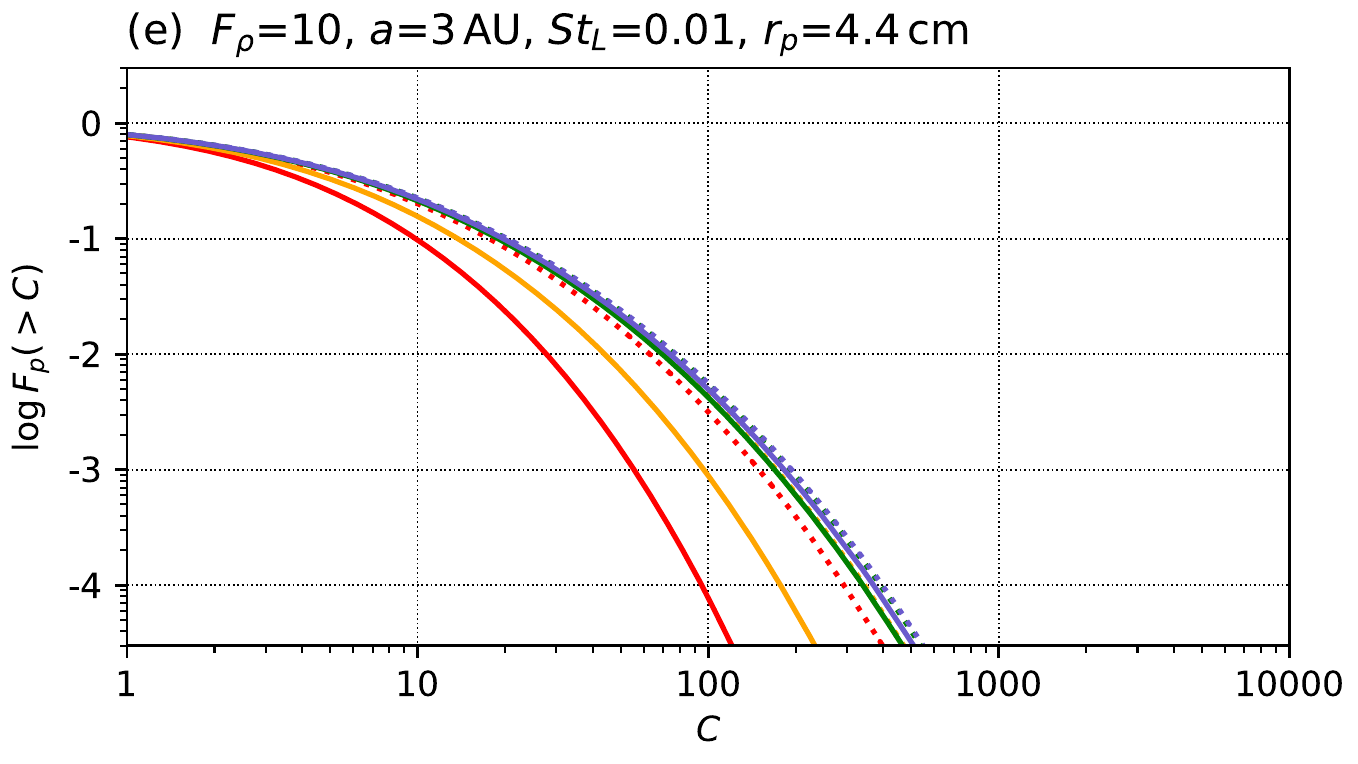}

   \vspace{-18pt}

   \includegraphics[width=\linewidth]{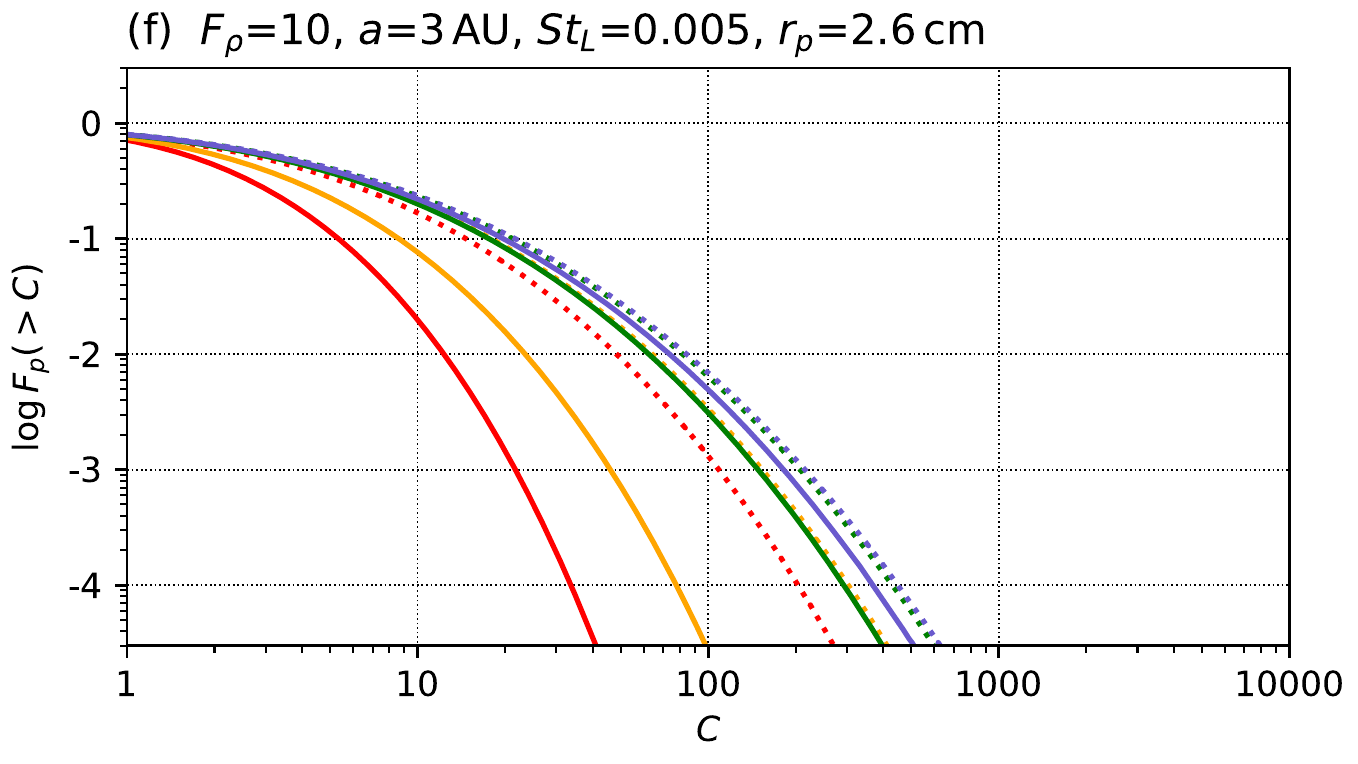}

   \vspace{-18pt}

   \includegraphics[width=\linewidth]{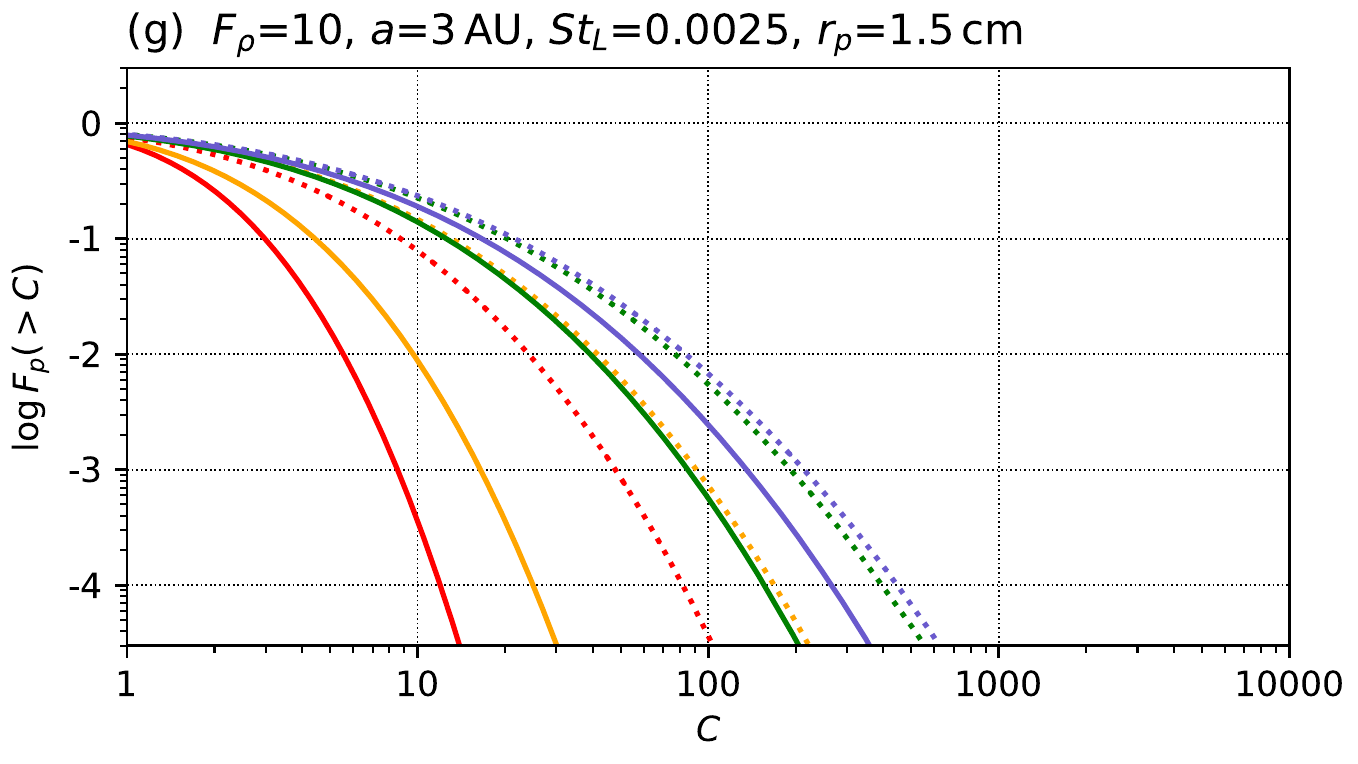}

   \vspace{-18pt}

   \includegraphics[width=\linewidth]{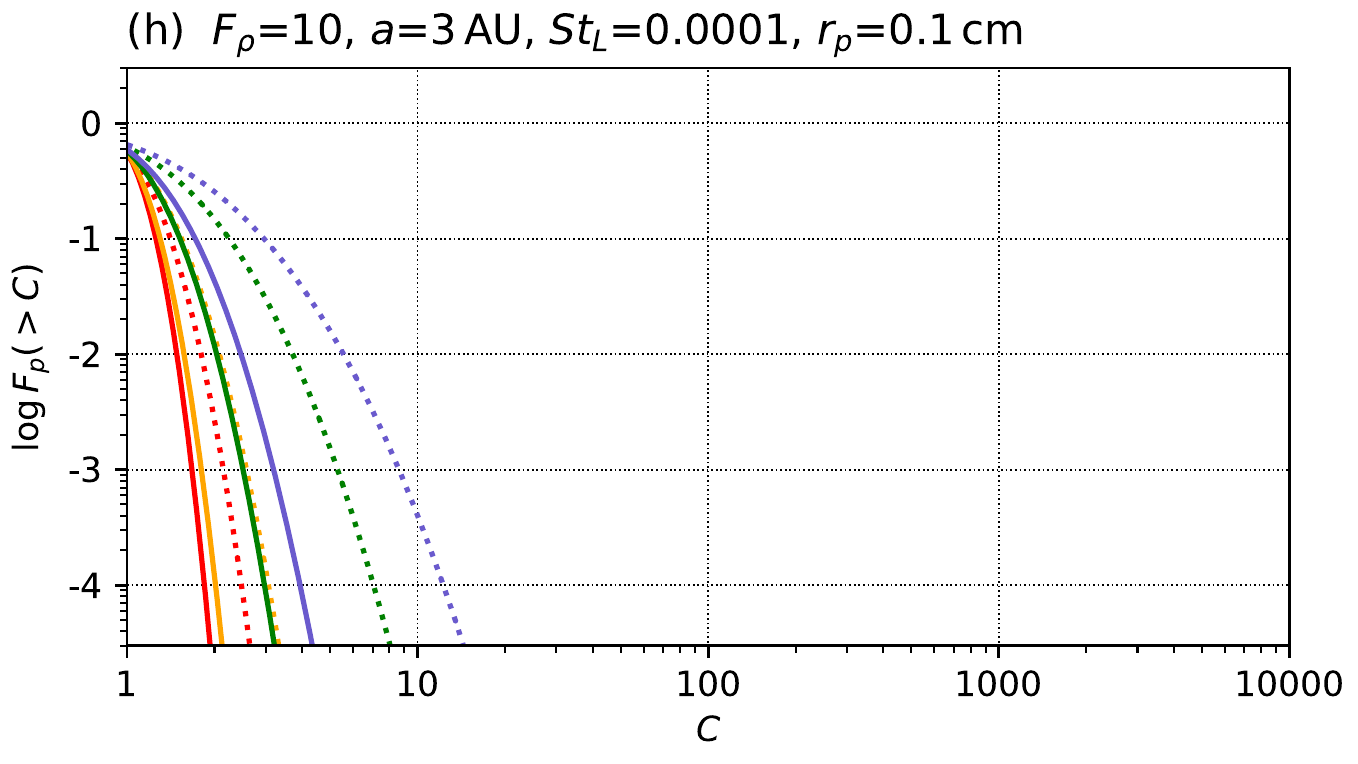}

   \end{minipage}

   \includegraphics[width=0.84\linewidth]{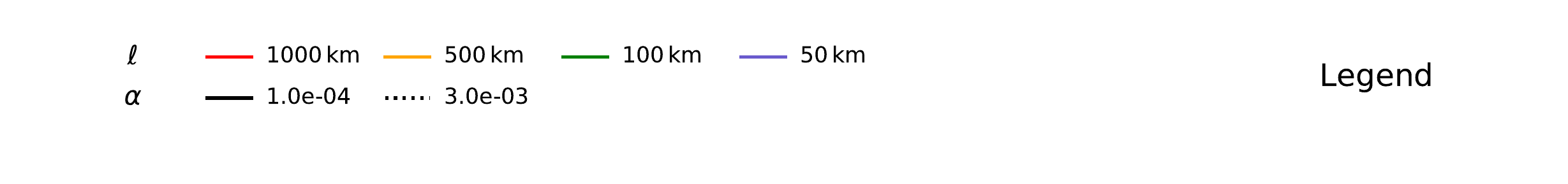}

   \vspace{-10pt}

   \caption{Cumulative fraction $F_p(>C)$ of particles lying in a region where the concentration is larger than some value $C$.
   In this context, concentration refers to the ratio of the local density (averaged on some length scale $\ell$) to its global average, and does not account for any other enhancement effects such as settling towards the midplane, or various kinds of radial enhancement.
   Results from our cascade model are shown for different Stokes numbers (panels (a)--(h)) with Stokes number and corresponding particle sizes noted above each panel, for a range of length scales $\ell$ within each panel, and two different values of $\alpha$.
   All results are for a $F_\rho=10$ nebula gas at 3\,AU.}
   \label{fig:ConcentrationPDFs}
\end{figure*}

\begin{figure}
   \centering
   \begin{minipage}{\reducedlinewidth}
   \includegraphics[width=\linewidth]{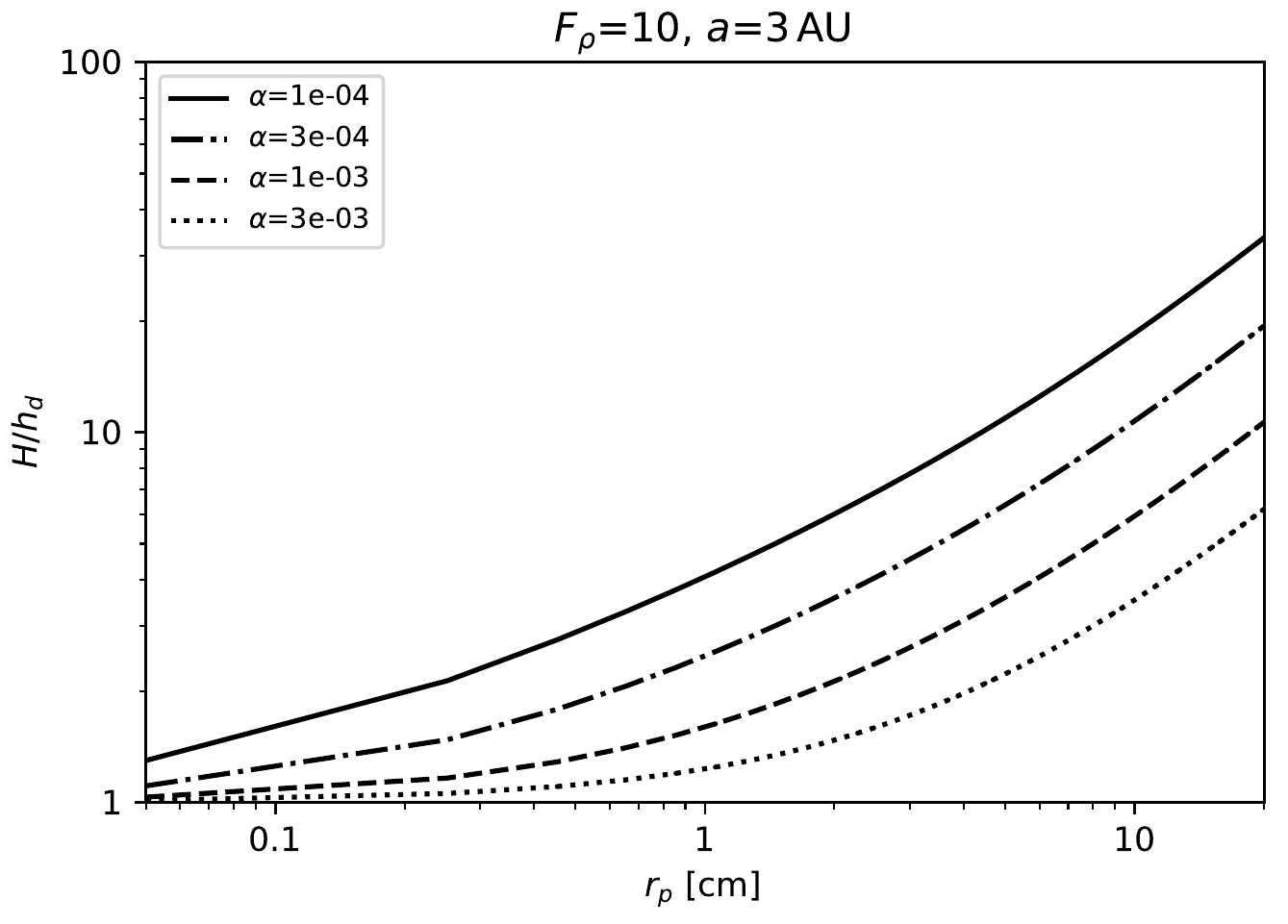}
   \end{minipage}
   \caption{Ratio between gas density scale height $H$ and solids density scale height $h_d$ as functions of particle radius $r_p$.
   Particles settle towards the midplane depending on their Stokes number and the turbulent intensity $\alpha$, leading to a scale height that is smaller than that of the gas (Equation~\ref{eqn:h_d}).
   The values shown are for a nebula with a gas density $10\times$ that of a MMSN, at a distance of 3\,AU.}
   \label{fig:SettlingFactor}
\end{figure}

Figure~\ref{fig:ConcentrationPDFs} shows our new PDFs of concentration $C \equiv \rho_p/\langle\rho_p\rangle$, the ratio of the local particle density (averaged on spatial scale $\ell$) to its global average, for individual particles of various sizes, from chondrule size, through ``pebble" size, to small ``boulder" sizes. The PDFs are given in the form of the cumulative fraction $F_p(>C)$ of particles lying in a region where the average concentration is larger than some value $C$; \citet{Cuzzietal2001} show that this quantity is equivalent to the fraction of time a given particle spends in such regions. The PDFs depend on the turbulent intensity $\alpha$, the particle Stokes number $St_L$, and the binning scale $\ell$. These PDFs replace the predictions shown in \citet{Cuzzietal2001, Cuzzietal2008, Cuzzietal2010} and \citet{HoganCuzzi2007}, which were based on an incorrect turbulent concentration model.

Based on meteorite evidence, we expect the larger size ``particles" in the range shown (in the inner solar system) to be {\it aggregates} of individual chondrules, chondrule precursors, or other bits of rock. In the outer solar system (even for CI chondrites), nature may not be so kind as to provide distinguishable macroscopic monomers, producing only grainy aggregates of different sizes and porosities.

These particle concentration PDFs are useful for understanding the formation environment of chondrules (by whatever heating mechanism), since the oxidation state of the ferromagnesian silicates that result, as manifested in their Mg/Fe ratios, is a sensitive indicator of the local solids/gas ratio.  Values of $10-30$ or so are not uncommon on lengthscales $\ell \sim 100-1000$\,km, and apply to any underlying background (which may already be enhanced in solids/gas ratio above the cosmic abundance of roughly 0.01). For example, particles tend to settle towards the nebula midplane by an amount that depends on their $St_L$ and the nebula $\alpha$, which enhances the near-midplane solids density (Figure~\ref{fig:SettlingFactor}) even before the concentration factors shown in Figure~\ref{fig:ConcentrationPDFs} are applied. Differential radial drift, evaporation fronts, pressure bumps, or even ``streaming overstabilities" can further enhance the local background above ``cosmic abundance" \citep{Estradaetal2016,Umurhanetal2019}.

Meteoritical studies find two markedly different ranges for the solids/gas ratios of regions containing chondrule-mass precursors, at the times and places where chondrules are formed by still mysterious ``flash heating" events \citep{ConnollyJones2016}. Studies of the silicate Mg/Fe ratios in most chondrules in a number of different carbonaceous chondrite groups infer background solids/gas enrichments over cosmic abundance by factors of $50-200$, while more Fe-rich, so-called Type-II ordinary chondrite chondrules (a few are even found in the carbonaceous chondrites) require enhancements of $2000\times$ \citep{Tenneretal2015, Tenneretal2017, Tenneretal2019, Hertwigetal2018}. The {\it highest-Fe} silicate grains \citep{GrossmanFedkin2012}, and alkali abundances in the Semarkona primitive ordinary chondrite, \citep{Alexanderetal2008, Hewinsetal2012sodium} call for local dust enhancements several orders of magnitude higher still.

For comparison, Table~\ref{table:concentrations} shows net enhancement factors for several representative values of particle radius $r_p$, for two different values of $\alpha$, on two different lengthscales $\ell$.  The table is intended to estimate total enhancement factors $C$ relative to cosmic abundance, by estimating a plausible non-TC enhancement to the background particle density due to vertical settling and radial concentration by drift or ``Incipient Streaming Instability" \citep{Umurhanetal2019}\footnote{A domain covering realistic turbulence and particle growth, which leads to saturation of particle density growth at $\rho_p \sim \rho_g$ without planetesimal formation.}. The values of $C$ are given for ``common" probability levels of $F_p(>C)$ = 50\% and 30\%, and a ``rare" probability level of $F_p(>C)$ = 1\%. As an example, chondrule aggregates of between 1.5 and 2.6cm radius are apparently seen in the primitive ordinary chondrite NWA5717 \citep{Simonetal2018}. According to Table~\ref{table:concentrations}, ``particles" of $2.6$\,cm radius, whether aggregates of chondrules or of chondrule precursors, are commonly found in $500-1000$\,km size regions with average concentrations of $50-140$. The lengthscale of chondrule formation is poorly known, but \citet{SahagianHewins1992} and \citet{CuzziAlexander2006} venture estimates ranging from $150-6000$\,km. On smaller scales of $50-100$\,km, they are found at lower probability with average concentrations of $1400-4900$. These values compare favorably with findings of \citet{Tenneretal2015, Tenneretal2017, Tenneretal2019} and \citet{Hertwigetal2018} for ``common" Type I chondrules, and ``rare" type II chondrules in carbonaceous chondrites. If the nebula gas density is some $10\times$ higher than a ``Minimum Mass Nebula" value, as many of our IMFs prefer, \citet{CuzziAlexander2006} also suggest a ``common" enrichment over cosmic abundance of $140-230 \times$, assuming shock heating that is accompanied by significant compression of the dust-gas mix. Notice that at $0.1$\,cm radius (typical for single chondrules perhaps), concentration values never get much more than $10\times$. Because the concentration PDFs $F_p(>C)$ (time spent by particles in regions of concentration larger than some $C$) show strong dependence on concentration and are scale dependent (Figure~\ref{fig:ConcentrationPDFs}), it is not hard to envision large-scale heating events that extend over a range of concentrations and might have simultaneously formed batches of chondrules with different oxidation states in the same heating event. This theory would predict that the more rare, denser concentrations would be found on smaller lengthscales, perhaps as dense cores of zones with more common concentrations.

Thus, we believe that if typical particles are indeed few-cm-size aggregates \citep{Simonetal2018}, whether of chondrule precursors or chondrules themselves, TC is probably capable of providing fairly common enhancements at the $\sim 100\times$ level, perhaps by operating on an already somewhat elevated background particle density layer. However, we feel that values as high as $10^4-10^6\times$ \citep{Alexanderetal2008, GrossmanFedkin2012, Hewinsetal2012sodium}, which are more than 100 times the local gas density, are hard to support given our current understanding. The cascades should be checked against larger-scale simulations however, as they may be conservative.

\begin{table*}[p]
\caption{This table combines all the various enhancement effects into a total enhancement of the solids/gas mass density ratio, for selected values from Figures~\ref{fig:ConcentrationPDFs} and \ref{fig:SettlingFactor}.
Table (a) lists concentration factors from vertical settling combined with an added factor from radial drift and Incipient Streaming Instability (assumed to be $A/A_o = 10\times$) for three particle sizes and two values of $\alpha$. These factors are independent of spatial scale.
Tables (b) and (c) list the additional factors from turbulent concentration (TC) at the 50\% , 30\%  and 1\% probability level, and the resulting ``total" enhancements to the solids/gas ratio compared to cosmic abundances of 0.01.
TC is scale dependent, and tables (b) and (c) list the values of $C$ at these levels for spatial scales $500-1000$\,km and $50-100$\,km, respectively.
For $r_p$=1.5 and 2.6cm, at the $F_p(>C)=30-50$\% level, TC only produces $2-6\times$ at the scales shown, but combined with $10\times$ (assumed) from the combination of radial effects, and another $2-6\times$ from settling, gives $30-240\times$ (50\%) to $60-500 \times$ (30\%) which covers the observed range for ``common" concentrations (Section~\ref{sec:concentrations}).
At the $F_p(>C)=1$\% level, we get concentrations of $1000-3000\times$. This is where TC dominates ($10-80\times$).
Red numbers in the tables are suspect, since Incipient Streaming Instability on top of radial drift cannot lead to $\rho_p > \rho_g$, i.e. factors $> 100$ ($r_p=11$\,cm, table (a)).
All quoted values are for a nebula gas with $10\times$ the gas density of a MMSN ($F_\rho=10$) at 3\,AU (which does not affect the concentration values for a given Stokes number, but affects the particle size $r_p$), and have been rounded to 2 significant digits.
}
\label{table:concentrations}
\centering

\footnotesize

\vspace{-10pt}

\input{concentration_tables.tex}

\end{table*}

\pagebreak

\section{Conclusions} \label{sec:conclusions}

We report results on two different effects of the concentration or clustering of small particles in turbulent nebula gas, both directly relevant to different aspects of primitive body formation: (a) the concentration PDF for ``pebble"-size particles, with sizes between chondrules and their aggregates (in the inner nebula), describing ``typical" local solids densities under which chondrules might be likely to be formed; and (b) the modal diameter and formation rates for primary accretion of $10-100$\,km diameter planetesimals -- the planetesimal Initial Mass Function or IMF -- a process which requires larger local particle concentrations, on larger spatial scales, that are statistically more rare. The implications of the concentration PDFs themselves (a) were discussed immediately above in Section~\ref{sec:concentrations} and will not be repeated here.

Regarding planetesimal formation (b), we have modeled turbulent concentration combined with vertical settling in turbulence, based on a new, validated, cascade model of the statistics of particle concentration and gas vorticity (enstrophy) as a function of spatial scale, and simple physical threshold criteria. We find the process leads to planetesimal Initial Mass Functions (IMFs) with well-defined modal diameters, instead of powerlaws as is typically found for ``incremental growth" and ``streaming instability" mechanisms. The ``fossil asteroid belt" and the KBO population both appear to have such modal distributions, where the modal size or mass lies at the ``knee" between two powerlaws (Section~\ref{sec:intro}). Our predicted modal diameters (which are not weighted by mass) vary from roughly $10-100$\,km, as nebula properties are varied across a plausible range ($\alpha =10^{-4}-10^{-3}$, gas densities $1-30\times$ MMSN, local solid/gas ratios $1-30\times$ cosmic abundance). This mechanism thus produces planetesimals that are ``born big" \citep{Morbidellietal2009BornBig} directly from small, freely-floating, nebula particles. As discussed in \citet{Cuzzietal2010} in more detail, planetesimals formed by this mechanism are expected to form as internally homogeneous sandpiles \citep[see also][]{Johansenetal2015}.

There is one highly significant difference between these results and those of \citet{Cuzzietal2010}. The current model, using our revised and updated cascades,  no longer supports making sizeable ($10-100$\,km diameter) objects directly from individual chondrules (in the ice-free inner nebula). To do this, the constituent particles must be larger, ``pebble" or even ``cobble" sized particles of $1-10$\,cm radius, almost certainly aggregates of chondrules or unmelted precursor objects of similar mass (in the inner nebula). IMFs that result under most plausible nebula conditions, for particles in this size range, typically peak at tens of km diameter, perhaps a bit small for the $100$\,km fossil asteroid diameters (care should be taken to distinguish between number-weighted and mass-weighted IMFs). Inner nebula IMFs that {\it do} peak at $100$\,km diameter require constituent particles that are perhaps $20$\,cm radius (Figures~\ref{fig:InnerNebulaPlanetesimals}--\ref{fig:InnerNebulaStLDependence}), perhaps because the larger particles concentrate on larger spatial scales, and are more settled towards the midplane to start with. In the outer nebula, particles from mm- to few-cm- radius produce planetesimals in the $10-100$\,km diameter range (Figures~\ref{fig:OuterNebulaPlanetesimals}--\ref{fig:OuterNebulaStLDependenceF03}).

Current models of growth by sticking (at least in the inner, silicate-dominated nebula) tend to see growth frustrated by bouncing in the cm-size range \citep{Birnstieletal2011, Estradaetal2016}. However, actual observations are telling us that the current models, based on laboratory sticking measurements, may be missing something. \citet{Simonetal2018} have analyzed a very primitive ordinary chondrite, which has the unusual property of containing two visually distinct (dark and light) ``lithologies" which on closer examination are, apparently, aggregates of chondrules formed in two very different regions, as reflected in their very different chemical and isotopic compositions (and slightly different particle sizes, even). Somehow, nature is making several-cm-diameter aggregates of chondrules even if our models are not yet doing so \citep[cf.][]{Arakawa2017}. It is natural to wonder if the building blocks of other -- maybe all -- chondrites may also be similar aggregates, but generally indistinguishable because they are all made of similar chondrule monomers. One hopes that this speculation can be tested in the future. Another possible clue might be found in the so-called ``cluster chondrules" \citep{Metzler2012, Metzleretal2012, MetzlerPack2016, HewinsZanda2012}. These are rare clumps of semi-molten chondrules found all smashed together -- a possible outcome for an aggregate of chondrules that was not broken up before being melted in a chondrule formation event.

The predicted planetesimal formation {\it rates} in Figures~\ref{fig:InnerNebulaPlanetesimals}--\ref{fig:OuterNebulaStLDependenceF03} span a range plus or minus two orders of magnitude around the estimated nominal value. Combined with the current uncertainty in nebula properties, this means the theory is not predictive of exact planetesimal sizes or rates. However, in a general way, it satisfies the observational constraints of typical size and formation rate. Indeed the statistically low probabilities of the dense clumps needed to trigger planetesimal formation give this process an extended, drawn-out nature that is in agreement with observations, which indicate that planetesimal formation, while it started early, continued for several Myr as the nebula continued to evolve. The environment is highly conducive to post-primary-accretion growth by pebble accretion \citep{VisserOrmel2016}, because there is nothing {\it but} primary planetesimals and pebbles around. By contrast, traditional linear instability is either inoperative, or all over in a moment.

In spite of the slow trickle of planetesimal formation, once a dense clump is triggered (in clumps with particle densities that allow them to be bound), it sediments into a planetesimal on a timescale of approximately $t_{\rm dyn}^2/t_s$, in the regime where $t_s < t_{\rm dyn}$, which is not much different from the orbit time \citep[Section~\ref{sec:model:thresholds},][]{Cuzzietal2008, ShariffCuzzi2015}. The sedimentation time is thus thousands of orbits for $St \sim 0.001$, in agreement with thermal evolution models of planetesimal interiors that favor ``rapid" accretion of a given planetesimal, once it starts \citep{Ghoshetal2003, Ghoshetal2006, Vernazzaetal2014, Pedersenetal2019}. The extended sedimentation time for these rotating loose clumps of particles may allow for bifurcation into binaries of comparable sizes, which are found to be very common in the KBOs \citep{StephensNoll2006}. \citet{Nesvornyetal2010} have modeled such a scenario, but in the limit of clumps composed of large particles where gas drag is unimportant and collapse occurs on a dynamical time $t_{\rm dyn}$. In the more general case where gas drag is important, ``collapse" is slower \citep{ShariffCuzzi2015} but fission into binaries will probably still be a potential outcome. More study is surely needed of this stage.

Primary accretion of planetesimals by turbulent concentration would probably not work alone. The slightly settled particle layer in which this process occurs would support more complex collective ``peloton" effects such as seen in streaming instabilities (SI), extending and amplifying the process, even though the layer itself, in the absence of perturbations by TC, might not be unstable to SI and would produce no planetesimals \citep{Umurhanetal2019}. Such a triggered or nonlinear instability, which we have called ``clustering instability" \citep{Cuzzietal2017a}, is beyond the scope of this work and will require large-scale numerical models to study. Also, $10-100$\,km size sandpile planetesimals will incur mutual eccentricities \citep{Gresseletal2012}, which may allow gentle collisions, suitable for further growth without destruction or significant erosion, to reach sizes at which pebble accretion can begin.

{\it Much work needed to be done:}  Clearly, more work is needed to understand growth by sticking in terms of the potential for growing observed aggregates of chondrules in the inner nebula (and probably aggregate pebbles of small grains in the outer nebula) that appear to be significantly larger than result from the most recent models using bouncing and fragmentation outcomes based on current experimental work \citep{Estradaetal2016}.

More observational work is needed on large slab samples of chondrites to explore whether large (several cm diameter) aggregates of chondrules are the rule or an exception \citep{Simonetal2018}. Evidence for aggregate formation in the outer nebula will be harder to find, because of the likely absense of easily-distinguishable chondrule monomers, and the general lack of samples of any kind, but similar physics must be at work.

Regarding numerical fluid dynamical models, direct numerical simulations are needed of turbulent concentration in which particle mass loading feedback on the gas is included, to see how it affects the process. More highly resolved simulations at higher Reynolds numbers (a real challenge) would be desirable to check the cascade parameters for enstrophy and concentration. Of course, combined numerical models showing how global collective effects may amplify triggering perturbations from TC alone (the clustering instability) are critical.

Because turbulence excites eccentricities in the orbits of small bodies, which can lead to collisions \citep{Idaetal2008,Gresseletal2011,Gresseletal2012}, a possible second stage of growth (mutual collisions between loose, primary sandpile planetesimals, at low relative velocity and favoring growth over erosion/destruction), would be valuable to study as well, followed perhaps by pebble accretion. Nevertheless, it may be that a generalized turbulent concentration process, such as the clustering instability, may be able to provide the first ``seeds", or primary planetesimals, to start the process and to keep it going for several Myr.

Finally, studies should be conducted of binary formation by fission of rotating, sedimenting clumps of particles with a range of $St$, extending the gas-drag-free, rapidly collapsing models of \citet{Nesvornyetal2010}.

\acknowledgments
We are happy to acknowledge the National Aeronautics and Space Administration (NASA) Origins of Solar Systems Program, and the Emerging Worlds Program, for support of this work.
We thank Noriko Kita and Travis Tenner for helpful conversations, and are grateful to Karim Shariff and Debanjan Sengupta for reviewing a draft version of the manuscript and their helpful comments and suggestions.
We also thank the journal reviewer for their detailed review and insightful questions that helped improve the paper.

\bibliographystyle{aasjournal}
\bibliography{bibliography}



\end{document}

%% file: concentration_tables.tex
\vspace{1em} 
  
\begin{tabular}{L{1cm}|R{1.5cm}R{1.5cm}} 
\multicolumn{3}{c}{(a) Settling and radial factors combined} \\ 
\hline 
\hline 
$r_p$ & $\alpha$=1e-04 & 3e-03 \\ 
\hline 
0.1\,cm & \textcolor{black}{13} & \textcolor{black}{10} \\ 
1.5\,cm & \textcolor{black}{50} & \textcolor{black}{13} \\ 
2.6\,cm & \textcolor{black}{72} & \textcolor{black}{16} \\ 
11\,cm & \textcolor{red}{200} & \textcolor{black}{38} \\ 
\hline 
\hline 
\end{tabular} 
  
\vspace{1em} 
  
\begin{tabular}{L{1cm}|R{1.5cm}R{1.5cm}|R{1.5cm}R{1.5cm}|R{1.5cm}R{1.5cm}} 
\multicolumn{7}{c}{(b) $\ell=500-1000$\,km} \\ 
\hline 
\hline 
 & \multicolumn{2}{c}{TC, $C$: $F_p(>C)$=50\%} & \multicolumn{2}{c}{TC, $C$: $F_p(>C)$=30\%} & \multicolumn{2}{c}{TC, $C$: $F_p(>C)$=1\%} \\ 
$r_p$ & $\alpha$=1e-04 & 3e-03 & 1e-04 & 3e-03 & 1e-04 & 3e-03 \\ 
\hline 
0.1\,cm & 1.0 & 1.1 & 1.1 & 1.2 & 1.5 & 2.0 \\ 
1.5\,cm & 1.4 & 2.5 & 2.1 & 4.6 & 7.3 & 32 \\ 
2.6\,cm & 1.9 & 3.1 & 3.3 & 6.2 & 17 & 56 \\ 
11\,cm & 2.6 & 2.6 & 4.9 & 5.1 & 37 & 41 \\ 
\hline 
\hline 
 & \multicolumn{2}{c}{total} & \multicolumn{2}{c}{total} & \multicolumn{2}{c}{total} \\ 
$r_p$ & $\alpha$=1e-04 & 3e-03 & 1e-04 & 3e-03 & 1e-04 & 3e-03 \\ 
\hline 
0.1\,cm & \textcolor{black}{13} & \textcolor{black}{11} & \textcolor{black}{15} & \textcolor{black}{12} & \textcolor{black}{20} & \textcolor{black}{20} \\ 
1.5\,cm & \textcolor{black}{72} & \textcolor{black}{33} & \textcolor{black}{110} & \textcolor{black}{61} & \textcolor{black}{360} & \textcolor{black}{430} \\ 
2.6\,cm & \textcolor{black}{140} & \textcolor{black}{51} & \textcolor{black}{240} & \textcolor{black}{100} & \textcolor{black}{1200} & \textcolor{black}{910} \\ 
11\,cm & \textcolor{red}{510} & \textcolor{black}{99} & \textcolor{red}{980} & \textcolor{black}{190} & \textcolor{red}{7400} & \textcolor{black}{1600} \\ 
\hline 
\hline 
\end{tabular} 
  
\vspace{1em} 
  
\begin{tabular}{L{1cm}|R{1.5cm}R{1.5cm}|R{1.5cm}R{1.5cm}|R{1.5cm}R{1.5cm}} 
\multicolumn{7}{c}{(c) $\ell=50-100$\,km} \\ 
\hline 
\hline 
 & \multicolumn{2}{c}{TC, $C$: $F_p(>C)$=50\%} & \multicolumn{2}{c}{TC, $C$: $F_p(>C)$=30\%} & \multicolumn{2}{c}{TC, $C$: $F_p(>C)$=1\%} \\ 
$r_p$ & $\alpha$=1e-04 & 3e-03 & 1e-04 & 3e-03 & 1e-04 & 3e-03 \\ 
\hline 
0.1\,cm & 1.1 & 1.2 & 1.3 & 1.7 & 2.2 & 4.5 \\ 
1.5\,cm & 2.9 & 3.5 & 5.7 & 7.4 & 47 & 80 \\ 
2.6\,cm & 3.3 & 3.6 & 6.8 & 7.5 & 69 & 83 \\ 
11\,cm & 2.7 & 2.7 & 5.1 & 5.1 & 41 & 42 \\ 
\hline 
\hline 
 & \multicolumn{2}{c}{total} & \multicolumn{2}{c}{total} & \multicolumn{2}{c}{total} \\ 
$r_p$ & $\alpha$=1e-04 & 3e-03 & 1e-04 & 3e-03 & 1e-04 & 3e-03 \\ 
\hline 
0.1\,cm & \textcolor{black}{14} & \textcolor{black}{12} & \textcolor{black}{17} & \textcolor{black}{17} & \textcolor{black}{29} & \textcolor{black}{46} \\ 
1.5\,cm & \textcolor{black}{140} & \textcolor{black}{47} & \textcolor{black}{290} & \textcolor{black}{100} & \textcolor{black}{2400} & \textcolor{black}{1100} \\ 
2.6\,cm & \textcolor{black}{240} & \textcolor{black}{58} & \textcolor{black}{490} & \textcolor{black}{120} & \textcolor{black}{4900} & \textcolor{black}{1400} \\ 
11\,cm & \textcolor{red}{530} & \textcolor{black}{100} & \textcolor{red}{1000} & \textcolor{black}{190} & \textcolor{red}{8200} & \textcolor{black}{1600} \\ 
\hline 
\hline 
\end{tabular}